\documentclass[conference]{IEEEtran}
\IEEEoverridecommandlockouts
\def\BibTeX{{\rm B\kern-.05em{\sc i\kern-.025em b}\kern-.08em
    T\kern-.1667em\lower.7ex\hbox{E}\kern-.125emX}}
\usepackage{cite}
\usepackage[T1]{fontenc}
\usepackage{graphicx}
\usepackage[utf8]{inputenc}
\usepackage{amsmath,amssymb,amsfonts,bm}
\usepackage{epsfig}
\usepackage{mwe}
\usepackage{lipsum}
\usepackage{multicol}
\usepackage{graphicx}
\usepackage{graphics}
\usepackage{xcolor}
\usepackage{balance}
\usepackage{textcomp}
\usepackage{comment}
\usepackage{algorithm}
\usepackage[noend]{algpseudocode}
\algblockdefx[Foreach]{Foreach}{EndForeach}[1]{\textbf{for each} #1 \textbf{do}}{\textbf{end for}}
\usepackage[bf]{subfigure}

\usepackage{enumitem}
\usepackage{times}
\setlist{nolistsep,leftmargin=*}
\def\fm{\textsc{Km}\xspace}
\def\fg{\textsc{FoodGraph}\xspace}
\def\fmplus{\textsc{FoodMatch}\xspace}
\DeclareMathOperator*{\argmin}{argmin}
\DeclareMathOperator{\atantwo}{atan2}

\algnewcommand{\IIf}[1]{\State\algorithmicif\ #1\ \algorithmicthen}
\algnewcommand{\EndIIf}{\unskip\ \algorithmicend\ \algorithmicif}
\newtheorem{theorem}{\textbf{Theorem}}
\newtheorem{definition}{\textbf{Definition}}

\newtheorem{example}{\textbf{Example}}
\newtheorem{lemma}{\textbf{Lemma}}

\newtheorem{problem}{\textbf{Problem}}

\def\CV{\mathcal{V}}

\def\maxload{\textsc{MaxI}\xspace}
\def\maxorders{\textsc{MaxO}\xspace}

\def\@secnumfont{\bfseries}
\begin{document}

\title{Batching and Matching for Food Delivery in Dynamic Road Networks
}

\author{
    \IEEEauthorblockN{Manas Joshi\IEEEauthorrefmark{3}, Arshdeep Singh\IEEEauthorrefmark{1}, Sayan Ranu\IEEEauthorrefmark{1}, Amitabha Bagchi\IEEEauthorrefmark{1}, Priyank Karia\IEEEauthorrefmark{2}, Puneet Kala\IEEEauthorrefmark{2}}

   \IEEEauthorblockA{\IEEEauthorrefmark{3}Dept. of Computer Science and
	Engineering, IIT, Delhi, India. manasjoshi241@gmail.com}
	\IEEEauthorblockA{\IEEEauthorrefmark{1}Dept. of Computer Science and
	Engineering, IIT, Delhi, India.
   \{cs5160625, sayanranu, bagchi\}@iitd.ac.in}
    		\IEEEauthorblockA{\IEEEauthorrefmark{2}Swiggy, Bangalore, India.
    \{puneet.k,priyank.karia\}@swiggy.in}
}

\maketitle

\begin{abstract}
Given a stream of food orders and available delivery vehicles, how should orders be assigned to vehicles so that the delivery time is minimized? Several decisions have to be made: (1) assignment of orders to vehicles, (2) grouping orders into batches to cope with limited vehicle availability, and (3) adapting to dynamic positions of delivery vehicles. 
 We show that the minimization problem is not only \emph{NP-hard} but \emph{inapproximable} in polynomial time. To mitigate this computational bottleneck, we develop an algorithm called \fmplus, which maps the vehicle assignment problem to that of \emph{minimum weight perfect matching} on a bipartite graph. To further reduce the quadratic construction cost of the bipartite graph, we deploy best-first search to only compute a subgraph that is highly likely to contain the minimum matching. The solution quality is further enhanced by reducing batching to a graph clustering problem and anticipating dynamic positions of vehicles through \emph{angular distance}. 
 Extensive experiments on food-delivery data from large metropolitan cities establish that \fmplus is substantially better than baseline strategies on a number of metrics, while being efficient enough to handle real-world workloads.

\end{abstract}

\section{Introduction and Related Work}
\label{sec:introduction}
The market for food delivery today is valued at above \texteuro{83} billion~\cite{mckinsey}. 
Several companies, such as Uber Eats, Takeaway.com, Swiggy, Zomato, etc. have a valuation above $1$ billion USD~\cite{forbes,toi}. 
Once the customer places an order, the food delivery process unfolds in four discrete steps:  {\bf (1)} assigning a delivery vehicle to an order (\emph{assignment time}), 
{\bf (2)} a delivery vehicle driving to the restaurant to pick up the order (\emph{first mile time}), {\bf (3)} preparation and packaging of the food by the restaurant (\emph{preparation time}) and {\bf (4)} dropping off the order at the customer's location (\emph{last
mile time}). Note that the food preparation time can progress in parallel with the first mile and assignment time. 
Mathematically, the time to deliver an order is expressed as follows.

\vspace{-0.20in}
\begin{alignat}{2}
\label{eq:dt}
\nonumber
\text{Delivery Time}=&\max \{\text{assignment time $+$ first mile time},\\ 
&\text{preparation time}\}+\text{last mile time}
\end{alignat}
\vspace{-0.15in}

Naturally, minimizing the delivery time is key to ensuring a positive customer experience. This minimization problem, however, is composed of several challenging subproblems.
\begin{itemize}
\item \textbf{Vehicle assignment: } Given a stream of unallocated orders and available vehicles, what is the best order assignment policy to vehicles so that the delivery time is minimized? 
\item \textbf{Batching: } A vehicle may have the capacity to carry multiple orders. How should we \emph{batch} orders so that they can be allocated to the same vehicle?

\item \textbf{Scalability: }
It is critical that the computation time of assigning orders to vehicles is faster than the rate at which new orders arrive. The proposed minimization problem is \emph{NP-hard} and \emph{inapproximable}. Can we design a heuristic that is efficient enough to be deployable on real-world load volumes and yet generate effective order assignments?
\end{itemize}

 \vspace{-0.05in}
 \subsection{Limitations of Existing Work}
 \label{sec:related}
 Yildiz et al.~\cite{yildiz_prove} propose an exact algorithm under the unrealistic assumption of perfect information about the arrival of orders. In reality, orders and vehicles come as a data stream. An approach applicable on data streams was proposed by Reyes et al.~\cite{mdrp}. However,  several simplifying assumptions are made that are not realistic. First, the distance between a source and  a destination is computed using the Haversine distance between the corresponding latitude and longitude values. In reality, the delivery time is dependent on the  road  network  distance. This assumption not only results in unrealistic distances, it also ignores  the  scalability  challenges  associated  with  computing shortest (or quickest) path distances in road networks. Second, batching  is  allowed  only  if  two  orders  are  from  the  same restaurant. Third, it lacks a comprehensive evaluation on real, high-volume food delivery data. 

Apart from the two papers mentioned above, there is little work that is closely aligned to ours. Ji et. al.~\cite{jiwww} focus on the problem of efficient geographical batching of orders for food delivery but do not solve the assignment and routing problem. 
Zeng et al.~\cite{last_mile} propose a solution for last mile delivery in a generic setting where there is no concept of food preparation time. Dai et. al.~\cite{mixed_drivers} focus on leveraging crowdsourcing to optimise the cost of delivery. 
 
\textbf{Connections to ride-sharing: } The need to assign vehicles to customer orders and batch multiple orders in a single vehicle also arises in the ride-sharing problem for taxi service~\cite{yuen-www:2019,vldbrideshare,share1,rp_insert1,rp_insert2,mdm,didi2}. Although ride-sharing has been extensively studied in the data mining community, there are several intricacies that are specific to food delivery, and thus techniques for ride-sharing do not transfer to our problem.

\begin{figure}[t]
\vspace{-0.20in}  
\centering
  \includegraphics[width=3.32in]{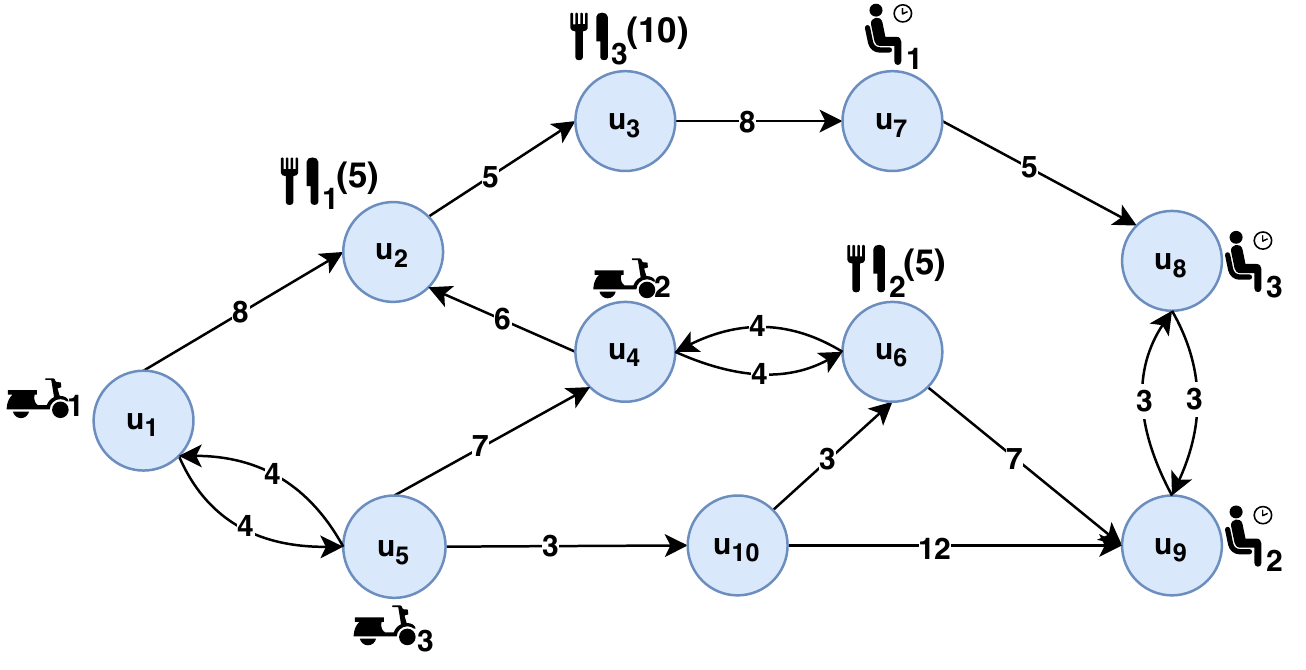}
\vspace{-0.15in}  
  \caption{An instance of the food delivery problem. The graph represents the road network. Edge weight represents road segment traversal time. Subscripts refer to order ID and vehicle ID. Food preparation time is in parentheses. The figure shows three order pick-up and drop-off points and three vehicles.}
  \label{fig:ex1}
\vspace{-0.20in}  
\end{figure}

 \textbf{1. Waiting time:} In ride-sharing, once a customer requests a cab, it is important to send a cab to the customer as soon as possible; otherwise, the customer's wait duration increases and the customer experience suffers. Thus, sending the nearest available cab is often a good solution. In food-delivery, this is often not the case. To elaborate, consider order 2 in Fig.~\ref{fig:ex1} where the food needs to be picked up from the restaurant in $u_6$ and delivered to the customer at $u_9$. The closest vehicle is $v_2$ at $u_4$. However, if $v_2$ is sent to $u_6$, it will incur a waiting time of $1$ minute, since the food preparation time is $1$ minute longer than the time it would take for the vehicle to reach the restaurant. Any idle time incurred by a vehicle is a loss of productivity and a better solution would be to send $v_3$. 

\textbf{2. Optimization function: }In ride-sharing the optimization function is proportional to the distance travelled by a customer. In food-delivery all deliveries typically incur the same delivery charge since only those restaurants are shown to a user that are within a distance radius. Thus, the optimization function is to enhance customer experience in terms of promptness of delivery rather than the revenue. 

\textbf{3. Serviceability: }In food-delivery, once an  order is placed, serviceability is non-negotiable; all orders must be delivered within a promised time limit (typically $45$ minutes). In ride-sharing, although not desirable, refusing cab to a customer due to low availability is one of the possible action choices.
\vspace{-0.05in}
\subsection{Contributions}
The above discussion highlights the need for a specialized and realistic algorithm for food delivery.  Our proposed technique \fmplus addresses these needs. Specifically, we make the following key contributions.
\begin{itemize}
\item We formulate the problem of minimizing food delivery time in road networks. This minimization problem is not only \emph{NP-hard} but also \emph{inapproximable}~(\S~\ref{sec:problem}).
\item To ensure scalability and high solution quality, we frame order assignment as a \emph{minimum weight perfect matching} problem on a bipartite graph. To overcome the quadratic cost of constructing the bipartite graph, we use {\em best-first search} to extract a subspace of the search space that is highly likely to contain the minimum weight matching. In addition, \fmplus incorporates several novel concepts such as \emph{batching} of orders through a delivery-optimized distance function, and \emph{angular distance} to adapt to dynamic locations of delivery vehicles (\S~\ref{sec:fm}). 
\item Through extensive experiments on food-delivery data provided by Swiggy, India's largest food-delivery company~\cite{swiggy}, we establish that \fmplus is scalable for real-world workloads in the largest cities of India. More importantly, \fmplus 
 imparts $30\%$ reduction in food delivery time when compared to baseline strategies (\S~\ref{sec:exp}). To the best of our knowledge, this is the first large-scale empirical evaluation of food-delivery strategies on real data.

\item As a by-product of our work, we release the first real, large-scale food-delivery data for further academic research.
\end{itemize}
 
\begin{table}[t]
    \centering
    \vspace{-0.20in}
    \caption{Summary of notations used.}
    \label{tab:notations}
    \vspace{-0.10in}
    {\scriptsize
    \begin{tabular}{|p{0.8in}|p{2.4in}|}
    \hline
         \textbf{Notation}&\textbf{Meaning}  \\
         \hline
         $\beta(e)$& Temporal edge weight of edge $e$ in road network (See Def.~\ref{def:road-network}).\\
         \hline         
         Order $o=\langle o^r, o^c, o^t, o^i, o^p\rangle$& See Def.~\ref{def:order}.\\
         \hline
         $O^v_t$ & Orders assigned to vehicle $v$ at time $t$.\\
        \hline
         A(o)& See Def.~\ref{def:assignment}.\\
         \hline
         \textsc{MaxO} and \textsc{MaxI} & Maximum number of orders and items that can be assigned to a vehicle respectively.\\
         \hline
         $\Omega$ & Rejection Penalty (See Prob\ref{prb:online}). \\
         \hline
         $\Delta$ & Length of accumulation window.\\
         \hline
         $O(\ell)$ and $\CV(\ell)$ & The unassigned orders and available vehicles at the end of the $\ell^{th}$ accumulation window.\\
         \hline
         $\pi$ and $\pi[1]$& $\pi$ represents a batch of orders and $\pi[1]$ represents the first order in $\pi$ that will be picked up in the quickest route plan.\\
         \hline
         $\eta$ & Terminating threshold for batching (See \S~\ref{sec:batch_hac})\\
         \hline
         $\gamma$& Weighing factor between angular distance and temporal distance (See Eq.~\ref{eq:dist}).\\
         \hline
    \end{tabular}}
    \vspace{-0.20in}
\end{table}
\vspace{-0.05in}
\section{Problem Formulation}
\label{sec:problem}
In this section, we define the preliminary concepts, formulate the problem and analyze its computational complexity. All notations used in our work are summarized in Table~\ref{tab:notations}.
\begin{definition}[Road Network] 
\label{def:road-network}
\textit{A road network is a weighted directed graph $G = (V, E , \beta)$, where $V$ is the set of nodes representing intersections of road segments, $E = \{(u, v): u,v \in V\}$ is the set of directed edges representing road segments and $\beta : (E,t) \mapsto \mathbb{R}^+$ is a function mapping each edge to its weight at time $t$. The weight of an edge $e \in E$ at time $t$ is the time required to traverse the corresponding road segment at that time of the day.}
\end{definition}

 The edge weights can either be fetched from Google Maps API, or extracted from the GPS pings of vehicles operated by the food delivery company. 

\begin{definition}[Food Order] 
\label{def:order}
\textit{Given a road network $G = (V,E, \beta)$, a {\em food order} $o$ is a tuple $\langle o^r, o^c, o^t, o^i, o^p\rangle$, where $o^r\in V$ is the restaurant location (pick-up node), $o^c \in V$ is the customer location (drop-off node), $o^t$ is the time of request, $o^i$ are the number of items associated with the order and $o^p$ is the (expected) preparation time for the order.}
\end{definition}

A food delivery company employs \emph{delivery vehicles} to pick-up and deliver orders. We use the notation $O^v_t$ to denote the orders assigned to vehicle $v$ at time $t$. In addition, $loc(v,t)$ denotes the node at which $v$ is located at time $t$. A vehicle may not be positioned exactly at a node. In such situations, we approximate its location to the closest node in the road network. 
 We assume all vehicles have a maximum carrying capacity of $\maxload$. Furthermore, at any time, no more than $\maxorders$ orders can be allocated to the same vehicle. 
To deliver the orders in $O^v_t$, $v$ follows the \emph{quickest route plan}.
\begin{definition}[Route Plan] 
\label{def:rp}
\textit{Given a road network $G = (V,E,\beta)$ and a set of orders $O^v_t = \{o_1,\ldots, o_m\}$, a route plan is a permutation of $\{o_i^r, o_i^c: 1 \leq i \leq m\} \subset V$ such that for each $i$, $o_i^r$ appears before $o_i^c$ in the permutation. }
\end{definition}

More simply, a route plan is a sequence of pick-up and drop-off nodes to fulfil the set of orders in $O^v_t$. In this work, we assume that any travel from one node to another follows the shortest (quickest) path. We use the notation $SP(u_i,u_{i+1},t)$ to denote the length of the \emph{shortest (quickest) path} from node $u_i$ to $u_{i+1}$ at time $t$. Thus, the \emph{length} of a route plan $RP=\{u_1,\cdots,u_m\}$ 
 is $\sum_{i=1}^{m-1} SP(u_i,u_{i+1},t)$. The \emph{quickest} route plan is the one with the smallest length among all feasible route plans for $O^v_t$. Any vehicle $v$ always follows the quickest route plan for its assigned order set $O^v_t$. Owing to this formulation, hereon, whenever we refer to a route plan for a vehicle, we assume it to be quickest route plan.
 
 We note that the number of feasible route plans grows exponentially with $|O^v_t|\leq\maxorders$. However, $\maxorders$ is typically small since vehicles in the food delivery paradigm tend to be riders on motorcycles or bicycles, e.g., Swiggy allows at most 3 orders per rider. Consequently, it is computationally feasible to try all permutations and compute the quickest route plan.

\begin{definition}[Order assignment] 
\label{def:assignment}
\textit{Given a set of unassigned orders $O$ and available vehicles $\CV$, an order assignment function $A: O\rightarrow \CV$ specifies the vehicle $v\in\CV$ to which $o\in O$ has been assigned. 
Order $o$ may be assigned to $v$ only if it satisfies the following properties: }
\begin{itemize}
\item \textit{$v$ has the capacity to carry $o$, i.e., $\sum_{\forall o_j\in O^v_t} o_j^i + o^i\leq \maxload$}.
\item $|O^v_t| < \maxorders$.
\end{itemize}
\end{definition}

If $o$ violates any of these constrains for every available vehicle then it is ``unassigned'' and $A(o)=NULL$. Furthermore, any assignment algorithm consumes some finite amount of time to map orders to vehicles. We use the notation $time(A(o))$ to denote the computation time of $A(o)$.

  Suppose $o$ is assigned to vehicle $v$, i.e. $A(o)=v$. 
  $firstMile(o,v)$ of order $o$ is the (temporal) distance from $v$'s current location $loc(v,o^t)$ to the restaurant pick-up location $o^r$ in the route followed by $v$. Similarly, the $lastMile(o,v)$ is the distance from $o^r$ to drop-off location $o^c$.
 
\begin{example}
\label{ex:firstmile}
\textit{Let us consider Fig.~\ref{fig:ex1}. Suppose vehicle $v_1$, located at $u_1$, has been assigned to order $o_1$. $o_1$ needs to be picked up from the restaurant at $u_2$ and dropped-off at $u_7$. The quickest route for this task is $RP=\{u_1,u_2,u_3,u_7\}$.  
Thus, $firstMile(o_1,A)=8$ and $lastMile(o_1,A)=13$.} 
\end{example}

\begin{definition}[Expected Delivery Time]
\label{def:edt}
\textit{The expected delivery time of order $o$ when assigned to vehicle $v$:}
\vspace{-0.05in}
\begin{alignat}{2}
\nonumber
EDT(o,A(o))&=EDT(o,v)\\
\nonumber
&=\max\left\{ time(A(o))+firstMile\left(o,v\right),o^p\right\}\nonumber \\
\label{eq:edt}
&+lastMile\left(o,v\right)
\end{alignat}
\vspace{-0.05in}
\end{definition}
\vspace{-0.20in}

$EDT(o,v)$ adds mathematical preciseness to Eq.~\ref{eq:dt}.
\begin{example}
\label{ex:edt}
\textit{For simplicity, we assume $time(A(o))$ for all orders in Fig.~\ref{fig:ex1}. Continuing from Ex.~\ref{ex:firstmile}, $EDT(o_1,v_1)=\max\{8,5\}+13=21$. On the other hand, if $A(o_2)=v_2$ the quickest route plan is $\{u_4,u_6,u_9\})$, and thus $EDT(o_2,v_2)=\max\{4,5\}+7=12$.}
\end{example}
\begin{definition}[Shortest Delivery Time]
\textit{The {\em shortest delivery time} for an order $o$ is $SDT(o)= o^p + SP(o^r, o^c, o^t)$.}
\end{definition}

$SDT(o)$ is achieved only if a vehicle $v$ is already waiting at $o^r$ or arrives just in time before the food is prepared. It serves as a natural lower bound on the expected delivery time. 
\begin{definition}[Extra Delivery Time (XDT)] \textit{The XDT of an order $o$ under a specific assignment $A$ is the difference between the time taken to deliver an order and the shortest delivery time, i.e., $XDT(o,A(o))=EDT(o,A(o)) - SDT(o)$.}
\end{definition}
\begin{example}
\label{ex:xdt}
\textit{Continuing from Ex.~\ref{ex:firstmile} and Ex.~\ref{ex:edt}, extra delivery times of $o_1$ and $o_2$ are $3$ and $0$ respectively. }
\end{example}

With the above definitions in place, our goal is now to minimize the aggregate extra delivery time. Note that it may not be possible to assign all orders to vehicles while satisfying the constraints in Def.~\ref{def:assignment}. 
In our formulation such rejected orders incur a large \emph{rejection penalty $\Omega$}. 
\begin{problem}[The food delivery problem (FDP)] 
\label{prb:online} 
\textit{Given a stream of orders $O$ and a stream of available vehicles $\CV$, if $\mathcal{A}$ is the set of all possible feasible assignments of $O$ to $\CV$, give an algorithm to find:}
\vspace{-0.05in}
\begin{align}
\label{eq:objective}
\nonumber
    \argmin_{A\in \mathcal{A}} & \sum_{o \in O} XDT(o,A(o))(1-\delta_{o}(A)) + \Omega\delta_{o}(A)
\end{align}
\vspace{-0.05in}

\textit{Here, $\delta_o(A)=1$ if $A(o)=NULL$, i.e., $o$ is rejected; otherwise, $\delta_o(A)=0$. $\Omega$ is a large positive constant.}
\end{problem}
When there are no rejections, minimizing the extra delivery time is equivalent to minimizing total expected delivery time.

\vspace{-0.05in}
\subsection{Problem Hardness}
\label{sec:nphard}
Problem~\ref{prb:online} is not just NP-hard but inapproximably hard. This follows from a result of Krumke~\cite{krumke:2002} on the inapproximability of the online Dial-a-Ride problem.
\begin{definition}[Single Server Online Dial-a-Ride with Flow time minimization (SS-OL-DARP-F)]
\label{def:darp}
\textit{
Suppose we are given an undirected, connected edge-weighted graph, $G= (V,E)$ with edge weight function $w$ representing the time taken to commute across an edge. An instance of SS-OL-DARP-F is a set of {\em orders} $\{(t_i, u_i, v_i) : i \geq 0\}$ where $t_i$ is the time of release of the $i^{th}$ order, $u_i \in V$ is the source of the $i^{th}$ order and $v_i$ is its destination. We assume that $t_i \leq t_{i+1}$ for all $i \geq 0$ and that order $i$ is made available as input at time $t_i$.  
}

\textit{
A single server has to fulfill all orders. An order $(t_i, u_i, v_i)$ is fulfilled when an object is picked up from $u_i$ at any time $t \geq t_i$ and delivered at $v_i$. Every object is assumed to be of the same size and the server is constrained to carry at most $C$ objects at a time for some $C > 0$. A schedule $\sigma$ is a sequence of moves made by the server on the graph in order to fulfill the received orders. Given a schedule $\sigma$ if the delivery time of the $i$th order is $\sigma_i$ then SS-OL-DARP-F seeks to minimize the aggregate flow time, i.e.,  $\sum_{i} \sigma_i - t_i$.}
\end{definition} 
From Def.~\ref{def:darp}, we deduce that for the food delivery problem, if we assume we have only one delivery vehicle and all food preparation times are 0, online FDP is exactly SS-OL-DARP-F, i.e., given an instance of SS-OL-DARP-F, we can create an instance of FDP by mapping every order $(t_j, u_j,v_j)$ of SS-OL-DARP-F to a food order $o_j$ of FDP such that $o_j^r = u_j$, $o_j^c = v_j$, $o_j^t = t_j$, $o_j^i = 1$ and $o_j^p = 0$.
Thus, FDP is a generalization of SS-OL-DARP-F. Hence, if SS-OL-DARP-F is inapproximable, so is online FDP. Krumke~\cite{krumke:2002} established the inapproximability of SS-OL-DARP-F. We state this inapproximability as a theorem with respect to our problem.

\begin{theorem}[Krumke~\cite{krumke:2002}]
\label{thm:inapproximable}
\textit{For any $\varepsilon > 0$, no polynomial time algorithm exists that can achieve an approximation ratio of $n^{1/2 - \varepsilon}$ for the FDP with order set $O$ such that $|O| = n$. }
\end{theorem}

Since no polynomial time algorithm for the FDP can achieve a reasonable approximation ratio we investigate heuristics.

\vspace{-0.05in}
\section{Baseline: The Greedy Approach}
\label{sec:baselines}

\label{sec:agreedy}
Given a stream of orders, we accumulate them over an \emph{accumulation window} of temporal length $\Delta$. After the accumulation period ends, all unassigned orders are allocated to available vehicles in a greedy manner, following which the processing of the next accumulation begins.

Let $O(\ell)$ be the set of unallocated orders at the end of the $\ell$-th accumulation window and $\CV(\ell)$ be the set of active vehicles. The greedy assignment policy picks the unassigned order-vehicle pair with the minimum \emph{marginal cost}. 
\begin{definition}[Marginal Cost]
\label{def:marginal}
\textit{Let $o$ be a new order that arrived at time $t$. The \emph{marginal cost} of assigning $o$ to vehicle $v$ is the increase in the extra delivery time of all orders already assigned to $v$.} 
\vspace{-0.10in}
\begin{alignat}{2}
mCost(o,v) =  Cost(v,O^v_t\cup\{o\}) - Cost(v,O^v_t)\\
\label{eq:cost}
\text{where, }Cost(v,O^v_t) = \sum_{\forall o\in O^v_t} XDT(o,v)
\end{alignat}
\vspace{-0.15in}
\end{definition}
\begin{example} \textit{Revisiting Fig.~\ref{fig:ex1}, $Cost(v_1,\{o_1\})=3$ (Recall Ex.~\ref{ex:xdt}) and $Cost(v_1,\emptyset)=0$. Thus, $mCost(o_1,v_1)=3$.} 
\end{example}
Greedy iteratively selects the best unassigned order-vehicle pair $(o^*,v^*)$ till no further assignments are possible. 
\vspace{-0.10in}
\[ (o^*,v^*) = \argmin_{(o,v) \in O(\ell)\times \CV(\ell)} mCost(o,v)\]
\vspace{-0.10in}

In each iteration, Greedy assigns $A(o^*)=v^*$.

\begin{example}
\textit{Let us revisit Fig.~\ref{fig:ex1}. The instance shows three available vehicles $v_1$, $v_2$ and $v_3$ and three unallocated orders $o_1$, $o_2$ and $o_3$ long with their preparation times. 
The greedy algorithm first assigns vehicle $v_2$ to order $o_2$ since it has the least marginal cost of $0$. Next, the algorithm assigns order $o_1$ to vehicle $v_1$ at a cost of $3$ units and finally assigns order $o_3$ to vehicle $v_1$ with a route plan $(o_1^r, o_3^r, o_1^c, o_3^r)$ for another $3$ units. This results in an overall cost of 6 units. }
\end{example}

\textbf{Limitations: }Greedy is prone to producing sub-optimal solutions due to making \emph{locally optimal} choices. For example, in Fig.~\ref{fig:ex1}, we can achieve better costs by performing an one-to-one assignment as $v_2$ to $o_1$, $v_3$ to $o_2$ and $v_1$ to $o_3$ for an overall cost of $5$ units. Here, we choose a \emph{suboptimal} assignment for $v_2$ to achieve an overall assignment with a better cost. 

\textbf{Time complexity: } Since we have to look at all permutations of at most $2 \maxorders$ locations to find an optimal route plan $\mathcal{O}(\maxorders\cdot(\maxorders!)\cdot q)$ time where $\mathcal{O}(q)$ is the time taken for shortest path computation.  We compute $mCost(o,v)$ for order-vehicle pair which takes $\mathcal{O}\left(\maxorders\cdot\maxorders!\cdot qmn\right)$ time where $m=|\CV(\ell)|$ and $n=|O(\ell)|$. This is the first round of the algorithm. Once $(o^*, v^*)$ has been found, we need to recompute the $mCost(o,v^*)$ for the remaining orders viz-a-viz $v^*$ before an optimal can be found in the next round and so on till the orders are all assigned. Hence the total time is $\mathcal{O}\left(\maxorders\cdot(\maxorders!)\cdot q(mn +n^2)\right)$. The $\mathcal{O}$ notation suppresses a logarithmic term required to find the min marginal cost at each step. If Dijkstra's algorithm is used $O(q)=O(|E|+|V|\log(|V|))$. However, through index structures~\cite{hhl}, this cost can be significantly lower in practice.	

\textbf{Selecting $\Delta$:} Increasing $\Delta$ has both  positive and negative impact on the quality and scalability. In terms of quality, the negative impact comes from the fact that an order remains unallocated till the end of the window. Thus, the assignment time increases with increase in $\Delta$. On the other hand, since we accumulate more data and then the assignment happens, the minimum marginal costs have a high likelihood of being smaller. In terms of scalability, the cost per window goes up. However, we need to process fewer windows. Furthermore, in each window, we need to fetch the current location of all vehicles. With fewer of windows, this cost reduces. In general, it is hard to provide theoretical guarantees on the quality or scalability with respect to $\Delta$. We revisit this question again during our empirical evaluation in \S~\ref{sec:exp}.

\section{\fmplus} 
\label{sec:fm}
As illustrated in the previous section, locally optimal choices by Greedy may not converge to the globally optimal solution. Rather, we should minimize the \emph{cumulative} marginal costs across all assignments. To achieve this objective, we organize the assignment space in the form of a \emph{weighted bipartite graph}, which we call the \fg. On the \fg, we compute the \emph{minimum weight perfect matching}.

\vspace{-0.05in}
\subsection{Matching in \fg}
\label{sec:bp}
\fg is constructed as follows. Let $O(\ell)=\{o_1, \ldots, o_k\}$ be the set of unassigned orders at the end of the $\ell^{th}$ accumulation window, and $\CV(\ell)$ the set of all active vehicles. The two partitions of the \fg are the {\em order partition} $U_1=O(\ell)$ and the {\em vehicle partition} $U_2 = \CV(\ell)$. 
We add an edge between $o \in U_1$ and $v\in U_2$ with weight $w(o,v)$.

\vspace{-0.20in}
\begin{equation}
\nonumber
w(o,v)=
\begin{cases}
\Omega &\text{if $(o,v)$ violates the}\\
       &\text{constraints of Def.~\ref{def:assignment}}\\
min(mCost(o,v), \Omega) & \text {otherwise}
\end{cases}
\end{equation}

On this \fg, we compute the minimum weight perfect matching through the following minimization problem:
\vspace{-0.15in}
\begin{alignat}{3}
\nonumber
\text{Minimize }&\sum_{\forall o,v} w(o,v)\cdot x_{o,v}\\
\nonumber
\text{Subject to }& \sum_o x_{o,v} \leq 1& o\in U_1\\
\nonumber
& \sum_v x_{o,v} \leq 1& v\in U_2 \\
\nonumber
& \sum_{\forall o,v} x_{o,v}=min(|U_1|,|U_2|) &
\end{alignat}

where $x_{o,v}=1$ if $o$ is assigned to $v$; otherwise, $x_{o,v}=0$.

Minimum weight perfect matching in a bipartite graph is computed through Kuhn-Munkres algorithm~\cite{km1}. 

\begin{example}
\label{ex:fm}
\textit{Fig.~\ref{fig:fullgraph} shows the bipartite graph formed for the problem instance shown in Fig.~\ref{fig:ex1}. The solid edges indicate the mappings that will be selected by Kuhn-Munkres algorithm, i.e. $x_{o,v}=1$. The cumulative cost of the assignments is $5$ units, which is $1$ unit better than Greedy.}
\end{example}

\noindent
\textbf{Time Complexity:} 
Let $n=|O(\ell)|$ and $m=|\CV(\ell)|$.
As discussed in the analysis of Greedy, computing $mCost(o,v)$ consumes $\mathcal{O}(\maxorders\cdot(\maxorders!)q)$ time. Constructing the complete bipartite graph requires  $\mathcal{O}(nm\maxorders\cdot(\maxorders!)q)$ time. Let $k_{\bot}=\min(n,m)$ and $k_{\top}=\max(n,m)$. Kuhn-Munkres algorithm computes a maximum weighted matching in $\mathcal{O}(k_{\top}^2 k_{\bot})$ time, giving an overall complexity of $\mathcal{O}(nm\maxorders\cdot(\maxorders!)q + k_{\top}^2 k_{\bot})$.

\noindent
\textbf{Limitations: }Although better than Greedy, there is scope to improve further.
\begin{itemize}
\item \textbf{Batching: } By definition, no two edges will be incident on the same node in a minimum weight perfect matching. Hence, batching is not feasible. Furthermore, if $|\CV(\ell)|<|O(\ell)|$, $|O(\ell)|-|\CV(\ell)|$ orders would remain unassigned. 

\begin{figure}
\centering
\vspace{-0.20in}
\includegraphics[width=0.4\linewidth]{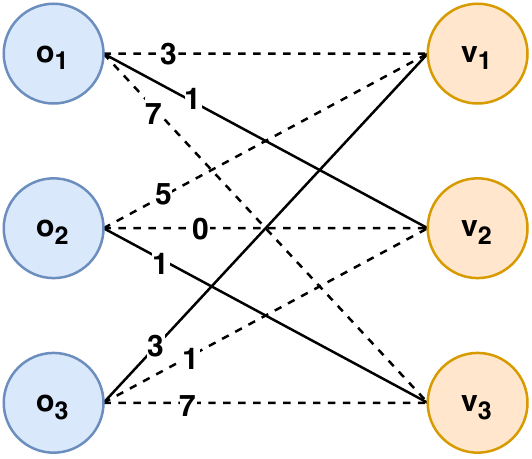}
\vspace{-0.15in}
    \caption{The bipartite graph formed for the instance in Fig.~\ref{fig:ex1}.
    The cost of each edge is shown as edge weight. Bold lines indicate the assignment found by Kuhn-Munkres algorithm.}
   \label{fig:fullgraph}
\vspace{-0.20in}
\end{figure}

\item \textbf{Scalability: }The $\mathcal{O}(nm\maxorders\cdot(\maxorders!)q + k_{\top}^2 k_{\bot})$ time complexity is too expensive to handle real-world workloads. 

\item \textbf{Dynamic Environment: }The matching process assumes that the marginal costs (edge weights in \fg) remain static while the minimization problem is being solved. In reality, it is a dynamic environment where the vehicles are moving. Thus, the marginal costs may become stale by the time the assignments are made.
\end{itemize}

In the subsequent sections, we address these limitations.

\begin{figure*}[t]
    \centering
\vspace{-0.20in}
    \subfigure[]{\label{fig:order_graph_1}\includegraphics[width=2.1in]{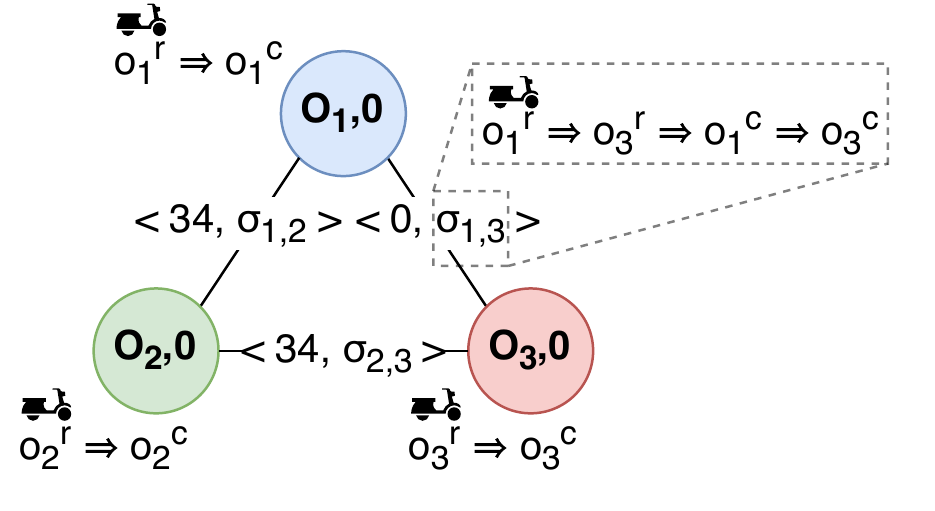}}
    \hfill 
    \subfigure[]{\label{fig:order_graph}\includegraphics[width=1.2in]{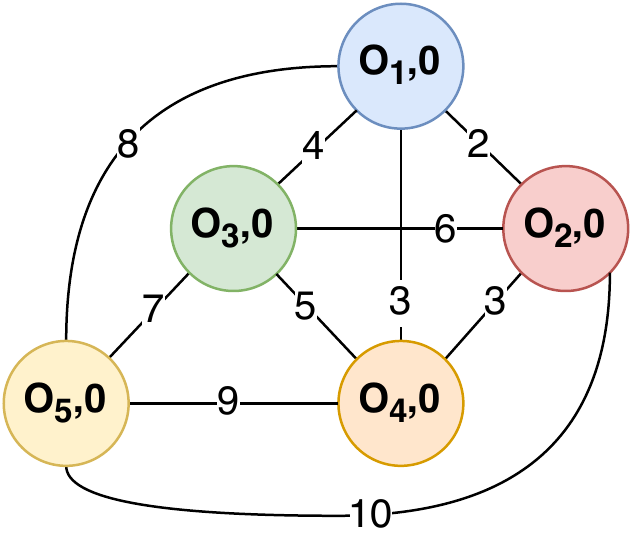}}
    \hfill 
    \subfigure[]{\label{fig:cl1}\includegraphics[width=1.2in]{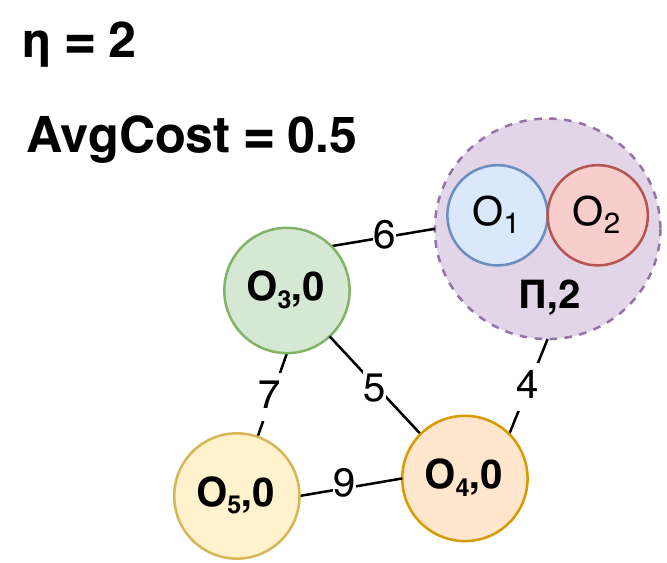}}
    \hfill 
    \subfigure[]{\label{fig:cl2}\includegraphics[width=1.1in]{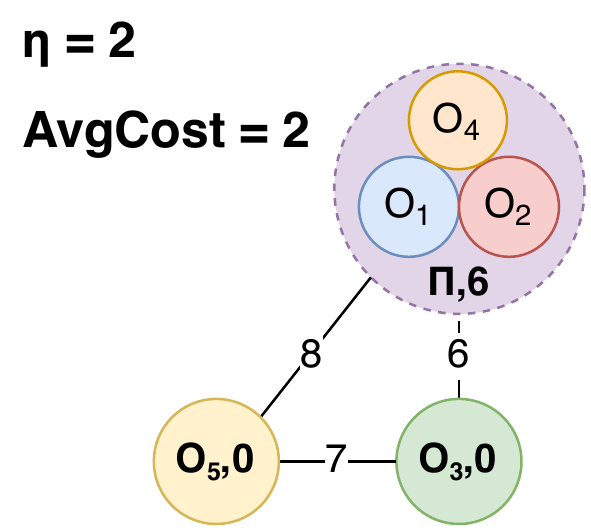}}
    \hfill 
    \subfigure[]{\label{fig:cl3}\includegraphics[width=1.1in]{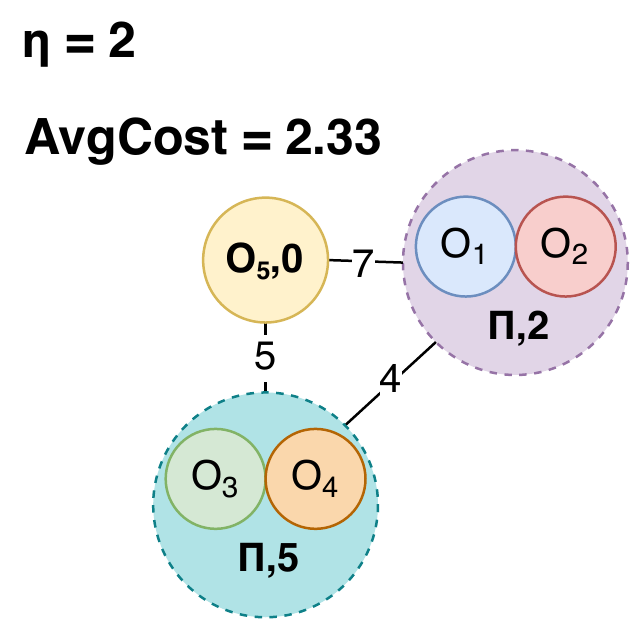}}
\vspace{-0.15in}
    \caption{(a) Order graph of Fig.~\ref{fig:ex1}. The edge weight and the optimal route plan are shown beside each edge. (b) An example order graph. (c) First iteration of clustering for order graph in Fig.~\ref{fig:order_graph}. (d) Second iteration of clustering for order graph in Fig.~\ref{fig:order_graph} for $\maxorders = 3$. (e) Second iteration of clustering for order graph in Fig.~\ref{fig:order_graph} for $\maxorders = 2$.}
    \label{fig:gc}
\vspace{-0.30in}
\end{figure*}
\begin{algorithm}[b!]
    \caption{Batching }\label{algo:hac}
    {\scriptsize
    \begin{flushleft}
        \textbf{Input:} Accumulation Window $l$, Quality cutoff $\eta$\\
        \textbf{Output:} Order Partition $U_1$, corresponding route plans $\Sigma$ 
    \end{flushleft}
	\vspace{-0.10in}
    \begin{algorithmic}[1]
    \State $r \leftarrow 0$
    \State $\Pi(0) \leftarrow \{\langle \{o\}, v,\{o^r, o^c\} \rangle: \forall o \in O(l)\}$
    \State $W(0) \leftarrow \{w_{i,j}: o_i, o_j \in O(l) \times O(l) \}$
    \State $G_O(0) \leftarrow (\Pi(0), W(0))$ 
    \While {true} 
        \If { $AvgCost(G_O(r))/|\Pi(r)| > \eta$} \textbf{break} 
        \EndIf
        \State $i,j \leftarrow \argmin W(r)$
        \If {$|\pi_i^g| + |\pi_j^g| > \maxorders$} \textbf{continue}
        \EndIf
        \State $\sigma_m \leftarrow$ Compute optimal route plan for merged nodes $ \pi_i$ and $\pi_j$.
        \State $g_m \leftarrow \{o: o \in \pi_i^g \cup \pi_j^g \}$
        \State $\pi_m \leftarrow \langle g_m, v_m, \sigma_m \rangle$ 
        \State $\Pi(r+1) \leftarrow \Pi(r) \cup \{\pi_m\} \setminus \{\pi_i,\pi_j\} $
        \State $W(r+1) \leftarrow$ Update edges from $\pi_m$ to $\pi_k \in \Pi(r+1)$
        \State $W(r+1) \leftarrow$ Remove $\pi_i$ and $\pi_j$ 
        \State $G_O(r+1) \leftarrow (\Pi(r+1), W(r+1))$ 
        \State $r \leftarrow r+1$
    \EndWhile
    \State $U_1 \leftarrow \{g_i: \pi_i \in \Pi(r)\}$
    \State $\Sigma \leftarrow \{\sigma_i: \pi_i \in \Pi(r)\}$
    \State \Return $U_1,\Sigma$
    \end{algorithmic}}
\end{algorithm}

\vspace{-0.05in}
\subsection{Batching}
\label{sec:optimization:batching}
In this section, we describe our algorithm to create order batches. Once these batches are created then it is these batches that become nodes in partition $U_1$ of \fg rather than individual orders, and the cost of the edges are now the marginal cost of assigning a batch of orders to a vehicle. 

Multiple orders should be grouped into a single batch, if they can be delivered by a single vehicle such that none of these orders suffer a long detour due to being batched. Intuitively, batching is similar to clustering. However, each order is not a high-dimensional point and hence clustering algorithms for euclidean spaces cannot be applied. More importantly, we need to formulate a distance function that is appropriate for our problem. Empowered with these observations, we formulate batching as a \emph{graph clustering} problem.

\subsubsection{The Order Graph} 
The order graph, $G_{O(\ell)}$, is an edge-weighted graph where each node corresponds to a batch (set) of orders and two batches $\pi_i$ and $\pi_j$ are connected by an edge if it satisfies the following two conditions: (i) $|\pi_i|+|\pi_j|\leq \maxorders$ and (ii) $\sum_{\forall o\in \pi_i\cup\pi_j}o^i\leq\maxload$. Recall, $o^i$ denotes the number of items associated with order $o$.

The weight $w_{i,j}$ of the edge between $\pi_i$ and $\pi_j$ indicates the quality of batching orders in $\pi_i$ with $\pi_j$ into a single batch $\pi_{i,j}$. 
 To quantify $w_{i,j}$, we simulate three vehicles $v_i$, $v_j$, and $v_{i,j}$ with order sets $\pi_i$, $\pi_j$ and $\pi_i\cup \pi_j$ respectively. The edge weight $w_{i,j}$ is defined as follows:

\vspace{-0.15in}
{\small
\begin{alignat}{1}
\label{eq:wij}
w_{i,j}&=Cost(v_{i,j},\pi_{i,j})-(Cost(v_i,\pi_i)+Cost(v_j,\pi_j))
\end{alignat} }
\vspace{-0.20in}

Recall, $Cost(v,O)$ (Eq.~\ref{eq:cost}) quantifies the extra delivery time incurred by each order in set $O$ in the optimal route plan when allocated to vehicle $v$. The initial location of each simulated vehicle is considered to be the first location in the optimal route plan for its corresponding order set. Lower weights indicate better batching. 
 
 We note here that although
  the concept of order graphs has been  used in the literature ~\cite{vldbrideshare,grouping}, our usage of the order graph is much different. In particular, the edge weights (Eq.~\ref{eq:wij}) are specialized for the the food delivery problem. Furthermore, we go beyond both \cite{vldbrideshare,grouping} to introduce a novel iterative clustering method for constructing batches. This allows us to create batches of size $3$ or more in contrast to~\cite{vldbrideshare}, where batch size is limited to two, and is a more sophisticated method than that of~\cite{grouping}, where only cliques are allowed to form batches.
  
\subsubsection{Batching by Iterative Clustering}
\label{sec:batch_hac}

Alg.~\ref{algo:hac} presents the pseudocode. Initially, we have a node corresponding to each order $o\in O(\ell)$ and the edges are constructed as discussed above (line 2). Fig.~\ref{fig:order_graph_1} shows the initial order graph for Fig.~\ref{fig:ex1}. 

The clustering proceeds in an iterative manner. At iteration $r$, we have $G_{O(\ell)}(r)$, a (partially) clustered version of $G_{O(\ell)}$, with node set $\Pi(r) = \{\pi_1, \ldots,\pi_m\}$, such that $\forall i,\:\pi_i\subseteq O(\ell)$ and $\forall i,j,\: \pi_i\cap \pi_j=\emptyset$ (lines 5-16). To construct $G_{O(\ell)}(r+1)$, we merge the endpoints of the minimum weight edge in $G_{O(\ell)}(r)$ into a single cluster $\pi_{i,j}$ (lines (9-11) and remove $\pi_i$ and $\pi_j$. $\pi_{i,j}$ is connected to remaining clusters in $G_{O(\ell)}(r)$ based on the conditions described earlier and edge weights are computed as outlined in Eq.~\ref{eq:wij} (lines 12-15).  

\textbf{Stopping Criterion: } 
To determine when to stop, we track the \emph{quality} of each batch (node) in the order graph. We stop further batching, when the average quality of batches fall below a certain threshold. 
 Formally, we assign an \emph{AvgCost} to the entire order graph as the \emph{average} of the costs of the individual batches. 
 
\vspace{-0.20in}
\begin{equation}
\label{eq:avgcost}
AvgCost\left(G_{O(\ell)}(r)\right) = \frac{\sum_{i=1}^{|\Pi(r)|} Cost\left(v_i,\pi_i\right)} {|\Pi(r)|}
\end{equation}
\vspace{-0.10in}

When $AvgCost(G_{O(\ell)}(r))$ exceeds a pre-defined threshold $\eta$, we stop. It is worth noting that computing $Cost\left(v_i,\pi_i\right)$ does not add any computational burden. Specifically,
 for batch $\pi_{i,j}$ formed by merging $\pi_i$ and $\pi_j$, we have $Cost(v_{i,j},\pi_{i,j})=Cost(v_i,\pi_i)+Cost(v_i,\pi_i)+w_{i,j}(r)$. Here, all components in the right hand side of the equation are already known. Hence, computing $Cost(v_{i,j},\pi_{i,j})$ takes $O(1)$ time.

An important question arises at this juncture: \textit{Can we guarantee that $AvgCost$ is a monotonically increasing function?} Otherwise, convergence cannot be guaranteed.  Theorem~\ref{thm:monotonicity} answers this question.

\begin{theorem}\label{thm:monotonicity} \textit{$AvgCost\left(G_{O(\ell)}(r+1)\right) \geq AvgCost\left(G_{O(\ell)}(r)\right)$ for all $r\geq 1$.
Thus, convergence is guaranteed.}
\end{theorem}
\textsc{Proof.} In the definition of $AvgCost$ (Eq.~\ref{eq:avgcost}), the denominator decreases with $r$. Thus, it is sufficient to show that the numerator $\sum_{i=1}^{|\Pi(r+1)|} Cost\left(v_i,\pi_i\right)$ in the $(r+1)^{th}$ iteration is always larger than the numerator  $\sum_{i=1}^{|\Pi(r)|} Cost\left(v_i,\pi_i\right)$ in the $r^{th}$ iteration. From the definition of edge weights in Eq.~\ref{eq:wij}, if vertices $\pi_i$ and $\pi_j$ are clustered to form $G_{O(\ell)}(r+1)$ from $G_{O(\ell)}(r)$ then, 

\vspace{-0.15in}
\[\sum_{i=1}^{|\Pi(r+1)|} Cost\left(v_i,\pi_i\right) -  \sum_{i=1}^{|\Pi(r)|} Cost\left(v_i,\pi_i\right)= w_{i,j}(r).\]
\vspace{-0.15in}

Hence, if we can establish that $\forall i,j,\:w_{i,j}(r) \geq 0$, we are done. From Eq.~\ref{eq:wij}, this is equivalent to showing $Cost(v_{i,j},\pi_{i,j})\geq (Cost(v_i,\pi_i)+Cost(v_j,\pi_j))$. Assume that the optimal route plan for $v_{i,j}$ with order set $\pi_{i,j}=\pi_i\cup\pi_j$ is the sequence of pick-up/drop-off points $\sigma_{i,j}=\{a_1,a_2,\cdots, a_k\}$. We now partition $\sigma_{i,j}$ into two subsequences $\sigma_i$ and $\sigma_j$ such that $\sigma_i$ contains only those pick-up and drop-off locations from $\sigma_{i,j}$ that are to be visited to deliver the orders in $\pi_i$ and the same for $\sigma_j$ with respect to $\pi_j$. 
 With this setup, 
  the extra delivery time incurred by orders in $\pi_i$  through route plan $\sigma_i$ is less than or equal to the extra delivery time incurred through $\sigma_{i,j}$ since $\sigma_i$ is a subsequence of $\sigma_{i,j}$. The same holds for $\pi_j$ as well when delivered through $\sigma_j$. Therefore, 
  \vspace{-0.05in}
\[Cost(v_{i,j}, \pi_i \cap \pi_j,\sigma_{i,j}) \geq Cost(v_i, \pi_i, \sigma_i) + Cost(v_j,\pi_j,\sigma_j).\]
\vspace{-0.15in}

By definition, the optimal route plans for $\pi_i$ and $\pi_j$ will incur lower costs than the particular route plans $\sigma_i$ and $\sigma_j$ and thus $Cost(v_{i,j}, \pi_i \cap \pi_j) \geq Cost(v_i, \pi_i) + Cost(v_j,\pi_j).\hfill\square$

\begin{example}
\textit{We illustrate the clustering procedure on a larger order graph shown in Fig.~\ref{fig:order_graph}. For this example, we assume $\eta = 2$ and the order capacity $o^i$ for all orders is $1$. In the order graph depicted in Fig.~\ref{fig:order_graph}, the closest clusters are $O_1$ and $O_2$. They are merged to form a new cluster $O_{1,2}$. The cost of each batch is shown within the corresponding node. For example, the cost of $O_{1,2}$ is $2$ 
 as shown in Fig.~\ref{fig:cl1}.
If $\maxorders=3$, in the next iteration, the minimum weight edge connects the clusters $O_{1,2}$ to $O_4$ (Fig.~\ref{fig:cl2}). In case $\maxorders=2$,  $O_3$ and $O_4$ would be merged as shown in Fig.~\ref{fig:cl3}. The algorithm terminates when either the average batch quality exceeds threshold $\eta = 2$ (satisfied in Fig.~\ref{fig:cl3}) or no edges are left to merge (also satisfied in Fig.~\ref{fig:cl3}).}
\end{example}

\textbf{Time Complexity: }As in the previous analysis, we assume the cost of shortest path queries $\mathcal{O}(q)$, $n=|O(\ell)|$. Computing the order graph takes $\mathcal{O}(n^2q)$ time. The number of iterations is bounded by $n - 1$. In each iteration, we find the edge with lowest weight ($\mathcal{O}(\log n)$ time) and recompute the weights of the edges from the remaining nodes to the newly formed cluster ($\mathcal{O}(n\maxorders\cdot(\maxorders!)q)$ time). This gives an overall complexity of: $\mathcal{O}(n^2q\maxorders\cdot(\maxorders!))$.
\vspace{-0.05in}
\subsection{Scaling \fmplus}
\label{sec:scale}
The $\mathcal{O}(nm\maxorders\cdot(\maxorders!)q + k_{\top}^2k_{\bot})$ complexity of bipartite matching stems from two components: creating the bipartite \fg $\left(\mathcal{O}\left(nm\maxorders\cdot\left(\maxorders!\right)q\right)\right)$ and applying Kuhn-Munkres algorithm for minimum weight perfect matching $\left(O(k_{\top}^2k_{\bot})\right)$. Creating \fg is the dominant contributor to the computation cost since it involves computing the edge weight among each pair of order and vehicle. 
We now explore strategies to reduce this intensive computation. 

\begin{figure}[t]
\vspace{-0.20in}
\centering
    \subfigure[]{\label{fig:rank_orders}\includegraphics[width=0.32\linewidth]{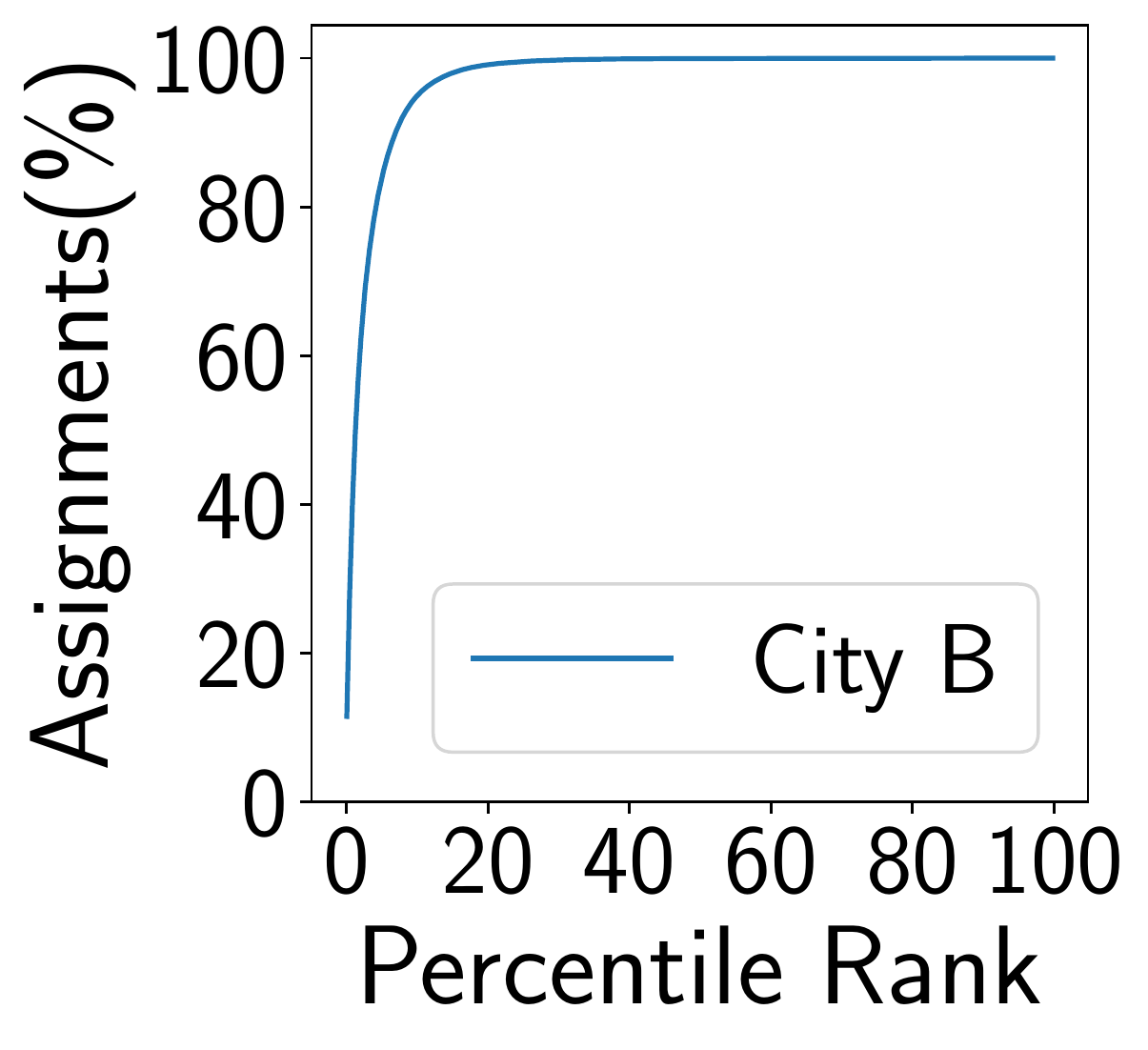}}
    \hspace{0.5in}
    \subfigure[]{\label{fig:angular}\includegraphics[width=0.35\linewidth]{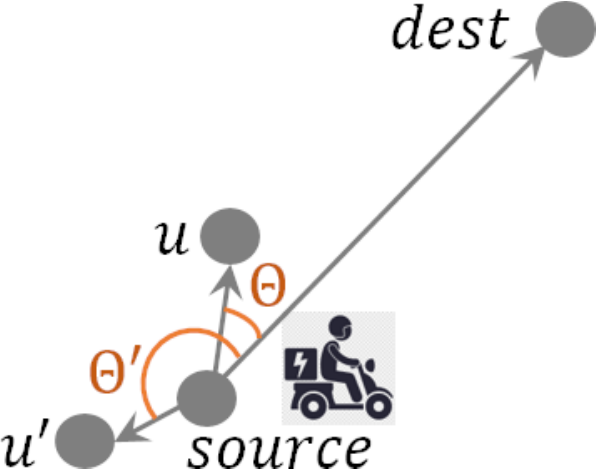}}
\vspace{-0.10in}
    \caption{ (a) Cumulative distribution of percentile ranks of assigned vehicles to orders. (b) Illustration of angular distance.}
\vspace{-0.20in}
\end{figure}

In practice, it is likely that the vehicle assigned to an order $o$ is located close to the pick-up location $o^r$. To verify this intuition, we analyze the assignments made by Kuhn-Munkres's algorithm on order and vehicle data from a large city in India (City B in Table~\ref{tab:dataset}). Specifically, for each vehicle $v$, we rank all orders $o\in O(\ell)$ in ascending order based on network distance $SP(loc(v,t),o^r,t)$, i.e., from the vehicle's location to the order's restaurant location. Next, we compute the percentile rank of the allocated order to $v$. Fig.~\ref{fig:rank_orders} presents the distribution of percentile ranks of allocated order.

Fig.~\ref{fig:rank_orders} reveals that for $95\%$ of all vehicles, the assigned order has a percentile rank below $10\%$. This behavior reveals that distant order-vehicle pairs rarely play a role in the eventual assignment and therefore, the corresponding edges in the \fg could be pruned. Incorporating this methodology, however, raises an intriguing question: \textit{How do we determine which vehicles are far away from an order without even computing the distances to them?}
  
\subsubsection{Efficient \fg construction} 
\label{sec:bp_topk}

First, we formalize our objective. Let $\Pi=\{\pi_1,\cdots,\pi_m\}$ be the set of batches obtained from Alg.~\ref{algo:hac}. For each batch $\pi_i$, we examine its (quickest) route plan and identify the order $\pi_i[1]\in \pi_i$ that needs to be picked up first from restaurant location $\pi_i[1]^r$. With this information,
we extract the subset of nodes $V_{\Pi}\subseteq V$, where $V_{\Pi}\subseteq V=\{u\in V\:\mid\:\exists \pi_i\in\Pi,\:\pi_i[1]^r=u\}$. $V_{\Pi}$ is the set of nodes where the route plan of at least one batch starts. Now, given an available vehicle $v\in\CV(\ell)$, our goal is to identify the $k$ closest nodes in $V_{\Pi}$ from $v$'s current location $loc(v,t)$. With this information, we want to construct a \emph{sparsified} bipartite \fg, where we add edges from vehicle $v$ to \emph{only} those batches that start from a node within the top-$k$ list of $v$. To all remaining batches, we would add an edge of weight $\Omega$ (c.f. Prob.~\ref{prb:online}) and avoid computing the true edge weight (marginal cost). $k$ is selected based on percentile information extracted from the data. For example, Fig.~\ref{fig:rank_orders} shows that $95\%$ of all vehicles get assigned an order that is within the top-$10\%$ percentile. Thus, $k$ could be set to $\frac{10}{100}\times |\Pi|$. We achieve this objective, through \emph{best-first} search.

Alg.~\ref{algo:fg} presents the pseudocode for our best-first-based method for constructing the \emph{sparsified} \fg. Given an available vehicle $v\in\CV(\ell)$, we extract its current location (node) $source=loc(v,t)$ (line 3). From $source$, we initiate a best-first search (lines 8-15). Specifically, we utilize a \emph{priority queue} $PQ$ to store candidate nodes that should be visited (line 4). PQ stores tuples $\langle u,\delta_u\rangle$, where $\delta_u=SP(source,u,t)$ is the length of the shortest path from $source$ to $u$. $PQ$ retrieves nodes in ascending order of $\delta_u$. Initially, $PQ$ contains only $source$ (line 5). Once the top node $u$ is popped from $PQ$ (line 9), we add an edge from vehicle $v$ to all batches that start from $u$ (if any) in the \fg (lines 12-15). The edge weight of assigning a batch to a vehicle is obtained by generalizing marginal cost (Def.~\ref{def:marginal}) to a set of orders. Specifically,

\vspace{-0.20in}
\begin{equation}
\label{eq:marginalset}
mCost(\pi,v) =  Cost\left(v,O^v_t\cup \pi\right) - Cost\left(v,O^v_t\right)
\end{equation}
\vspace{-0.20in}

Next, we mark $u$ as \emph{visited} and insert all of the unvisited neighbors $u'$ of $u$ in $PQ$ with $\delta_u'=\delta_u+\beta(u,u',t)$, where $\beta(u,u',t)$ is the edge weight of edge $(u,u')$ (average time taken) in the road network at time $t$ (line 11, 17-18). This process continues iteratively, till either $v$ attains a degree of $k$ in \fg or $PQ$ becomes empty (line 8). Once this condition is satisfied, we add edges of weight $\Omega$ from $v$ to all remaining batches that are not already connected (line 19). Finally, this entire process is repeated $\forall v\in\CV(\ell)$ (line 2). 

\begin{lemma}[Proof of Correctness]
\textit{If vehicle $v$ has an edge weight smaller than $\Omega$ to batch $\pi$ in the \fg obtained through Alg.~\ref{algo:fg}, then $\pi$ is among the top-$k$ closest batches to $v$. The distance between $v$ and $\pi$ is measured as the shortest path distance from $v$'s location to the first pick-up node in $\pi$.}
\end{lemma}
\begin{IEEEproof}
Alg.~\ref{algo:fg} explores nodes in ascending order of their network distance from the current location of the vehicle. Hence, if node $u_1$ is popped from the priority queue before node $u_2$, it is guaranteed that $SP(source,u_1,t)\leq SP(source,u_2,t)$. Since the while loop in Alg.~\ref{algo:fg} exits as soon as vehicle $v$ attains a degree of $k$, for any unvisited batch $\pi$, it is guaranteed that $k$ batches have been visited prior to $\pi$. This, in turn, means any of the visited batches start at a node closer to $source$ than $\pi$.
\end{IEEEproof}

\vspace{-0.10in}
\subsection{Managing Dynamic Environment}
\subsubsection{Handling dynamic vehicle positions}
\label{sec:angular}
Alg.~\ref{algo:fg} assumes that the location of a vehicle $v$ is static while the \fg is being constructed. In reality, $v$ moves according to its assigned route plan in the previous accumulation window. Thus, by the time \fg construction finishes, the assigned batches may no longer be the $k$ closest ones to $v$. Fig.~\ref{fig:angular} illustrates this issue more concretely. Consider vehicle $v$ that is located at position (node) $source$ at time $t$ and is moving towards location $dest$. We assume the \fg construction process starts at time $t$ and finishes at $t+\Delta$. Now, consider two candidate locations (nodes) $u$ and $u'$. $u'$ is closer to $source$ at time $t$. However, at time $t+\Delta$, there is a high likelihood that $u$ is closer to $v$ since $v$ is progressively moving away from $u'$ as $dest$ is located in an opposite \emph{direction} to $u'$. To mitigate this issue, we incorporate the novel idea of \emph{angular distance}. The angular distance of vehicle $v$ with node $u$ is proportional to the angle formed by the \emph{direction vectors} from $source$ to $dest$ with that of $source$ to $u$. In Fig.~\ref{fig:angular}, the angles are $\Theta$ and $\Theta'$ for $u$ and $u'$ respectively. Given the source and destination locations, the direction vector is captured through \emph{bearing}. 
\begin{algorithm}[t!]
    \caption{Sparsified \fg construction }\label{algo:fg}
    {\scriptsize
    \begin{flushleft}
        \textbf{Input:} Available vehicles $\CV(\ell)$, order batch $\Pi$, Maximum  degree $k$, current time $t$, road network $G(V,E,\beta)$\\
        \textbf{Output:} Sparsified \fg 
    \end{flushleft}
\vspace{-0.10in}
    \begin{algorithmic}[1]
    \State Initialize bipartite graph $B$ with node sets $\CV(\ell)$ and $\Pi$, and empty edge set $E_b$.
    \Foreach {$v\in\CV(\ell)$}
	\State $source \leftarrow loc(v,t)$
	\State $PQ\leftarrow$ Empty Priority Queue
	\State $PQ.insert(\langle source,0 \rangle)$
	\State Initialize $\forall u\in V,\:visited(u)\leftarrow false$
	\State $\Pi_v\leftarrow \emptyset$
    \While {$PQ.empty()=false$ \textbf{and} $degree(v,B)<k$}
	\State $\langle u, \delta \rangle\leftarrow PQ.pop()$
        \IIf {$visited(u)=true$} \textbf{continue}
	\State $visited(u)\leftarrow true$
	\State $I(u)\leftarrow \{\pi\in\Pi\:\mid\:\pi[1]^r=u \}$
	\Foreach {$\pi_i\in I(u)$} 
		\State Add edge from $v$ to $\pi_i$ in $E_b$ with edge weight given by Eq.~\ref{eq:marginalset}.
	\EndForeach
	\State $\Pi_v\leftarrow \Pi_v\cup I(u)$
	\State $N(u)\leftarrow (u'\:\mid\: (u,u')\in E,\:visited(u)=False,\text)$
	\State $\forall u'\in N(u),\:PQ.insert(\langle u',\delta+\beta(e=(u,u'),t)\rangle)$
    \EndWhile
	\State $\forall \pi_i\in\Pi\setminus \Pi_v$, add edge from $v$ to $\pi_i$ in $E_b$ with $\Omega$ edge weight.
    \EndForeach
    \State \Return bipartite graph $B(\CV(\ell),\Pi,E_b)$
    \end{algorithmic}}
\end{algorithm}

\begin{definition}[Bearing]
\textit{The bearing $\Theta_{s,t}$ captures the direction along a great circle between two points of $s(\varphi_s, \lambda_s)$ and $t(\varphi_t, \lambda_t)$ where $\varphi$ and $\lambda$ are the latitude and longitude in radians. Mathematically, it is computed as follows:}
\vspace{-0.05in}
\begin{alignat}{3}
\nonumber
\Theta(s,t) &= \atantwo(X,Y) \\
\nonumber
\text {where, }  X&=\cos(\varphi_t)\sin(\lambda_t-\lambda_s)\\
\nonumber
		Y&=\cos(\varphi_s)\sin(\varphi_t) - \sin(\varphi_s)\cos(\varphi_t)\cos(\lambda_t - \lambda_s))
\end{alignat}
\vspace{-0.25in}

\end{definition}
$\atantwo(X,Y) = \arctan(X/Y)$ is rendered in the range $[0,2\pi]$.
The \emph{angular distance} of vehicle $v$ to node $u$ at time $t$ is:
\vspace{-0.05in}
\begin{equation}
\nonumber
adist(v,u,t)=\frac{1-\cos\left(\Theta\left(loc(v,t),dest\right) - \Theta\left(loc(v,t),u\right)\right)}{2}
\end{equation}
\vspace{-0.10in}

Here, $dest$ denotes the next destination node in the route plan followed by $v$. The properties of the angular distance are as follows. {\bf (1)} The numerator ranges from $0$ to $2$, and hence we divide it by $2$ to obtain a distance value between $0$ to $1$. {\bf (2)} $0$ indicates that $u$ lies in the same direction as $dest$ whereas, $1$ indicates $u$ is located in a diametrically opposite direction. The closer the distance is to $0$, the more similar are the directions.

Angular distance allows us to estimate whether the distance to a node is likely to deteriorate with time or improve. Thus, instead of relying only on the average traveling time as the edge weight in the road network, we make the edge weight a function of both the travel time and the angular distance. Specifically, we definite a \emph{vehicle-sensitive} edge weight $\alpha(v,e,t)$ for edge $e=(u,u')$ as follows:

\vspace{-0.10in}
\begin{equation}
\label{eq:dist}
\alpha(v,e,t)=(1-\gamma)adist(v,u',t)+\gamma\frac{\beta(e,t)}{\max_{\forall e'\in E}\{\beta(e',t)\}}
\end{equation} 
\vspace{-0.10in}

Here $\gamma$ is a weighting factor. Recall from Def.~\ref{def:road-network} that $\beta(e',t)$ denotes the average traveling time in edge $e$ at time $t$.

\begin{figure}[t]
\centering
\vspace{-0.20in}
    \includegraphics[width=3.4in]{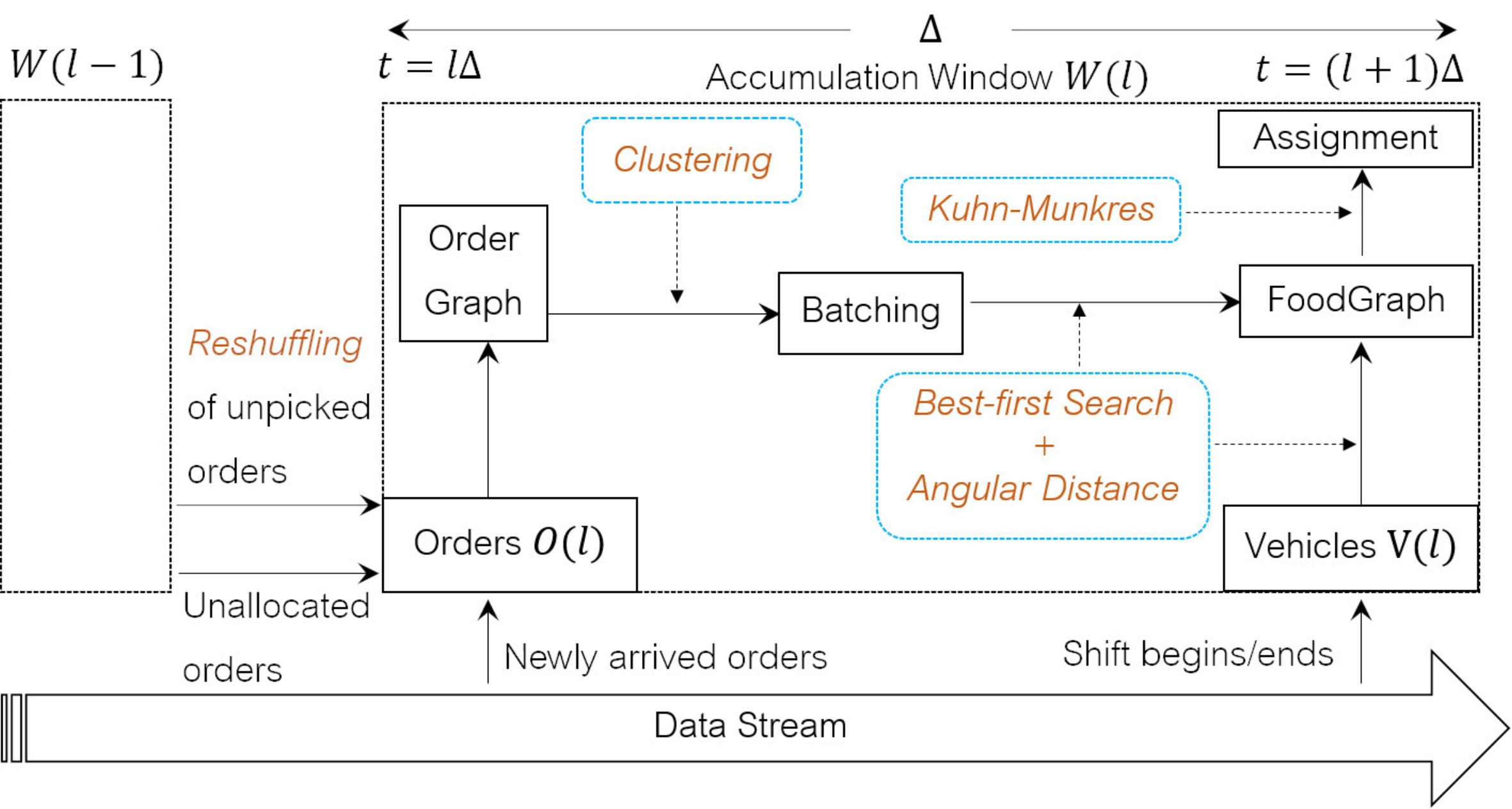}
\vspace{-0.15in}
    \caption{Flowchart of \fmplus.}
    \label{fig:flowchart}
\vspace{-0.15in}
\end{figure}

\subsubsection{Reshuffling}
\label{sec:reshuffling}

In food delivery, we have a buffer time until the food is prepared. We utilize this buffer by allowing \emph{reshuffling} of assignments done in previous accumulation windows. Specifically,  an order $o$ assigned to a vehicle $v$ is available for \emph{re-assignment} in subsequent windows if $v$ has not yet picked it up. Similarly, vehicle $v$ also gets added to $\CV(\ell)$ and may get assigned a new order. In other words, reshuffling takes into account the dynamic nature of the problem and exploits new data to find better assignments for previously made assignments that have not been picked up yet. 

\vspace{-0.05in}
\subsection{Putting it All Together}
Fig.~\ref{fig:flowchart} outlines the pipeline of \fmplus. Given a data stream of orders and available vehicles, we accumulate them in windows of length $\Delta$. Within the current window $W(\ell)$, $\CV(\ell)$ collects all available vehicles. $O(\ell)$, on the other hand, collects all unassigned orders as well as orders that have been assigned to a vehicle but not yet been picked up from their restaurants. $O(\ell)$ is next clustered to form batches. These batches and the available vehicles $\CV(\ell)$ are used to form the bipartite \fg. The construction of \fg is expedited through best first search. In addition, we incorporate dynamic movement of vehicles through angular distances. Finally, We perform minimum weight perfect matching on \fg using the \emph{Kuhn-Munkres} algorithm. 

\vspace{-0.05in}
\section{Experiments}\label{sec:exp}

In this section, we benchmark \fmplus and establish:
\begin{itemize}
\item \textbf{Quality: }\fmplus is effective in generating high-quality assignments and is, on average, $30\%$ better than the baseline strategies.
\item \textbf{Scalability: }\fmplus scales to the demands of food orders on the busiest Indian cities.
\item \textbf{Impact of optimization strategies: }\textit{Batching, reshuffling, best-first search} and \emph{angular distance} impart significant improvement in both assignment quality and efficiency.
\end{itemize}

Our code and datasets are available at \url{https://github.com/idea-iitd/FoodMatch} 

\begin{table}[t]
\vspace{-0.20in}
\centering
\caption{Summary of order history dataset}
\label{tab:dataset}
\vspace{-0.05in}
\scalebox{0.75} {
\begin{tabular}{lcccccc} 
    \
    {\bf City} & {\bf \# Rest.} & {\bf \# Vehicles} & {\bf \# Orders} & {\bf Food prep. time} & {\bf \# Nodes} & {\bf \# Edges}\\ 
         && {\bf (avg./day)} & {\bf (avg./day)} & {\bf (avg./min)}&&\\ \hline 
    GrubHub & $159$ & $183$ & $1046$ & $19.55$ &NA&NA\\
    City A & $2085$ & $2454$ & $23442$ & 8.45 & 39k & 97k\\
    City B  & $6777$ & $13429$ & $159160$ & 9.34 &116k&299k \\ 
    City C & $8116$ & $10608$ & $112745$ & 10.22 & 183k & 460k\\  
 \hline
\end{tabular}}

\vspace{-0.25in}
\end{table}
\vspace{-0.05in}
\subsection{Datasets}
\label{sec:dataset}
 Table~\ref{tab:dataset} presents the datasets used for our experiments. Grubhub is provided by Reyes et al.\cite{mdrp}. 
 The remaining datasets are provided by Swiggy - India's leading food delivery service~\cite{toi}. Due to business intelligence-related concerns, neither GrubHub nor Swiggy reveal the city names.
The Swiggy data represents
all orders delivered in these cities over a period of $6$-days.

\begin{figure*}[t]
\vspace{-0.20in}
  \centering
\subfigure[\#Orders/\#Vehicles]{
  \label{fig:dist}
	\includegraphics[width=1.25in]{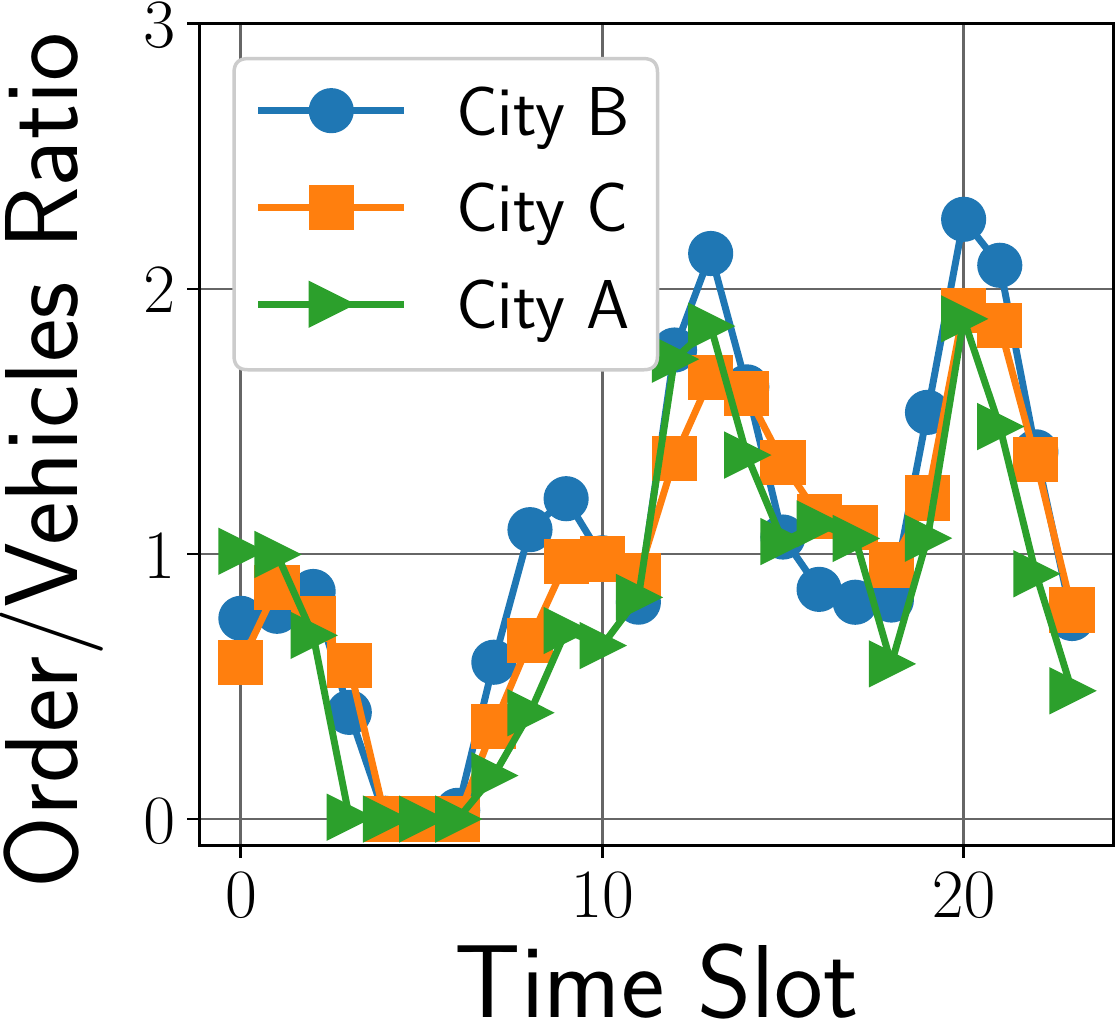}}
	\subfigure[XDT]{
  \label{fig:reyes_xdt}
	\includegraphics[width=1.32in]{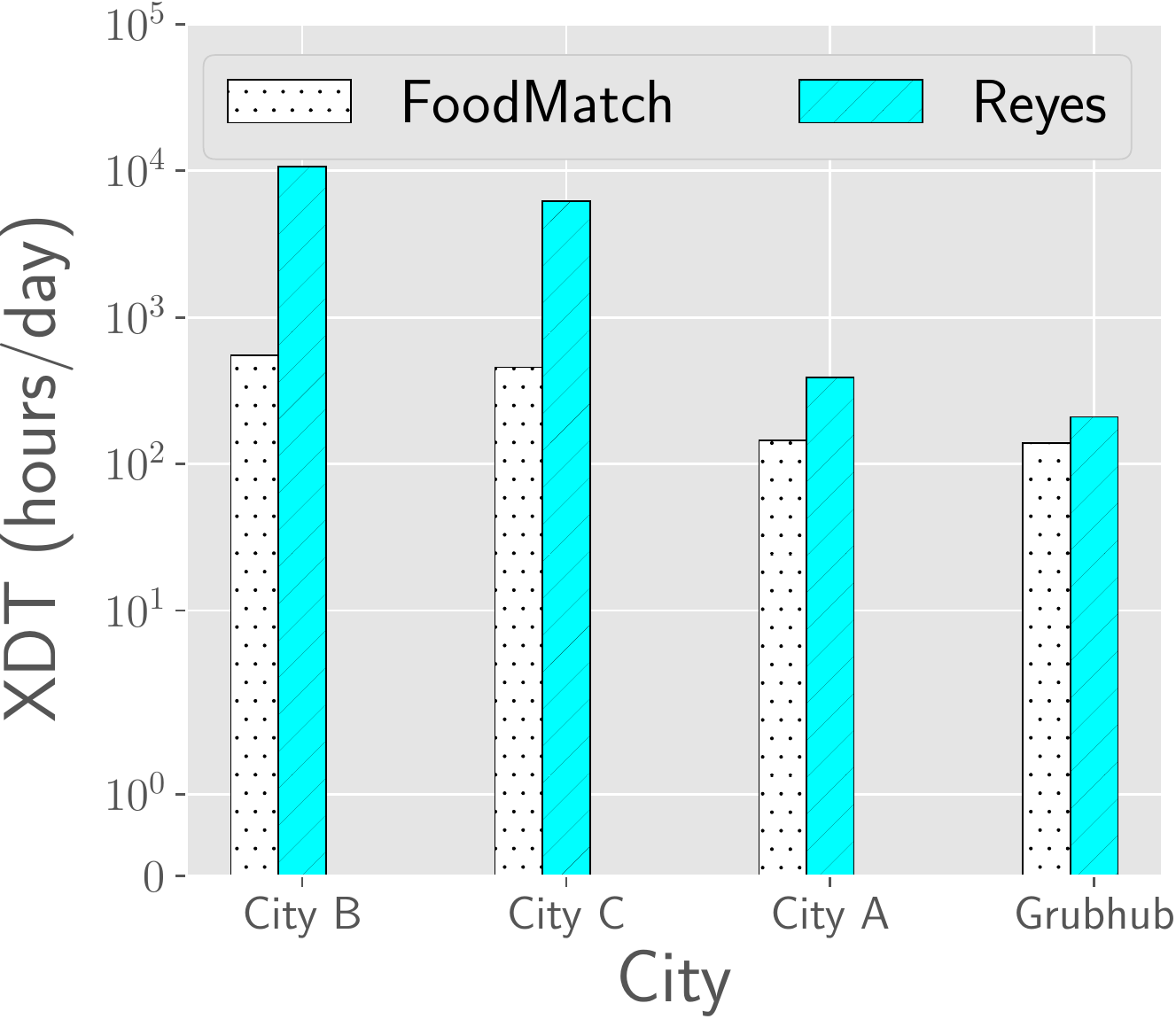}}
\subfigure[XDT]{
  \label{fig:improvement_edt}
	\includegraphics[width=1.30in]{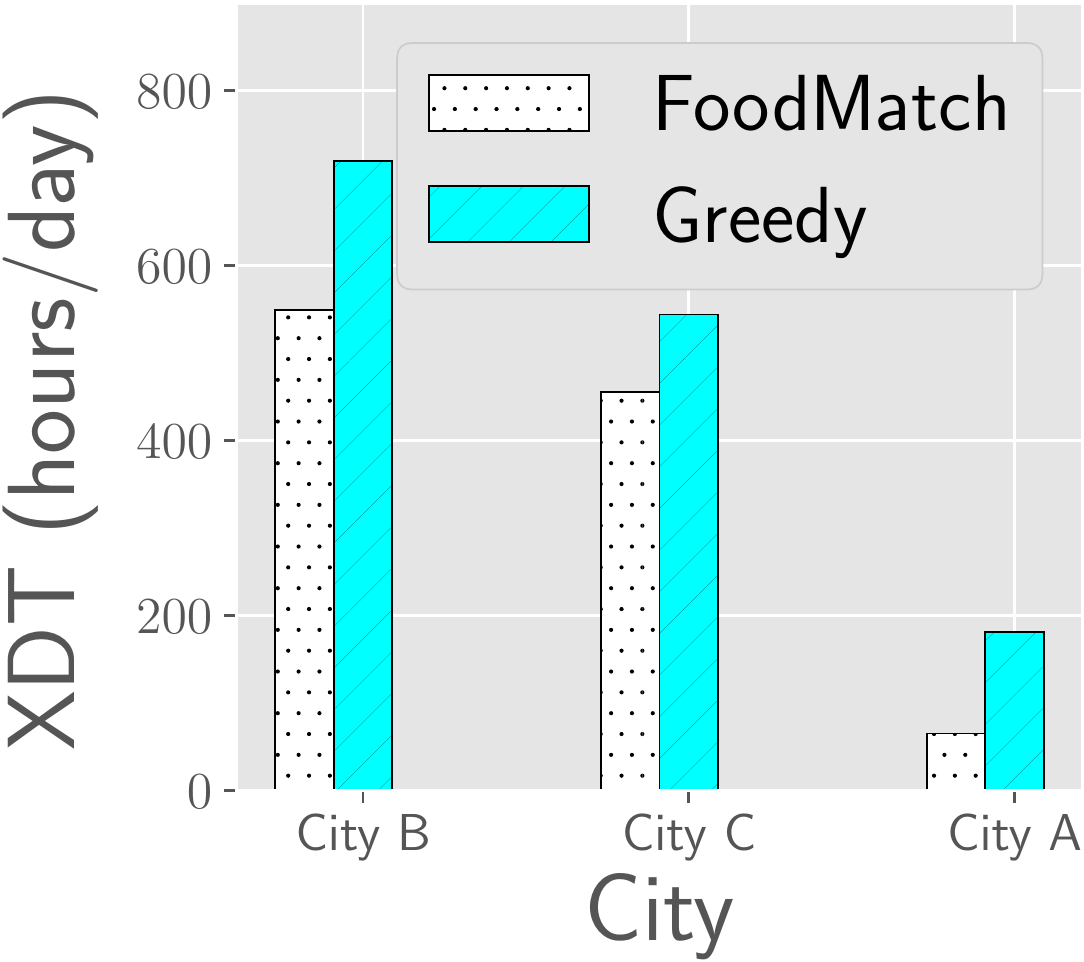}}
\subfigure[O/Km]{
  \label{fig:improvement_okm}
	\includegraphics[width=1.32in]{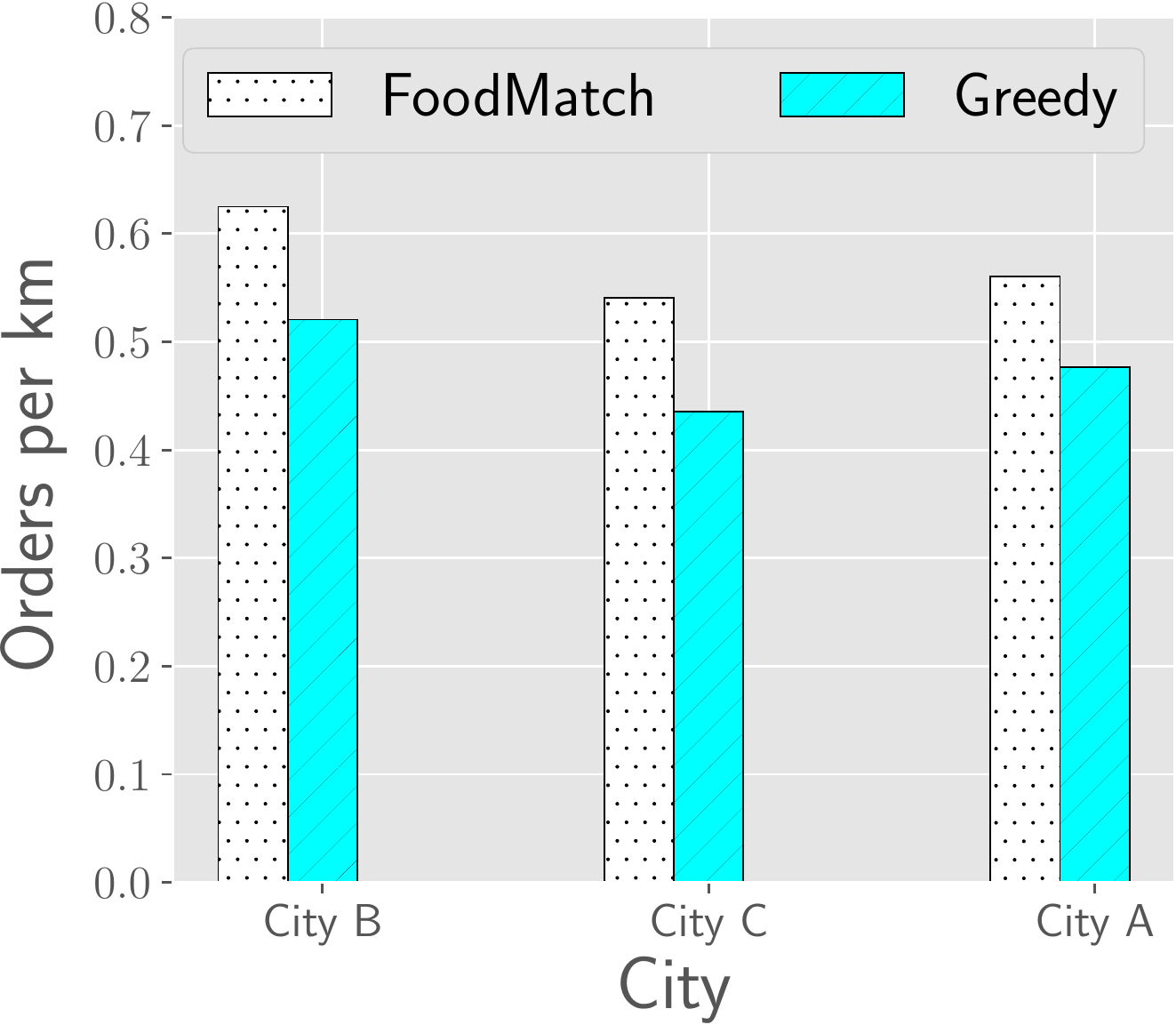}}
\subfigure[WT]{
  \label{fig:improvement_wt}
	\includegraphics[width=1.32in]{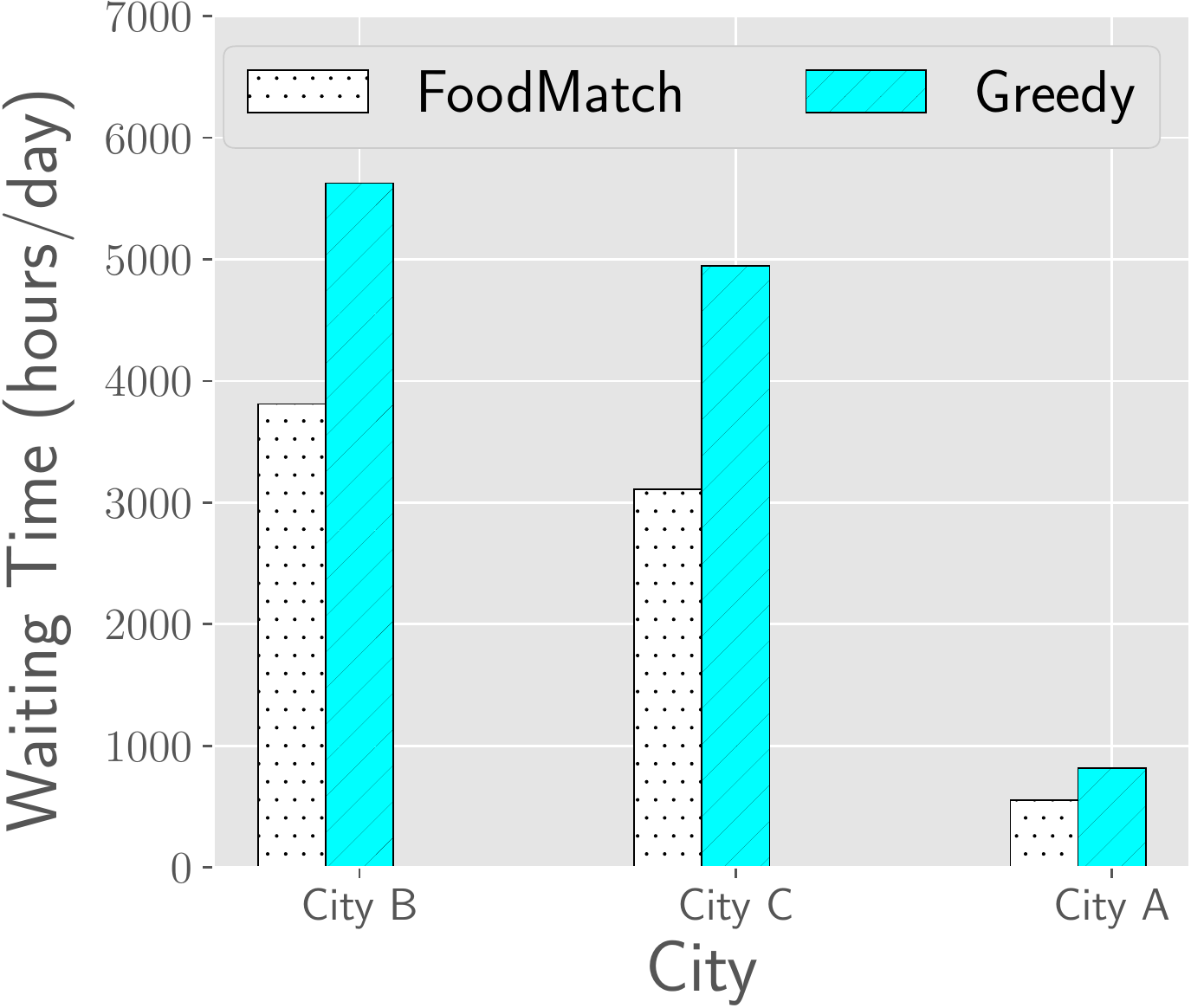}}\\
	\vspace{-0.10in}
\subfigure[Overflow (All)]{
  \label{fig:overflow}
	\includegraphics[width=1.12in]{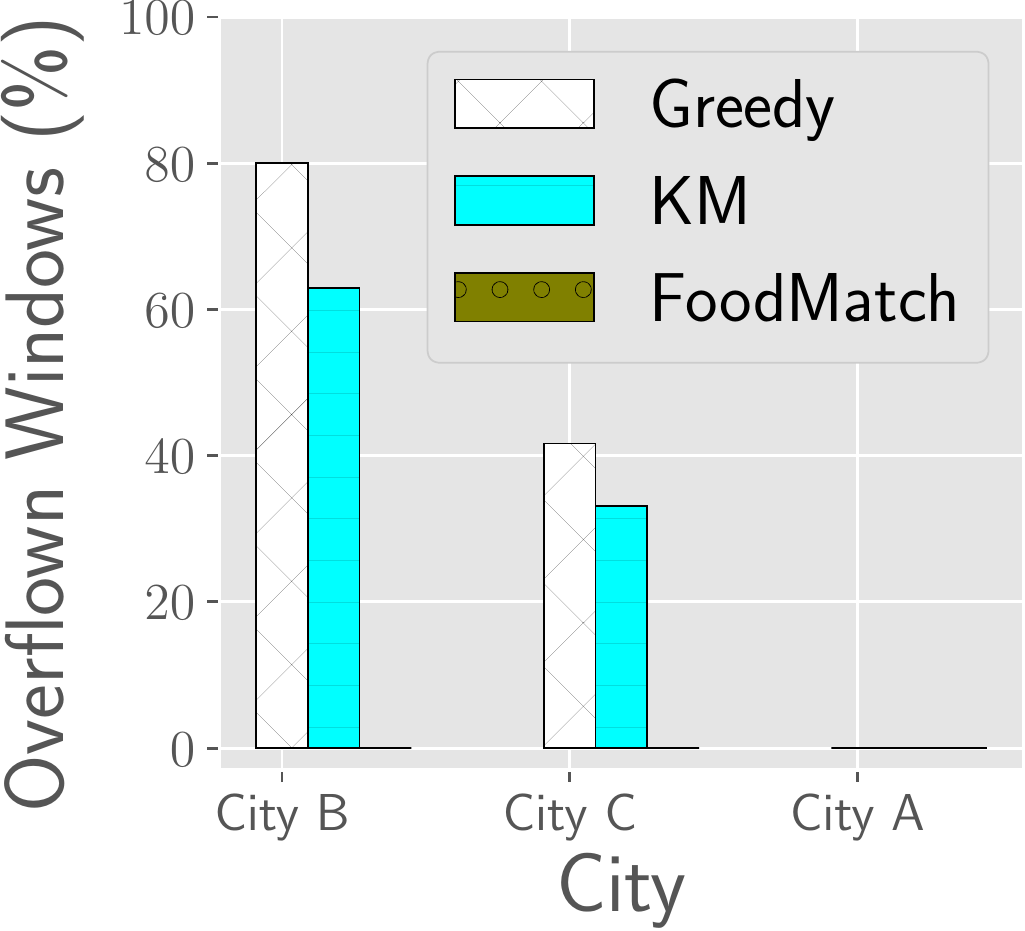}}
\vspace{-0.10in}
\subfigure[Overflow (Peak)]{
  \label{fig:overflow_peak}
	\includegraphics[width=1.16in]{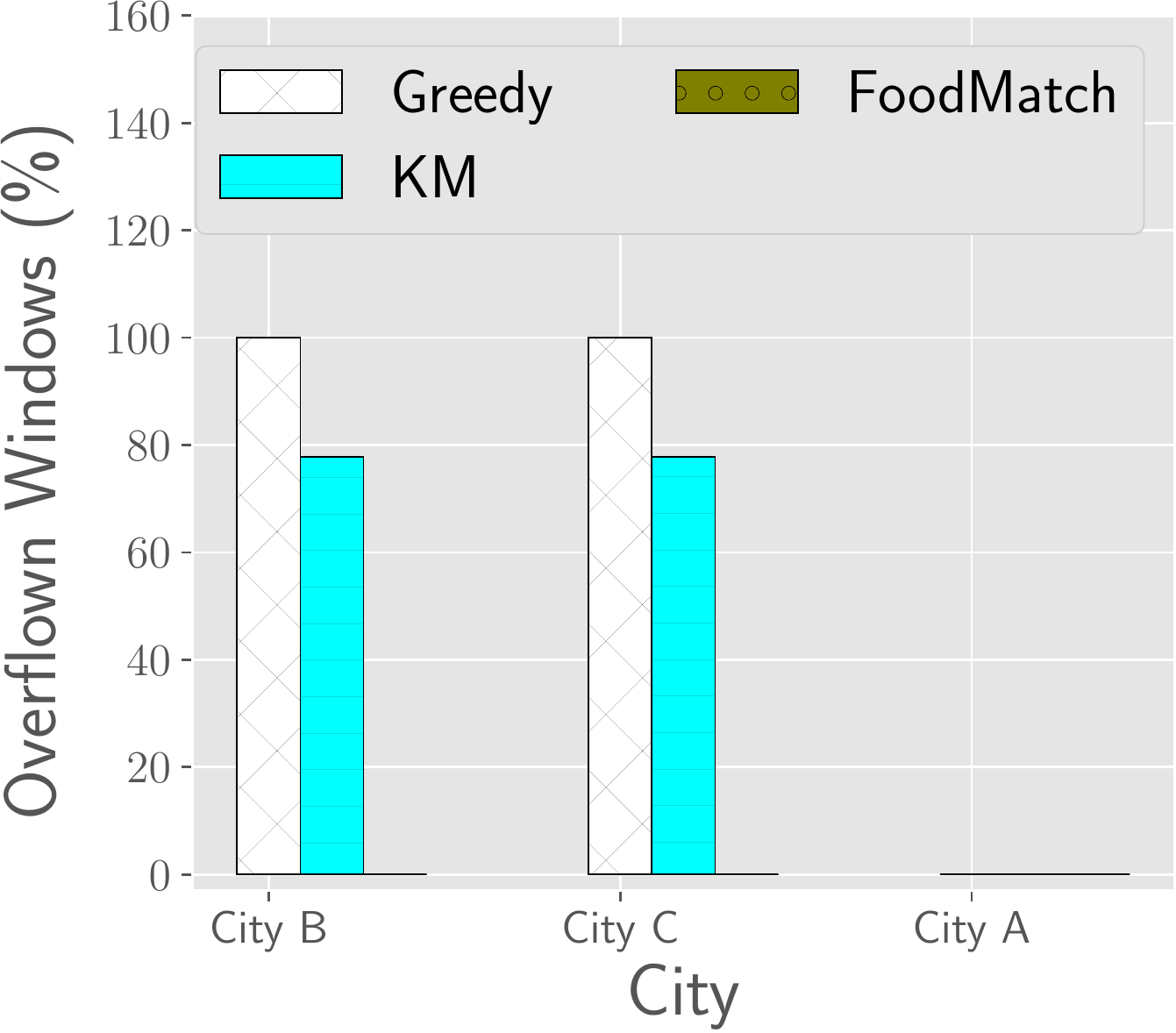}}
\subfigure[Time (All)]{
  \label{fig:time}
	\includegraphics[width=1.12in]{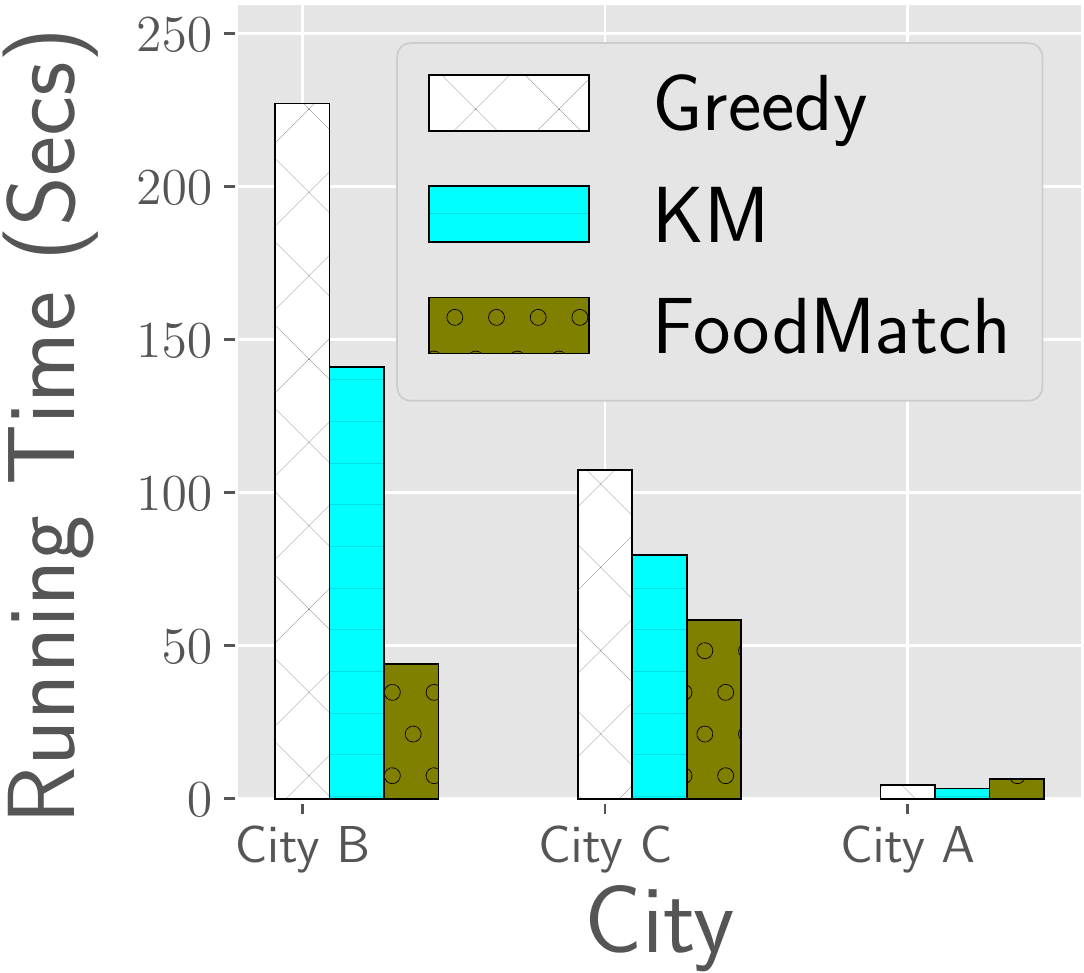}}
\subfigure[XDT]{
  \label{fig:timeslot_edt}
	\includegraphics[width=1.07in]{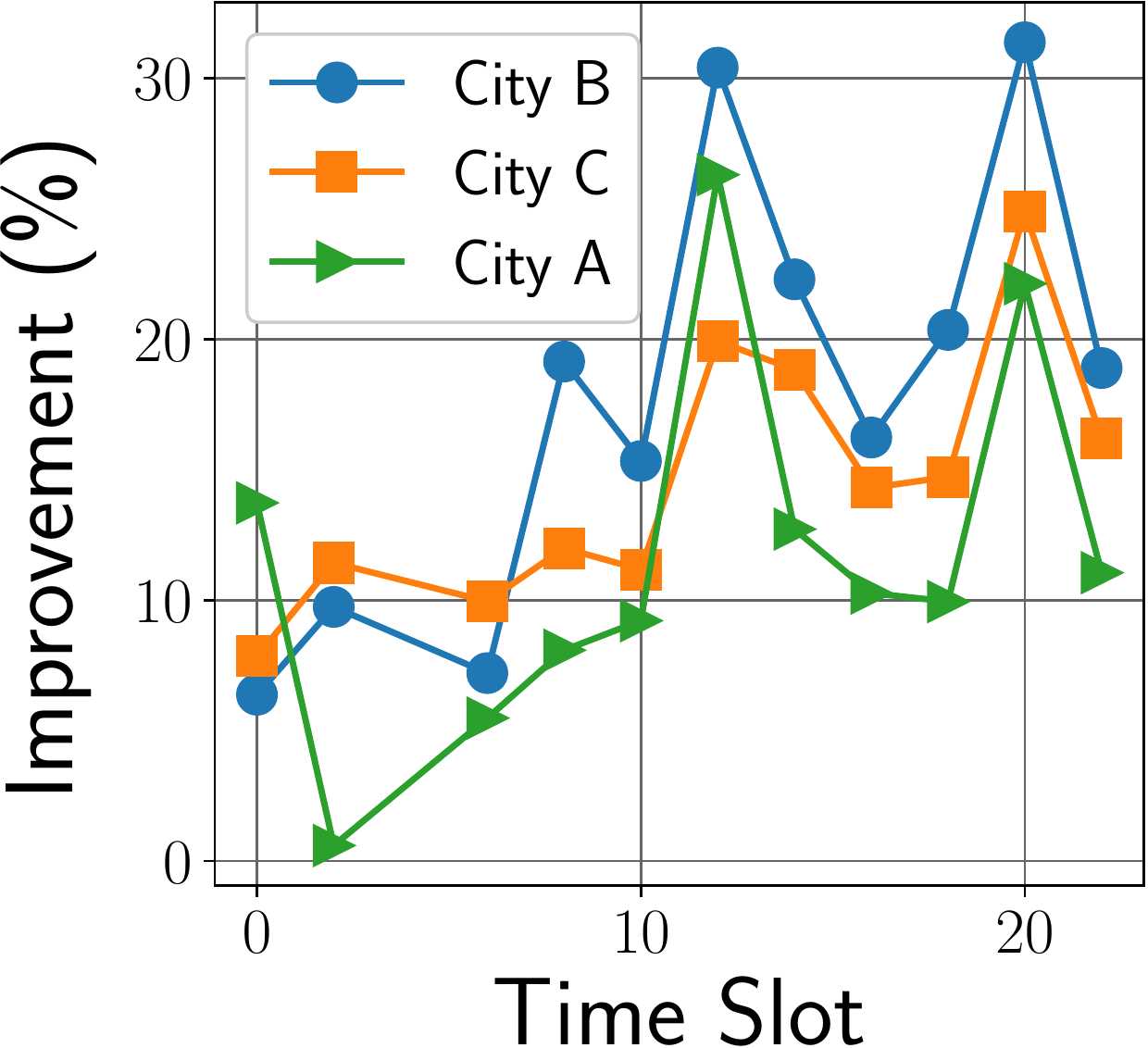}}
\subfigure[O/KM]{
  \label{fig:timeslot_okm}
	\includegraphics[width=1.07in]{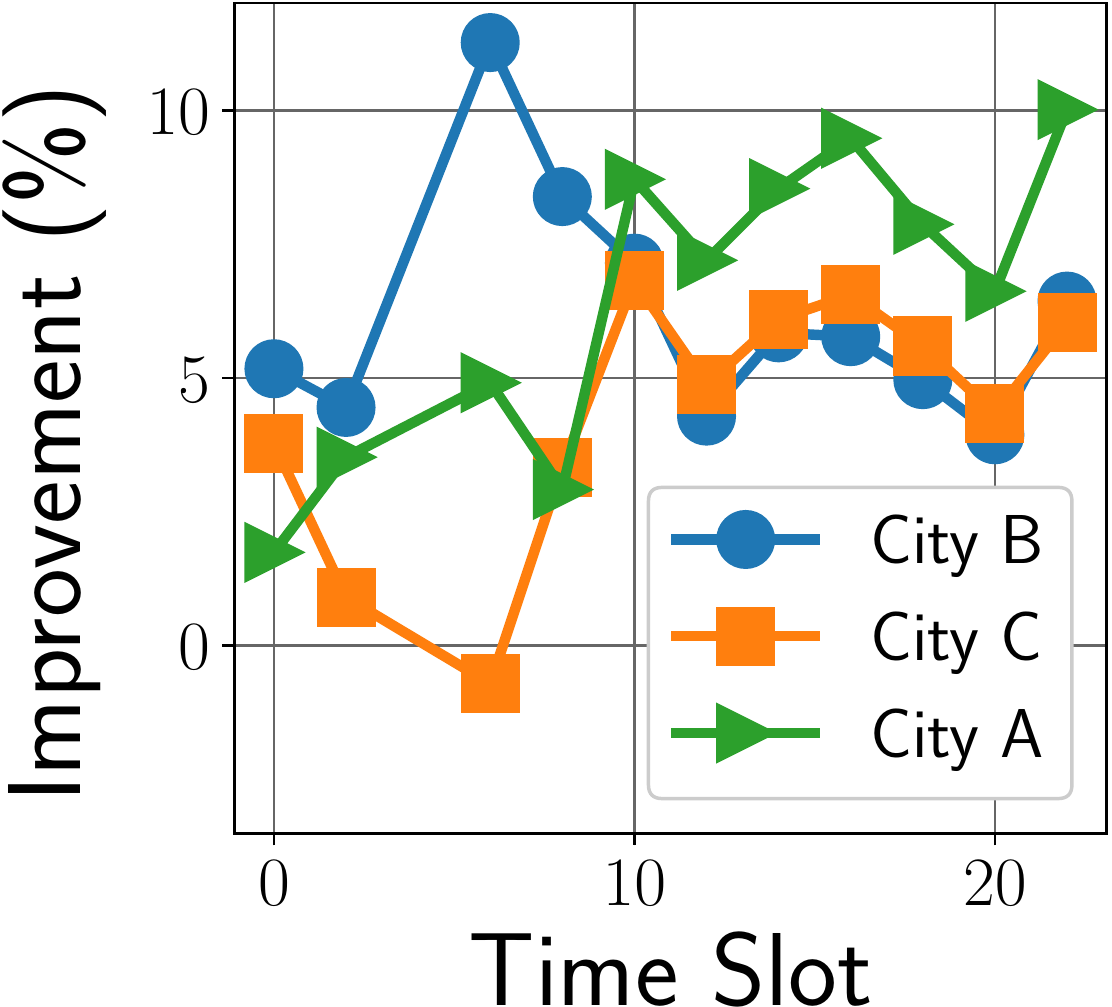}}
\subfigure[WT]{
  \label{fig:timeslot_wt}
	\includegraphics[width=1.06in]{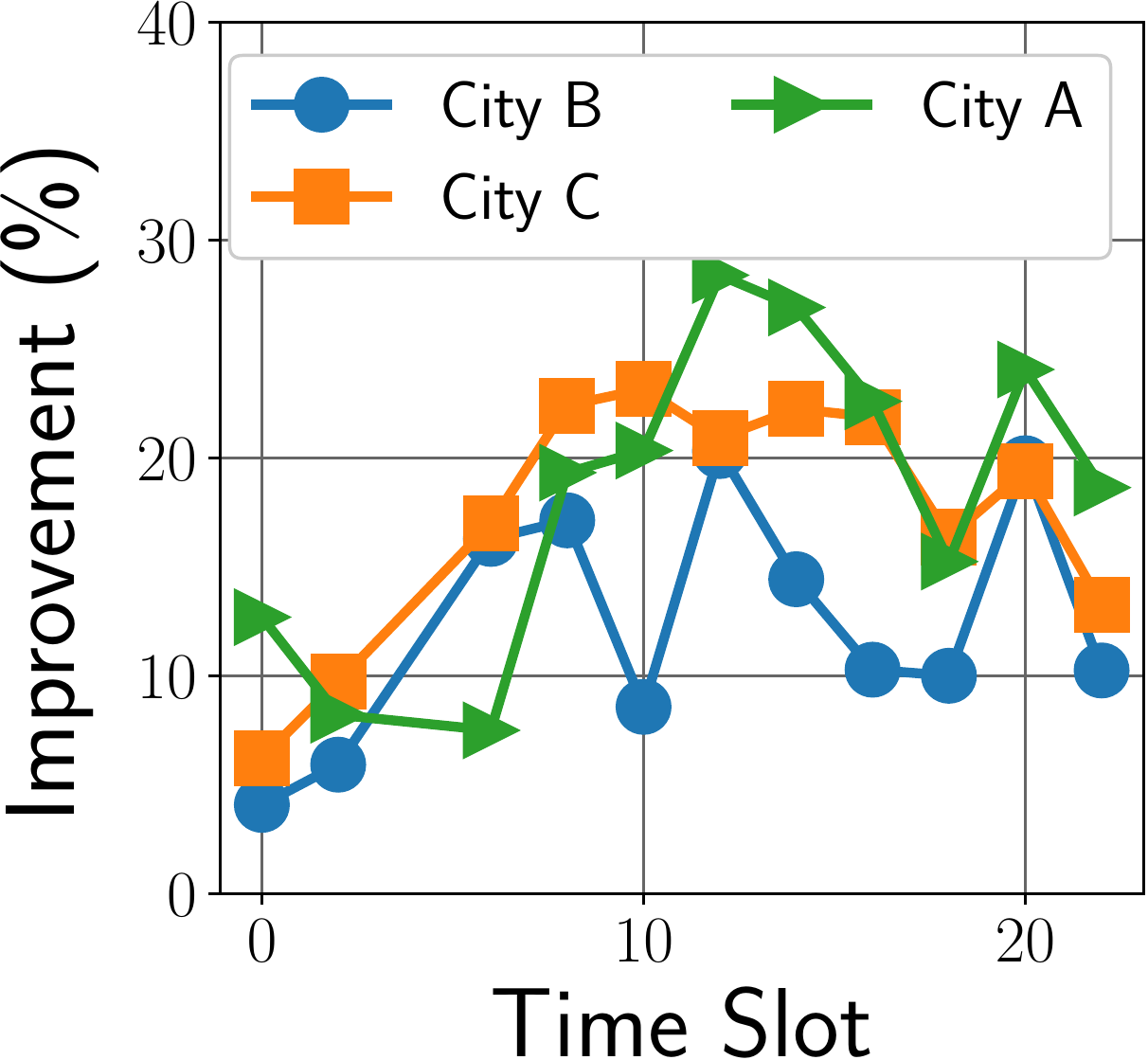}}
\vspace{-0.11in}
\caption{(a) Distribution of the order-to-vehicle ratio across timeslots. (b) Comparison with Reyes. (c-e) Comparison with Greedy. (f-g) Number of accumulation windows where the time taken for assignment is above $\Delta=3$ minutes across (e) all slots, and (f) peak slots. (h) Average running times of \fmplus,\fm, and greedy. (i-k) Improvement over \fm across timeslots in (i) XDT, (j) O/KM, and (k) WT.}
\label{fig:qw_time}
\vspace{-0.20in}
\end{figure*}

\textbf{City Characteristics: }City B and City C correspond to two large metropolitan cities in India, whereas City A is relatively smaller. Although City C has a larger number of restaurants, Swiggy fulfilled $41\%$ more orders in City B in the time period under consideration. Furthermore, Swiggy also employed about $27\%$ more vehicles to cope with the higher order volume. In Fig.~\ref{fig:dist}, we plot the ratio $\frac{No. of Orders}{No. of Vehicles}$ in each 1-hour timeslot of a city. Timeslot $1$ indicates the period 12AM-12:59PM, $2$ indicates 1AM-1:59AM and so on. A ratio above $1$ indicates that the number of vehicles that were available in that timeslot was lower than the number of orders received. We point out some of the key observations from Fig.~\ref{fig:dist}. {\bf (1)} The order-to-vehicle ratio is the highest across all cities during the lunch and dinner periods. {\bf (2)} The ratio is the highest in City B.

\textbf{Road Network:} The Swiggy datasets also come with the road networks of the the corresponding cities. They are obtained from OpenStreetMap~\cite{OpenStreetMap}. The vehicle GPS pings are \emph{map-matched} to the road network to obtain network-aligned trajectories~\cite{mapmatch}. The weight of each road network edge is set to the \emph{average travel time} across all of Swiggy's vehicles in the corresponding road. Instead of computing a single average speed across the entire 24-hour period, we create $24$ time slots corresponding to each hour of the day. The edge-weight is computed for each time slot separately.

 \textbf{Food preparation time:}
 Similar to edge weights, we partition the food preparation times of each restaurant into $24$ $1$-hour slots. The food preparation within an $1$-hour slot is next modeled as a \emph{Gaussian distribution} $N(\overline{\mu}_{R,T}, \overline{\sigma}_{R,T})$ for each restaurant $R$ in timeslot $T$.

\vspace{-0.05in}
\subsection{Experimental Setup}
All algorithms are implemented in C\texttt{++} and experiments are performed on a machine with Intel(R) Xeon(R) CPU @ 2.10GHz with 252GB RAM on Ubuntu 18.04.3 LTS.

\noindent
\textbf{Evaluation Framework:} All experiments on Swiggy datasets are reported based on \emph{$6$-fold cross validation}. We use $5$ days for training, where we learn the parameters such as travel times for each edge, food preparation times, etc. We benchmark the performance of all assignment strategies on orders received on the unseen $6^{th}$ day. This process is then repeated $6$ times taking each day as the test day and the remaining days for training. The initial position of vehicles on the test day is set to the location at which their first GPS ping is received.

In Grubhub, we use the entire dataset for testing since no parameters need to be learned. The network is absent and hence, spatial distance is used a proxy for network distance. The distributions of food preparation times in Grubhub are given to us separately.

\noindent
\textbf{Operational Constraints: } The food-delivery service needs to operate within several domain constraints. We use the same constraints used by Swiggy. If an order remains unallocated for $30$ minutes, it is \emph{rejected}.
The rejection penalty $\Omega$ (Recall Problem~\ref{prb:online}) is set to $7200$ seconds, i.e., 2 hours. Batching of more than 3 orders is rarely observed in real data, and hence  $\maxorders$ is set to $3$. The maximum item capacity of each vehicle, $\maxload$, is set to $10$.
Most delivery companies guarantee a maximum delivery time. For Swiggy, this time is $45$ minutes, which acts as a bound on the maximum possible distance between the vehicle and pick-up location. 
Thus, in \fg, if $SP(loc(v,t),\pi[1]^r,t)>45,\; mCost(\pi,v)=\Omega$.

\noindent
\textbf{Parameters: }The default size of accumulation window $\Delta$ is $3$ minutes for City B and City C and $1$ minute for City A. $\Delta$ is smaller in City A since the volume of orders is smaller in City A. We evaluate the impact of $\Delta$ in greater depth in \S\ref{sec:parameters} (see Fig.~\ref{fig:delta}). The weighting factor $\gamma$ is set to $0.5$ (Eq.~\ref{eq:dist}). $k$ is set $200\times \frac{|O(\ell)|}{|\CV(\ell)|}$ (Sec.~\ref{sec:bp_topk}). $\eta$ (Sec.~\ref{sec:batch_hac}) is set to 60 seconds (Sec.~\ref{sec:batch_hac}). 

\begin{figure*}[t]
\vspace{-0.20in}
  \centering
\subfigure[Optimizations]{
  \label{fig:optimizations}
	\includegraphics[width=1.32in]{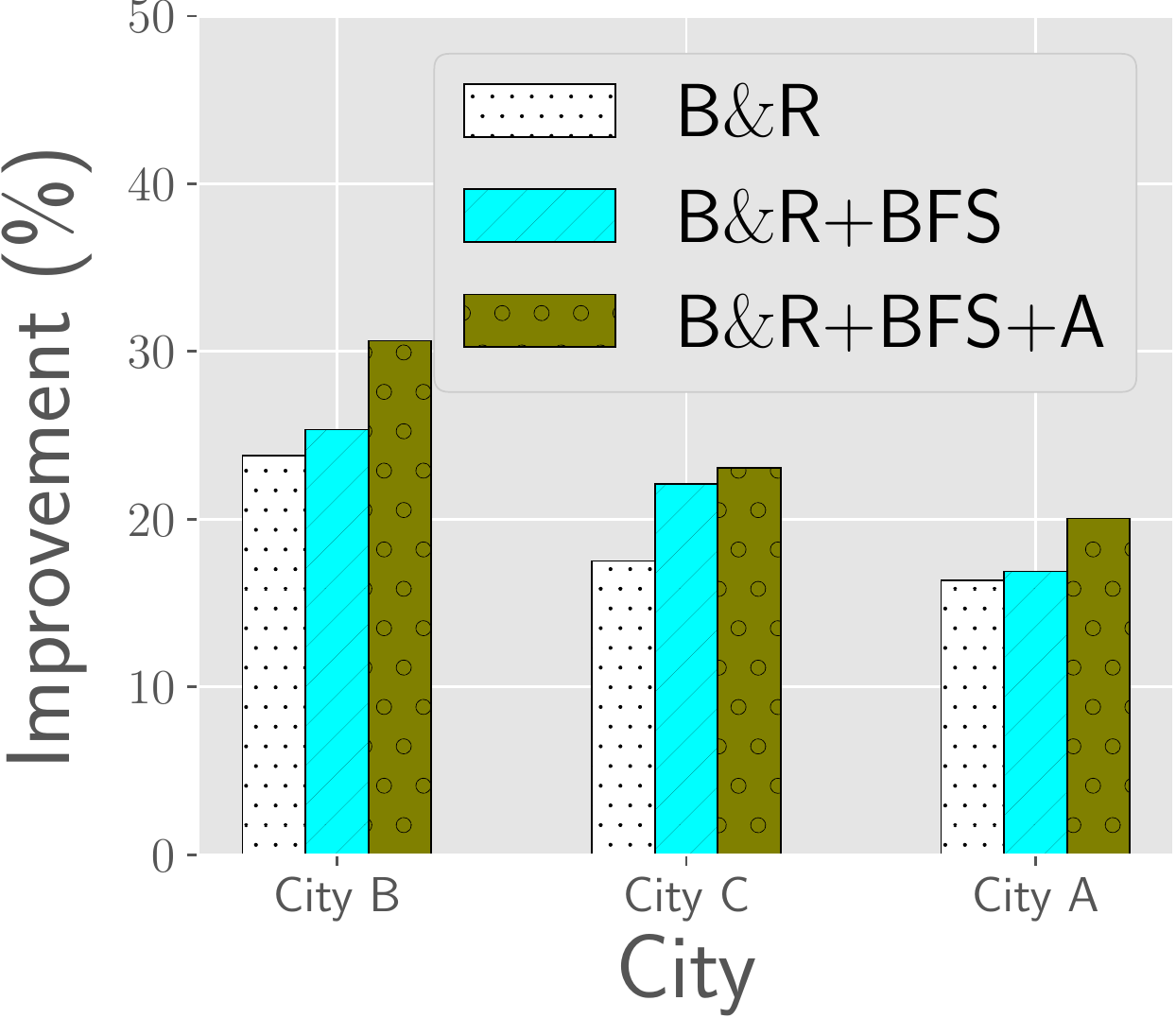}}
\subfigure[Extra Delivery Time]{
  \label{fig:drivervsedt}
	\includegraphics[width=1.32in]{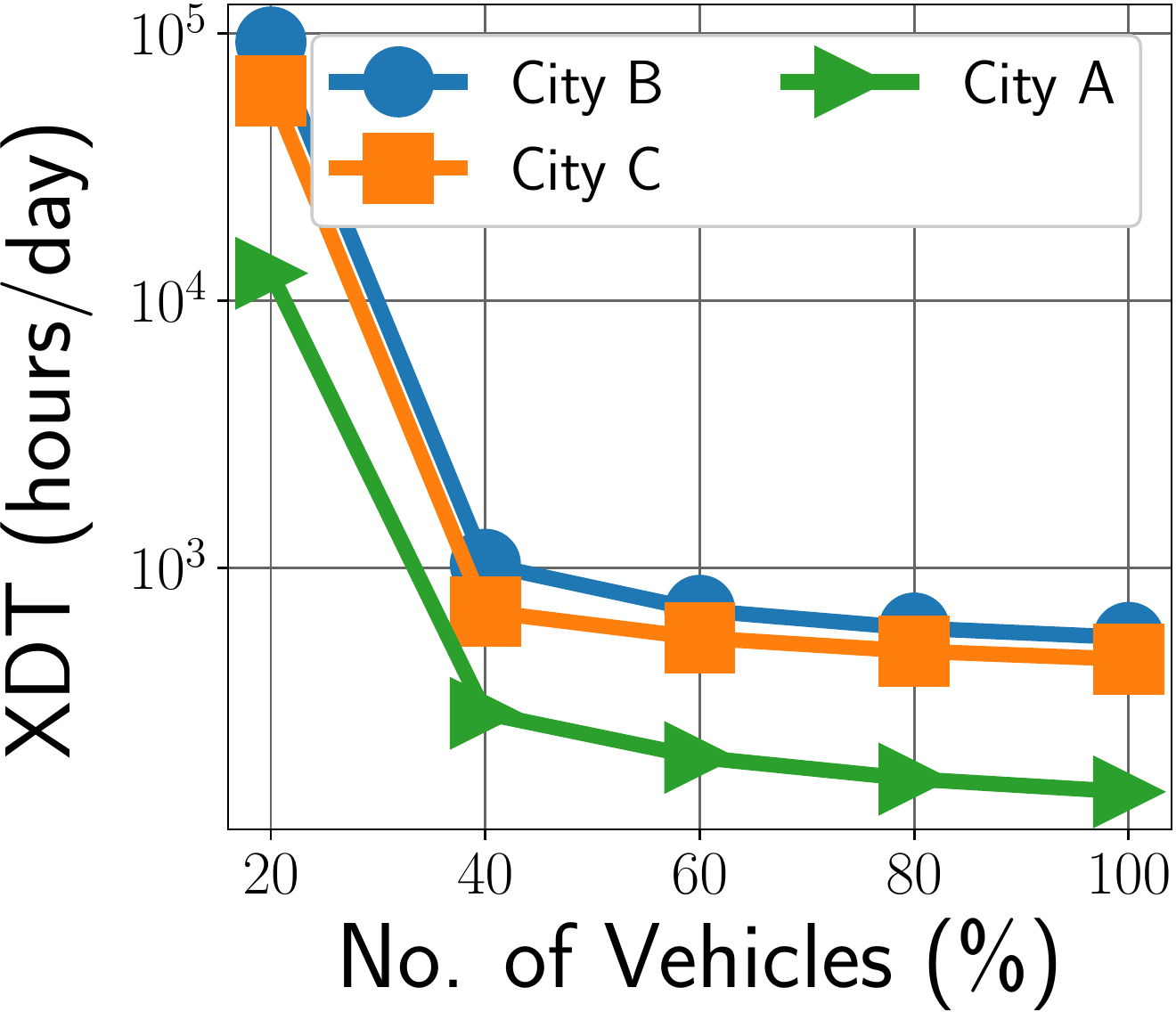}}
\subfigure[Orders/Km]{
  \label{fig:drivervsokm}
	\includegraphics[width=1.32in]{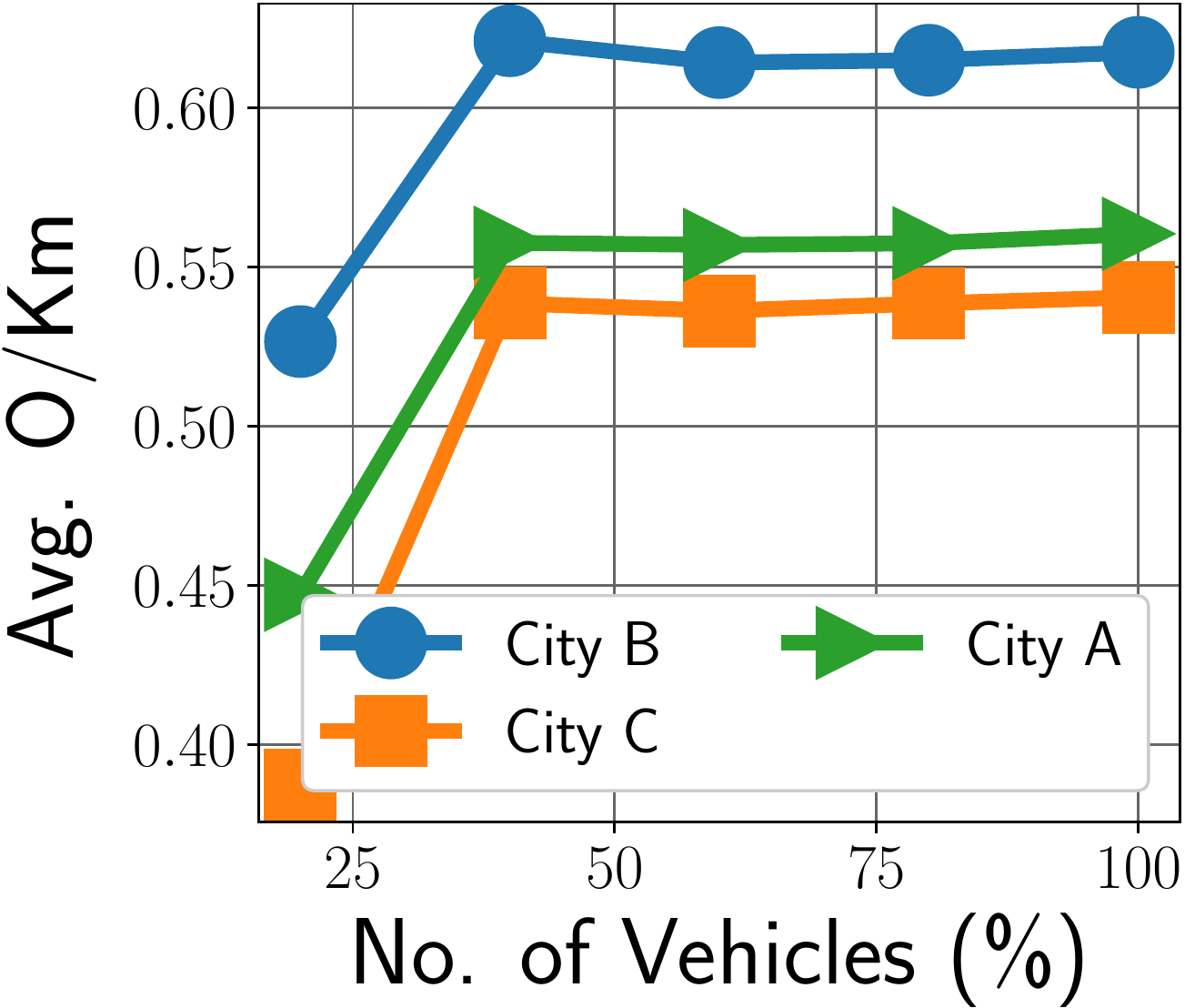}}
\subfigure[Waiting Time]{
  \label{fig:drivervswt}
	\includegraphics[width=1.32in]{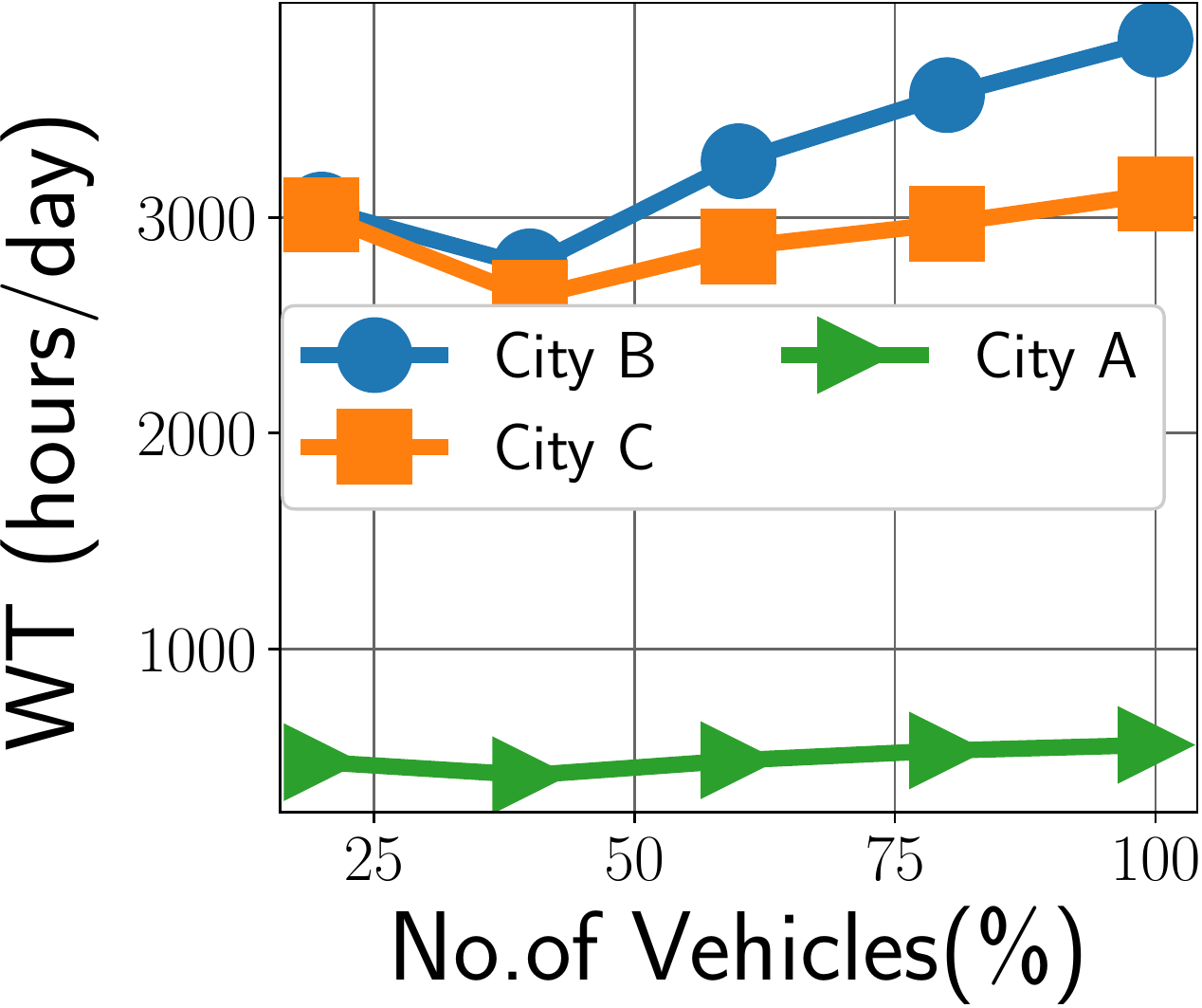}}
\subfigure[No. of Rejections]{
  \label{fig:drivervsrejections}
	\includegraphics[width=1.32in]{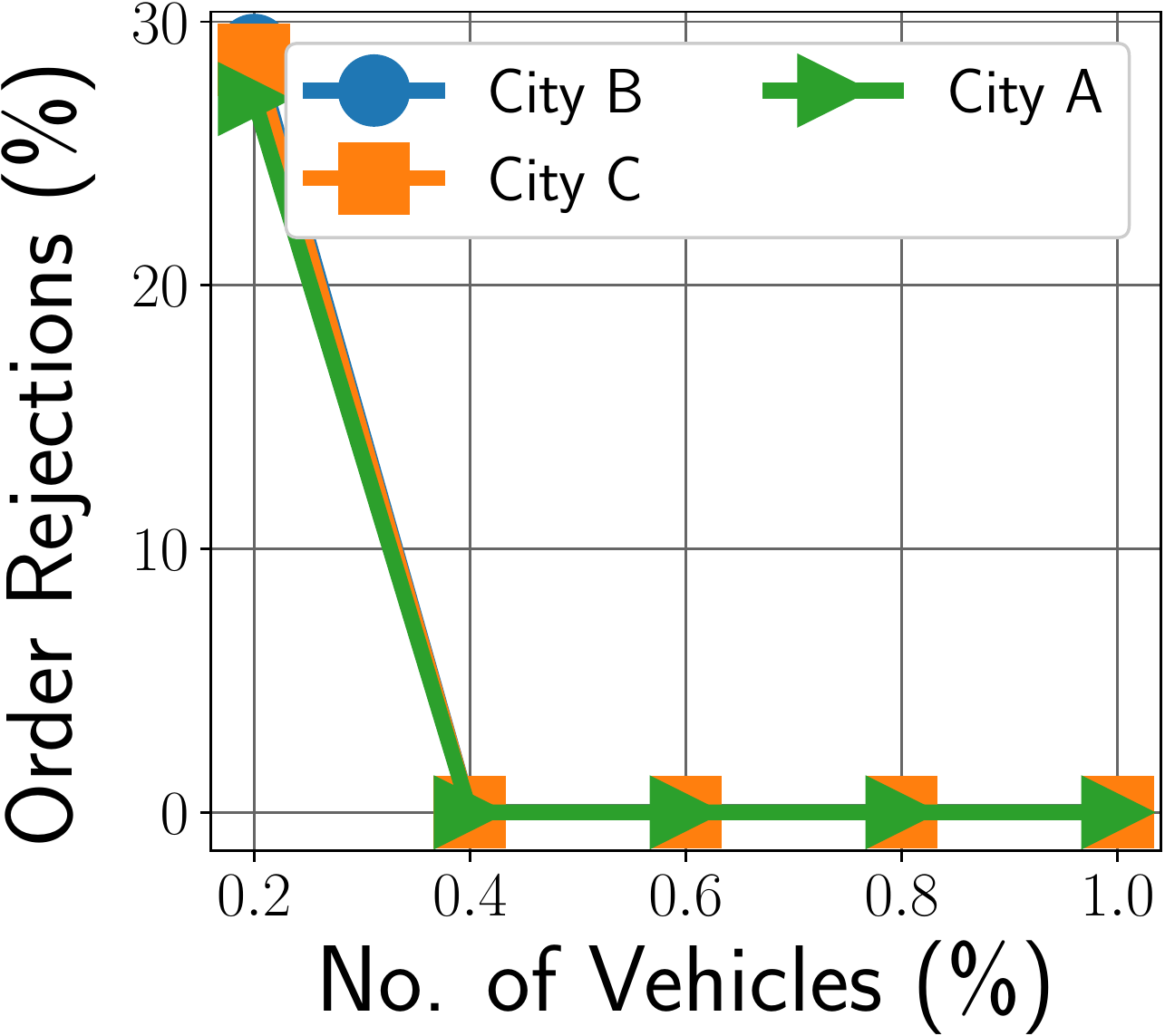}}
\vspace{-0.10in}
\caption{ (a) Impact of various optimization techniques on XDT. (b-e) Impact of number of vehicles on (b) XDT and (c) O/Km, (d) WT, and (e) percentage of order rejections.}
\vspace{-0.20in}
\end{figure*}

\noindent
\textbf{Baselines:} Reyes et al.\cite{mdrp} is the state of the art. Its limitation are discussed in detail in \S~\ref{sec:related}. In this section, we empirically compare its performance with \fmplus. In addition, we also benchmark against the Greedy approach (\S.~\ref{sec:agreedy}). We use hierarchical hub labeling~\cite{hhl} to index shortest paths queries in all benchmarked algorithms that operate on road networks.

\noindent
\textbf{Metrics: } The performance is quantified through:

 \textit{$\bullet$ Extra delivery time (XDT):} XDT is our objective function (Problem~\ref{prb:online}) and serves as the primary performance indicator.
 \textit{$\bullet$ Orders per kilometer (O/Km):} While our objective function is a customer-facing metric, O/Km quantifies operational efficiency in terms of distance travelled per order. Mathematically, let $D_k$ be the total distance covered by all vehicles while carrying $k$ orders. Then average orders per km is $\frac{\sum k\cdot D_k}{\sum D_k}$. 

We illustrate this metric through Fig.~\ref{fig:ex1}.

Assume the edge weights represent the distance in Km. Let vehicle $v_2$ start at $u_4$ and pick up $o_1$ and $o_3$ by travelling 6km and 5km respectively and finally deliver the two orders at $u_7$ and $u_8$ by travelling 8km and 5km respectively. The average number of orders per km is $\frac{0\times6 + 1\times5 + 2\times8 + 1\times5}{6+5+8+5} = 1.083$.

 \textit{$\bullet$ Average vehicle waiting time (WT):} Once a vehicle reaches its pickup location, it must wait until food is prepared. Since drivers are paid for every minute of duty, the WT (assignment time + firsmile - food preparation time) should be minimized. Similar to O/KM, WT is an indicator of operational efficiency.

\vspace{-0.05in}
\subsection{Comparison with Reyes}
\label{sec:reyes}
Reyes et al. does not incorporate the underlying road network to compute distances between locations. Instead, it uses Haversine distance. Furthermore, it uses a linear programming formulation that allows batching only if orders are from the same restaurant. Due to these unrealistic design choices, the performance suffers. As visible in Fig.~\ref{fig:reyes_xdt}, the number of man hours lost per day (XDT) is $10$ times higher on average in Reyes. In Grubhub, the difference is less due to two reasons. First, \fmplus does not have access to the road network. Second, the order volume in Grubhub is significantly lower, which reduces the impact of optimization strategies such as batching and reshuffling.

\vspace{-0.05in}
\subsection{Comparison with Greedy}
\label{sec:baseline}

Next, we compare the performance against Greedy.
Fig.~\ref{fig:improvement_edt} presents the results. The extra delivery time in \fmplus is, on average, $30\%$ less than Greedy. The improved performance of \fmplus indicates that optimizing a global objective through minimum weight perfect matching helps. Furthermore, through a dedicated batching step, \fmplus is better able to cope with scarcity of delivery vehicles. We also observe that for both \fmplus and Greedy, the XDTs are substantially higher in City B and City C. This is a direct result of the fact that they represent two large metropolitan cities in India with  substantially higher order volumes.
In Fig.~\ref{fig:improvement_okm} and Fig.~\ref{fig:improvement_wt}, we measure the improvement with respect to our secondary metrics O/km and WT. Recall that we do not directly optimize these metrics. However, they are important indicators of operational efficiency and hence good performance is desired. On both metrics, we observe a dramatic improvement. For example, the combined man hours wasted by delivery personnel waiting for food to be prepared is higher by $2000$ hours in both City B and City C when orders are assigned through Greedy. On the other hand, the orders delivered per km is $20\%$ higher in \fmplus. This is a direct consequence of having a dedicated batching component in the \fmplus pipeline. 
\vspace{-0.05in}
\subsection{Scalability }
A critical feature of \fmplus is its scalability as we see in Figs.~\ref{fig:overflow}-\ref{fig:overflow_peak}. As discussed earlier, if the rate of order arrival is higher than the rate of order assignment, then the algorithm is not efficient enough for real-time deployment. In this experiment, we partition the timeline into $3$-minute windows and find the percentage of windows where the assignment time of all orders that arrived within that window is higher than $3$ minutes. We call such windows \emph{overflown}. In addition to Greedy and \fmplus, we also include vanilla Kuhn-Munkres (\fm) without any of the other optimizations as a baseline in this experiment. This allows us to precisely measure the impact of best-first search on scalability. Fig.~\ref{fig:overflow} presents the percentage of overflown windows across all slots and Fig.~\ref{fig:overflow_peak} present the same statistics in the peak slots corresponding to lunch and dinner periods. As clearly visible, \fmplus is the only algorithm with $0\%$ overflows. In contrast, at least $80\%$ of the windows are overflown in Greedy and \fm during peak hours in City B and City C. In City A, scalability is not an issue for any of the techniques since the order volume is low. 

The same pattern is also visible in Figs.~\ref{fig:time}, where we examine the raw running times. We see that \fmplus is the fastest and Greedy is the slowest on average. 
\vspace{-0.05in}
\subsection{Impact of Optimizations}
\label{sec:optimizationimpact}
\fmplus builds on \fm by layering it with several optimizations such as Batching and Reshuffling (B\&R), sparsified \fg through best-first-search (BFS) and Angular distance (A). In this section, we systematically inspect the impact of these optimizations.

First, we compare the performance of \fmplus with just Kuhn-Munkres (\fm) that does not include any of the optimizations.
To enable comparison of performance across cities and metrics, we plot the \emph{improvement ($\%$)} obtained by \fmplus over \fm. For example, let $\fm_{XDT}$ and $\fmplus_{XDT}$ be the extra delivery times with algorithm \fm and \fmplus respectively. Then, improvement of \fmplus with respect to \fm is:

\vspace{-0.10in}
\begin{equation}
\label{eq:improvement}
\textsc{Imp}_{\fm,XDT}=\frac{\fm_{XDT}-\fmplus_{XDT}}{\fm_{XDT}}\times 100
\end{equation} 

For the metric of O/km, since higher values indicate better performance, we switch the numerator of Eq.~\ref{eq:improvement} to $\fmplus_{O/Km}$ $-\fm_{O/Km}$. Negative values indicate better performance by the baseline, whereas positive values indicate superior performance by \fmplus.

\begin{figure*}[t]
\vspace{-0.20in}
  \centering
  \subfigure[XDT]{
  \label{fig:etavsedt}
	\includegraphics[width=1.09in]{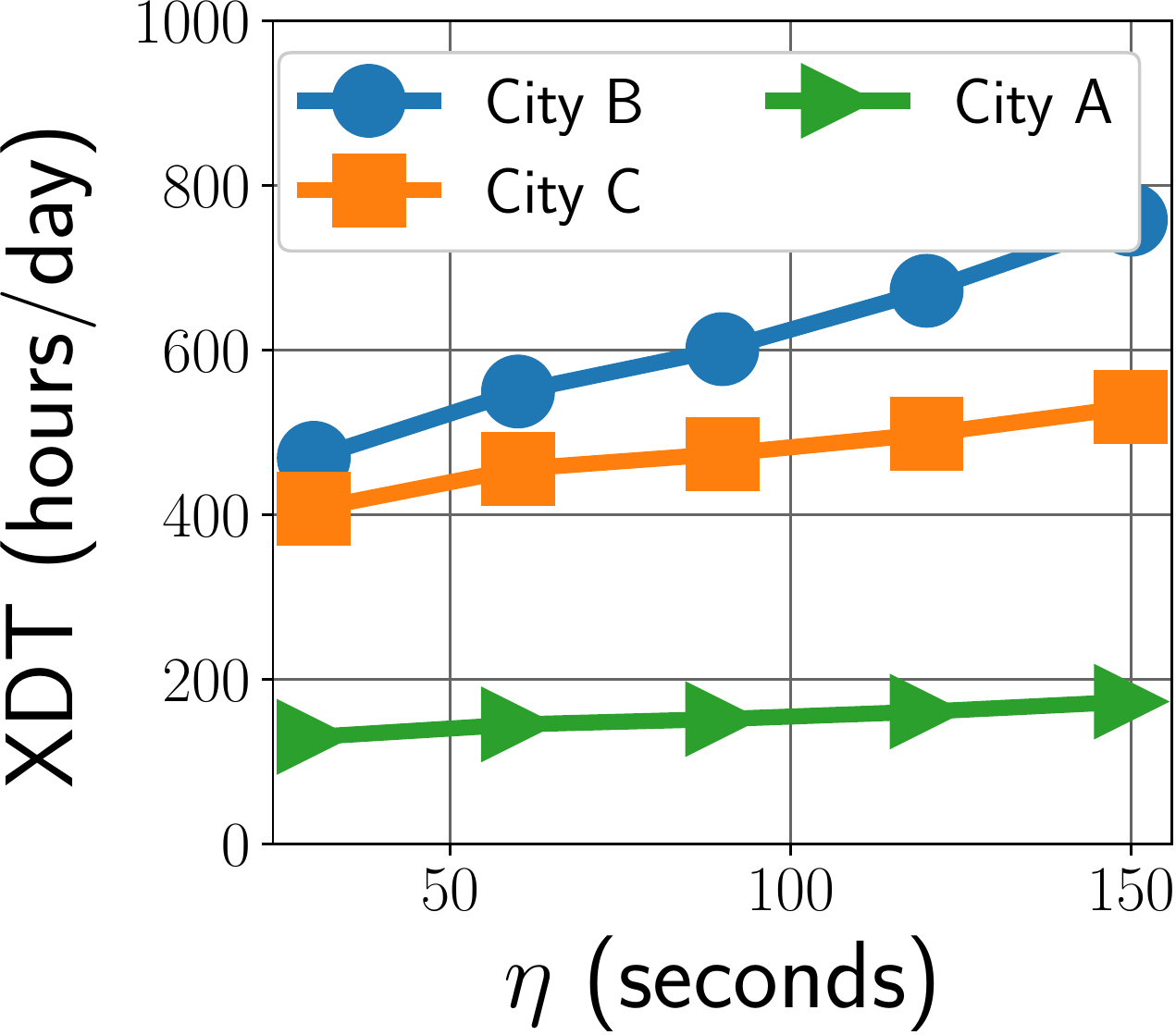}}
\subfigure[Orders/Km]{
  \label{fig:etavsokm}
	\includegraphics[width=1.09in]{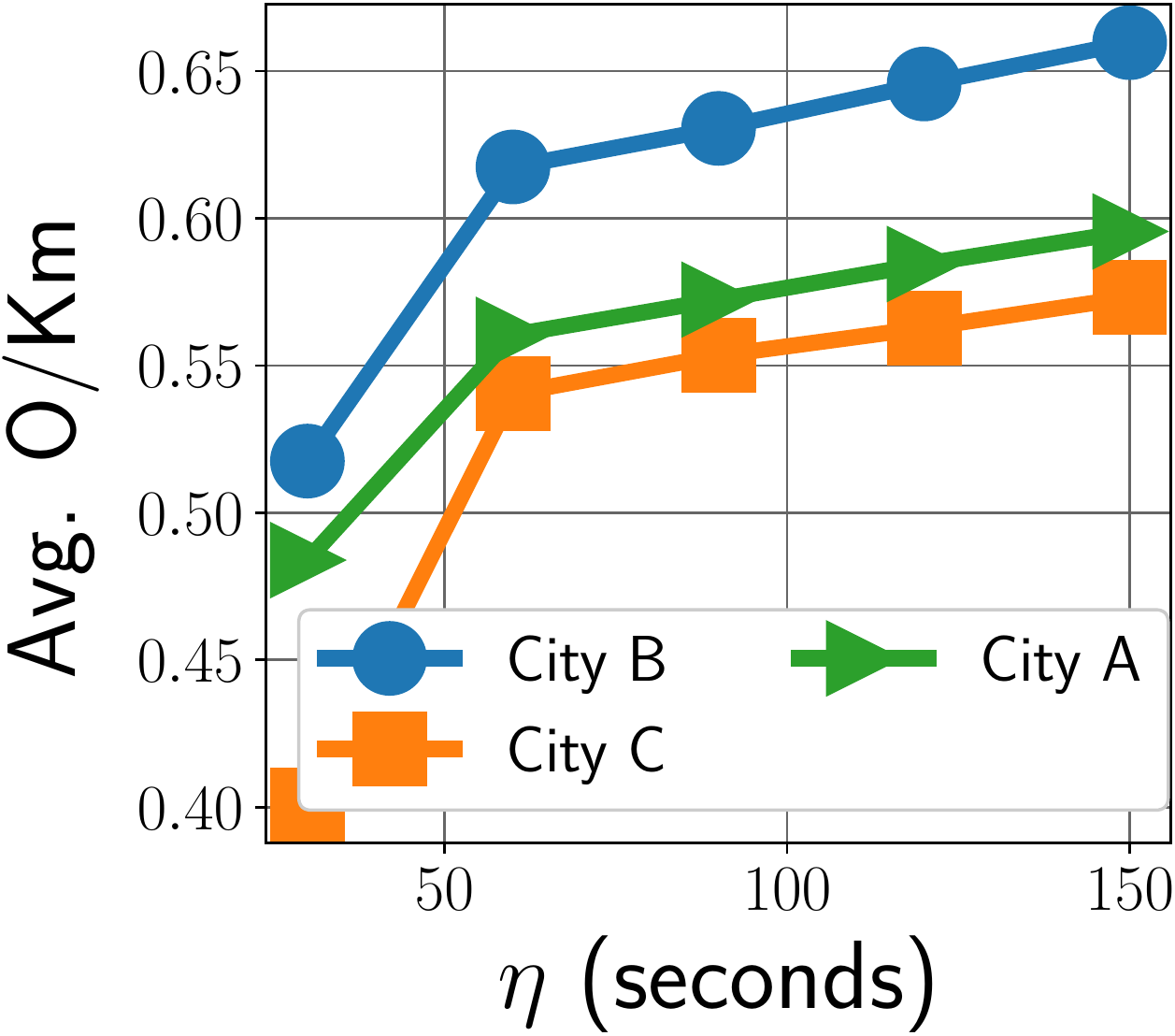}}
\subfigure[Waiting Time]{
  \label{fig:etavswt}
	\includegraphics[width=1.11in]{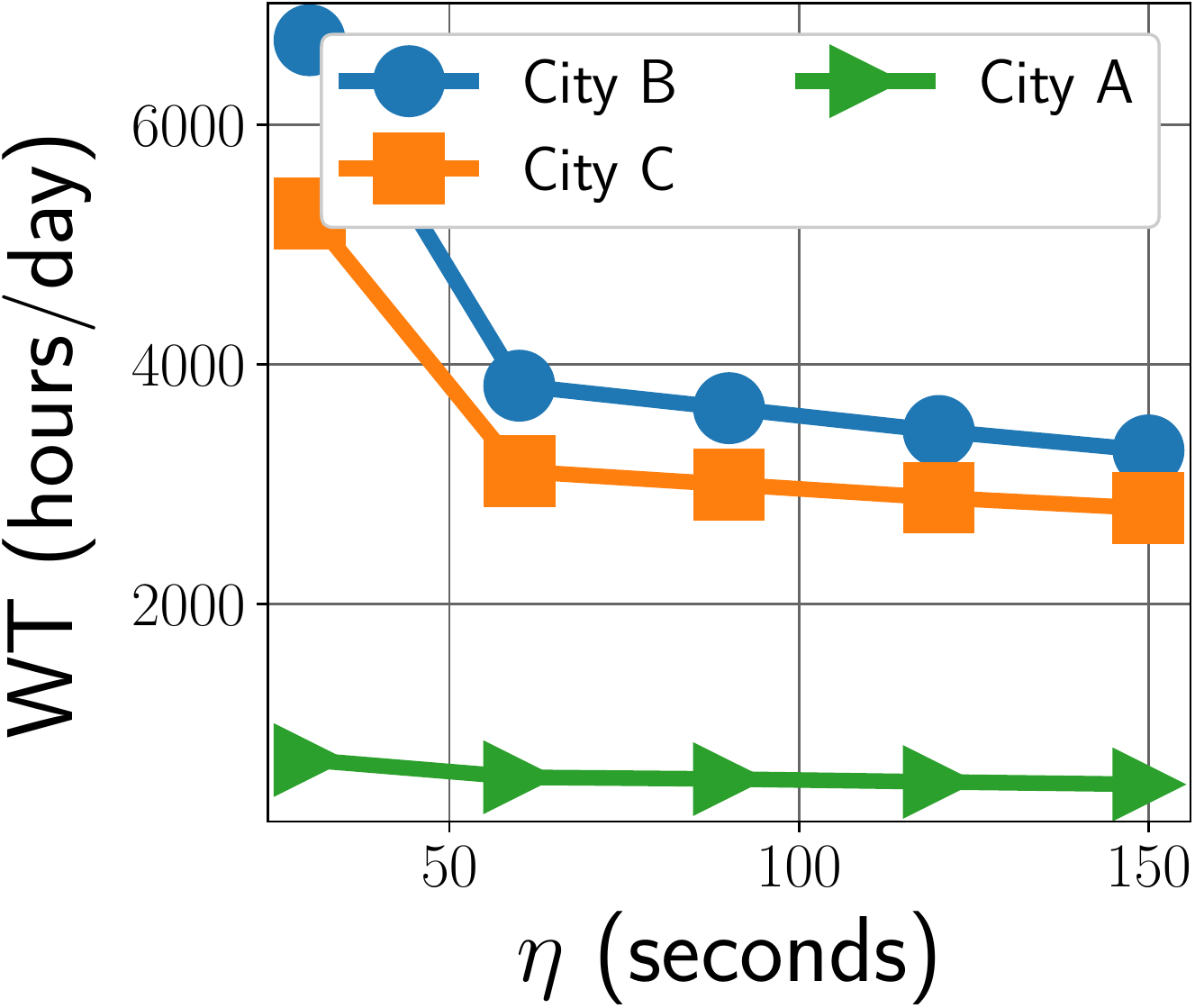}}
\subfigure[XDT]{
  \label{fig:deltavsedt}
	\includegraphics[width=1.08in]{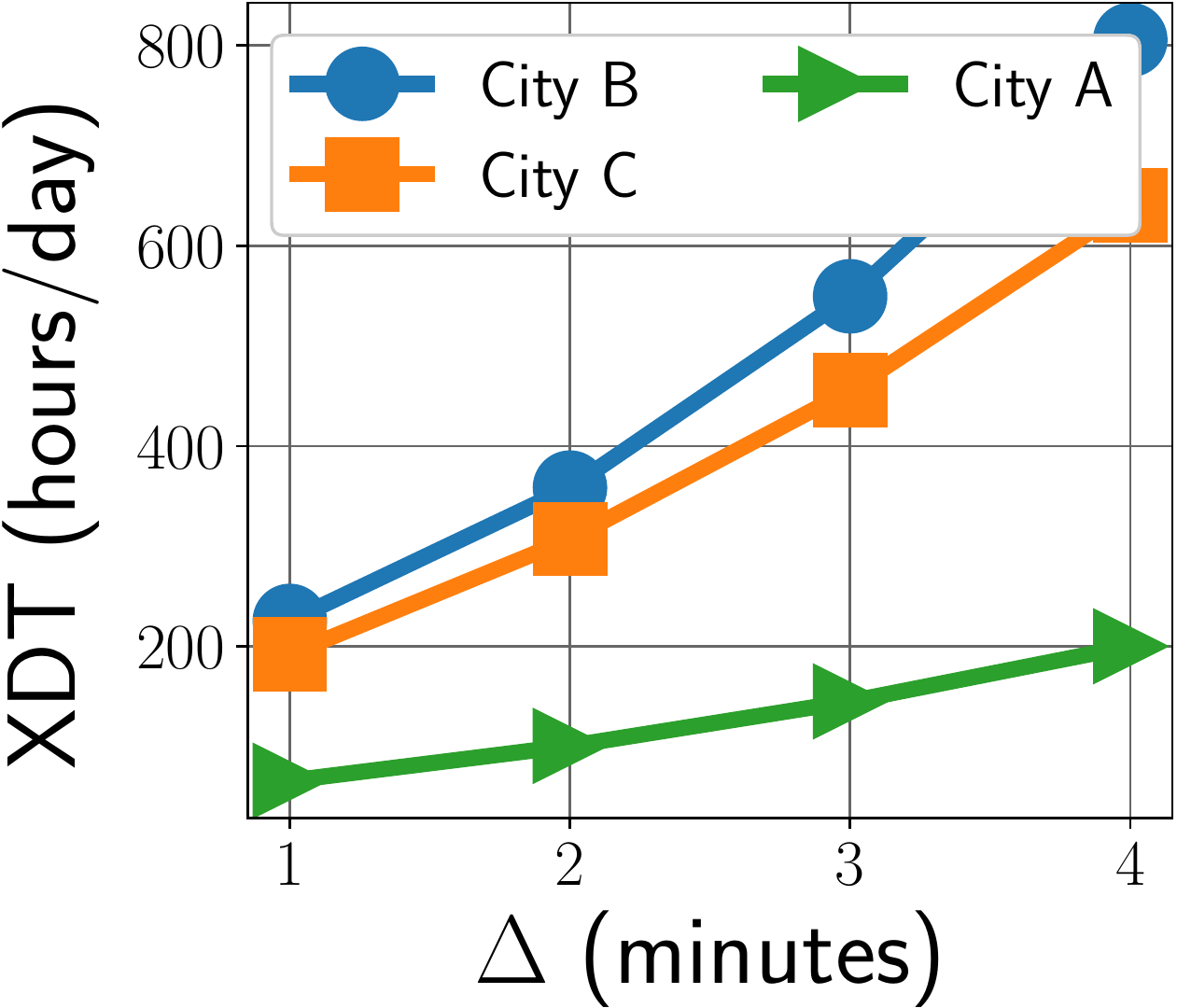}}
\subfigure[Orders/Km]{
  \label{fig:deltavsokm}
	\includegraphics[width=1.08in]{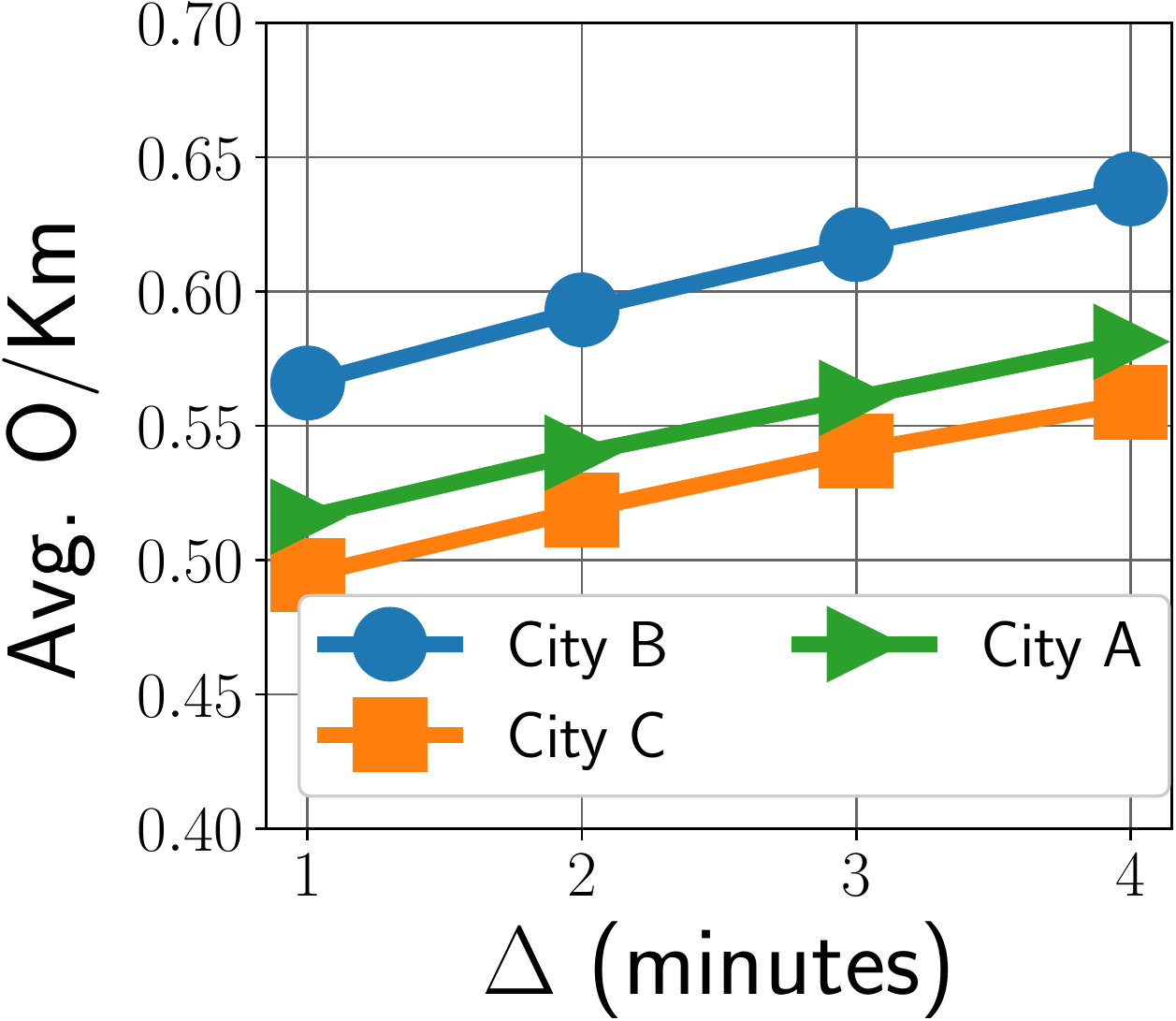}}
\subfigure[Waiting Time]{
  \label{fig:deltavswt}
	\includegraphics[width=1.08in]{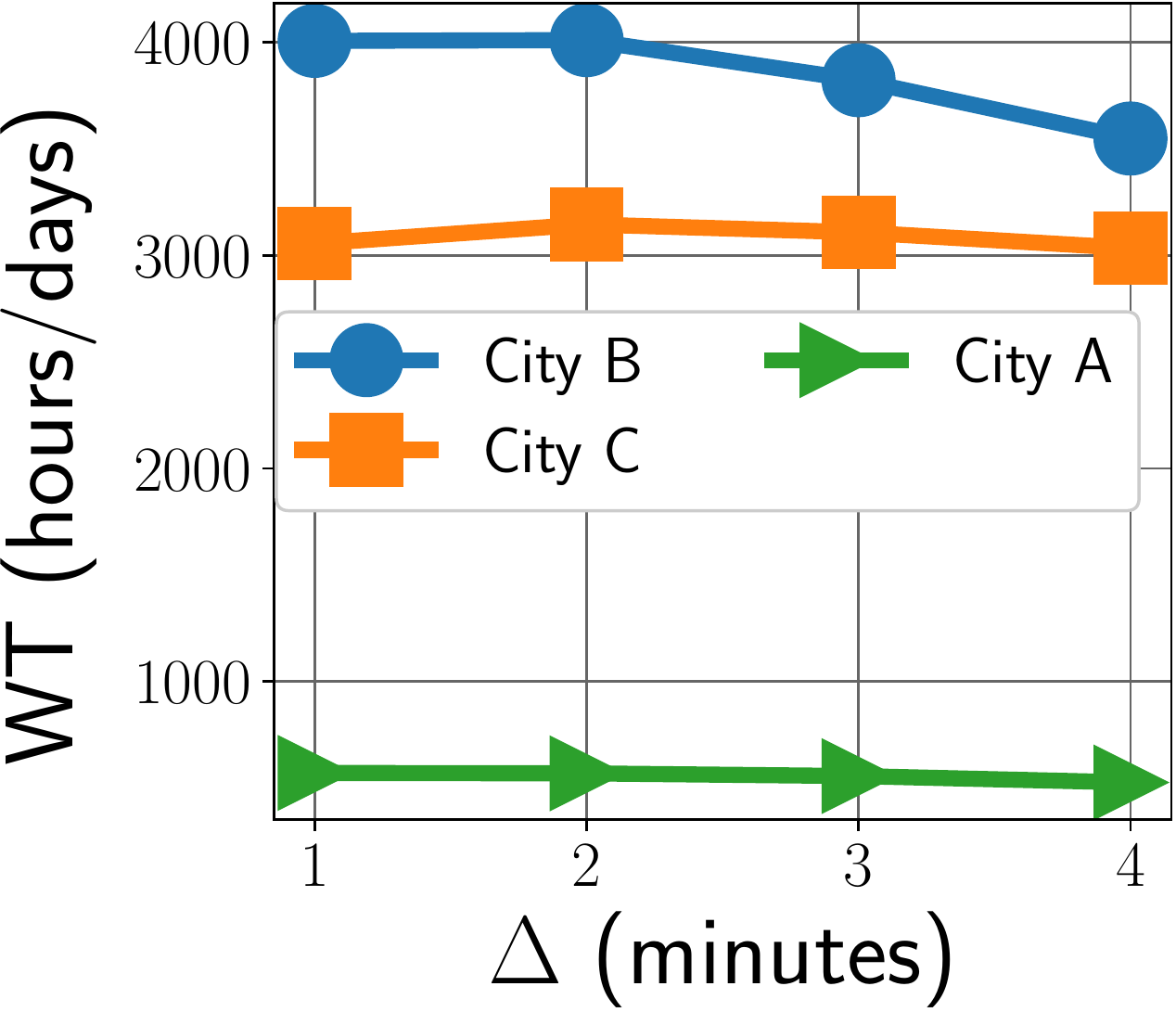}}\\
	\vspace{-0.10in}
\subfigure[Running Time (All)]{
  \label{fig:deltavstime}
	\includegraphics[width=1.27in]{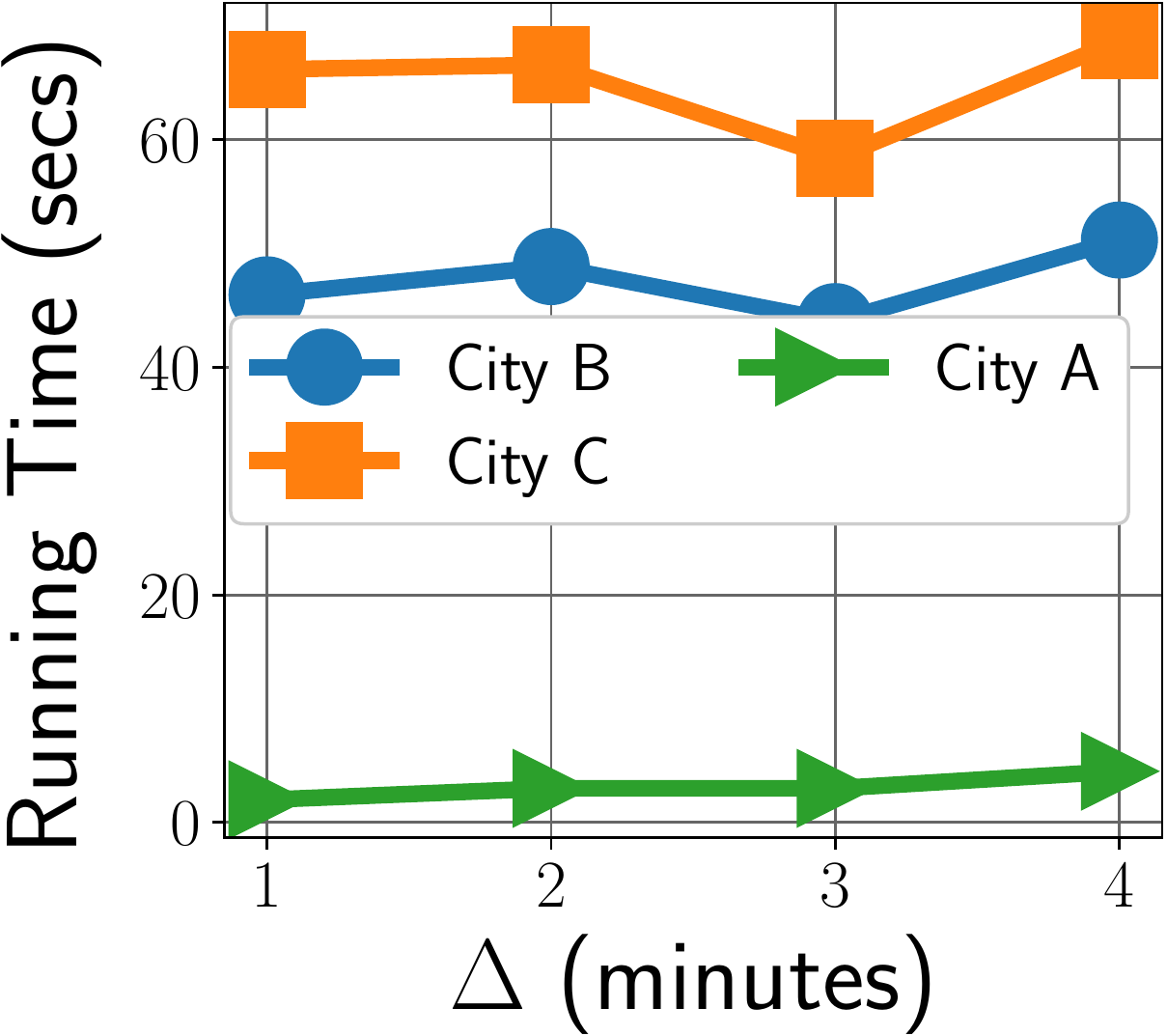}}
	\subfigure[XDT]{
  \label{fig:kvsedt}
	\includegraphics[width=1.32in]{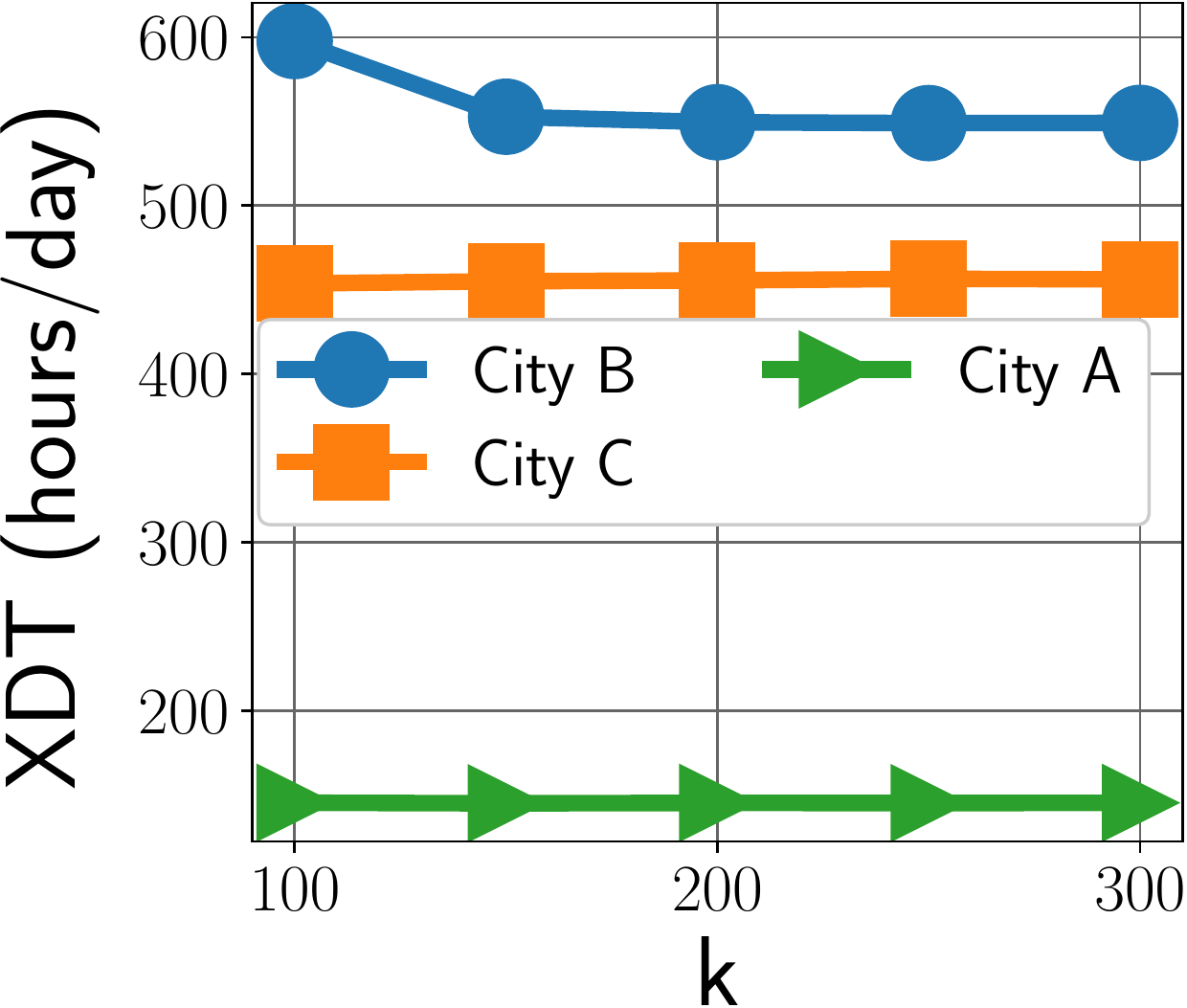}}
\subfigure[Orders/Km]{
  \label{fig:kvsokm}
	\includegraphics[width=1.27in]{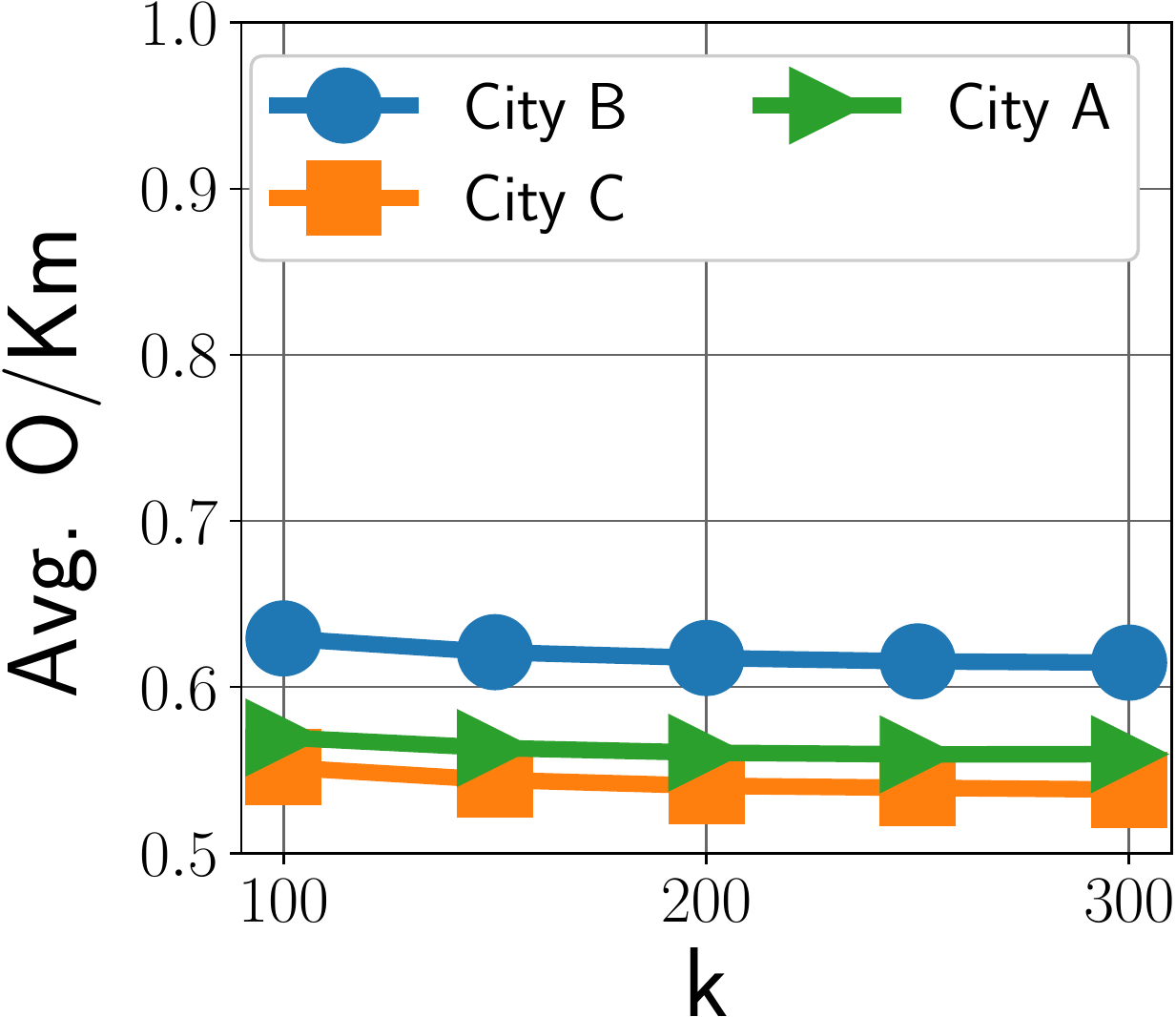}}
\subfigure[Waiting Time]{
  \label{fig:kvswt}
	\includegraphics[width=1.26in]{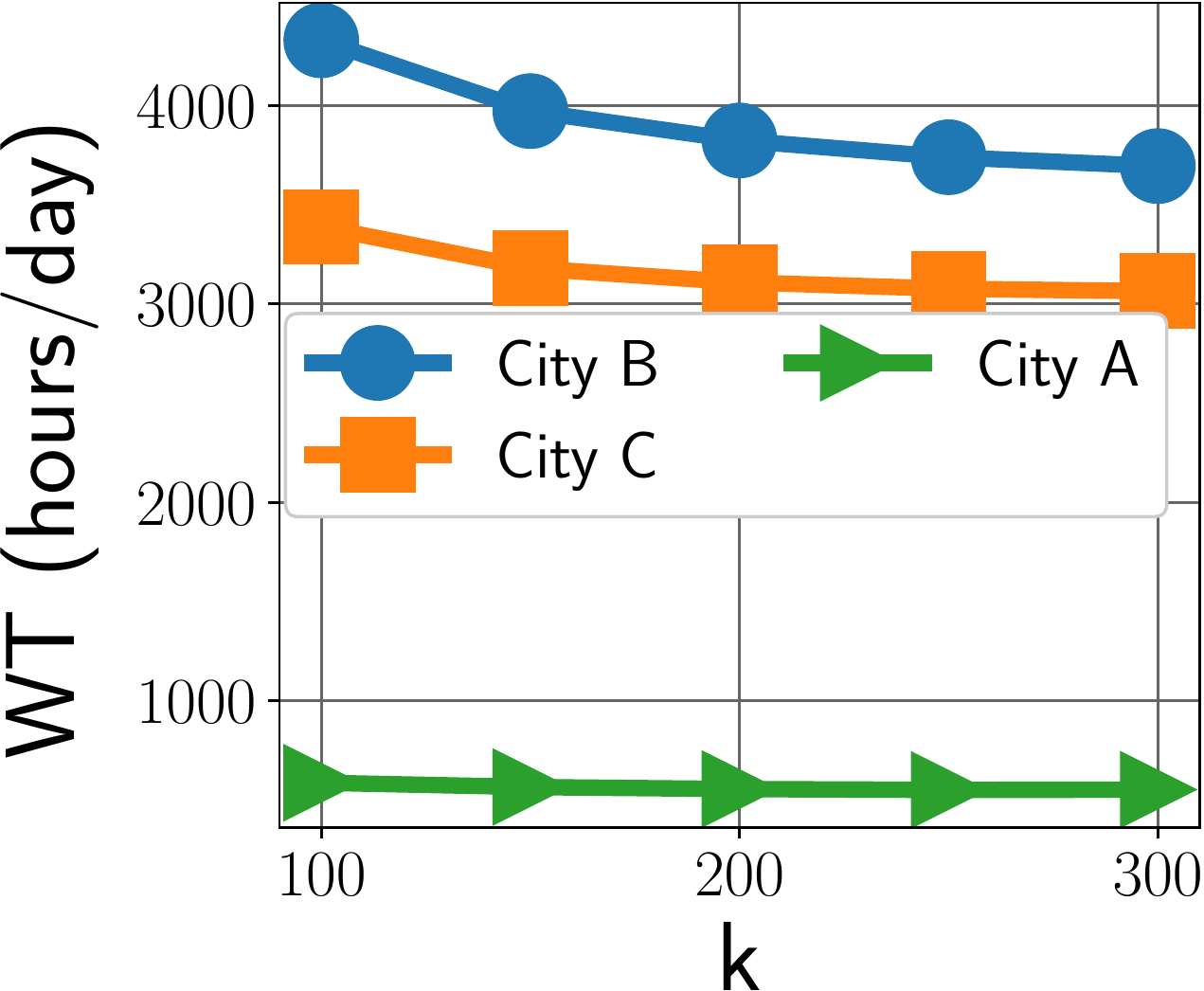}}
\subfigure[Running Time (All)]{
  \label{fig:kvstime}
	\includegraphics[width=1.32in]{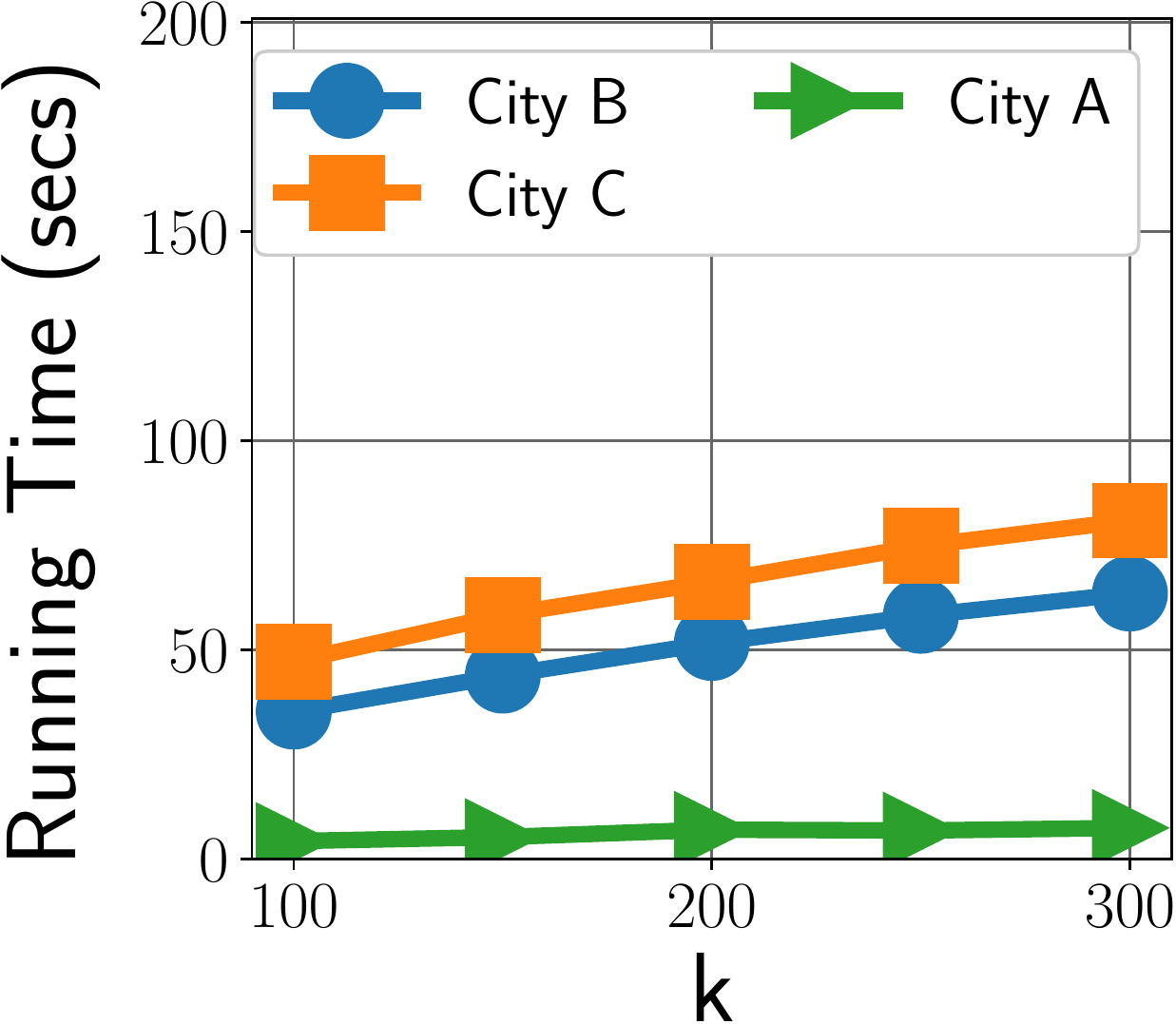}}
\vspace{-0.10in}
\caption{Impact of $\eta$ on (a) XDT, (b) O/Km, and (c) WT. Impact of accumulation window $\Delta$ on (d) XDT, (e) O/Km, (f) WT, and (g) running time. Impact of $k$ on (h) XDT, (i) O/Km, (j) WT, and (k) running time.}
\label{fig:delta}
\vspace{-0.15in}
\end{figure*}
\textbf{\fm Vs. \fmplus } In Figs.~\ref{fig:timeslot_edt}-\ref{fig:timeslot_wt}, we measure the improvement over \fm across various timeslots of the day. The timeslot partitioning is done in the same manner as in Fig.~\ref{fig:dist}. In XDT, we observe two peaks in Fig.~\ref{fig:timeslot_edt}. These peaks correspond to the lunch and dinner periods (Recall Fig.~\ref{fig:dist}), and indicates that as order volumes increase, the advantage of \fmplus becomes even more pronounced. In the other two metrics (Figs.~\ref{fig:timeslot_okm} and \ref{fig:timeslot_wt}) as well, we observe slight increase in improvement during lunch and dinner periods.

\textbf{Individual impact of optimizations: }In Fig.~\ref{fig:optimizations}, we measure the individual impact of each of the optimizations employed by \fmplus. 
Specifically, we add each optimization in a layered manner, and compute the improvement obtained over \fm. In Fig.~\ref{fig:optimizations}, \textsc{B\&R} represents Batching and Reshuffling on top of \fm, \textsc{B\&R+BFS} represents Batching, Reshuffling and best-first search, and \textsc{B\&R+BFS+A} further adds angular distance.

As visible, the highest impact comes from Batching. This is natural since \fm cannot batch two orders from the same accumulation window. In contrast, \fmplus performs an in-depth analysis of \emph{batchability} through the proposed clustering algorithm. In the clustering process, the quality of all batches are carefully monitored and the process continues only if the average cost of all batches is above a threshold $\eta$.
Next, we note that although in BFS we sparsify the \fg, the improvement increases (B\&R+BFS). This seems counter-intuitive. To understand this behavior better, we deep-dive into the data related to this experiment. We observe that removal of edges in \fg leads to non-assignment of batches that are not close to any of the available vehicles. Thus, they feed into the subsequent accumulation window, in which many of them get allocated to a better vehicle match. When working with the full \fg, some far-away pairings do happen, which increases XDT. Finally, increase in improvement($\%$) due to the addition of angular distance showcases the need to adapt to the dynamic vehicle positions.

\vspace{-0.05in}
\subsection{Impact of Vehicles}
So far we have used the same number of vehicles that were available in the real world in the corresponding timeslot. \textit{What happens if the number of vehicles is significantly smaller? Furthermore, is there scope to improve the economics of food-delivery business by reducing the number of vehicles?} Our next experiment studies these questions. Specifically, we \emph{subsample} the number of delivery vehicles and observe its impact on performance. Figs.~\ref{fig:drivervsedt}-\ref{fig:drivervswt} present the average XDT per day, O/KM, and WT per day. 

As expected, the XDT reduces with more vehicles. However, the rate of reduction is small beyond $40\%$, which indicates that the number vehicles can be significantly reduced without causing any noticeable impact on the customer experience. The impact on O/Km and WT appears odd at $20\%$ vehicles. Specifically, we would expect O/Km to decrease and WT to increase with increase in vehicles. While this pattern is visible in the region $[40\%,100\%)$, in the range of $[20\%,40\%)$, an opposite behavior is observed. To understand this phenomenon better, we look into the order rejection rate as a function of the number of vehicles (Fig.~\ref{fig:drivervsrejections}). We notice that at $20\%$, close to $30\%$ of the orders are rejected due to violating various operational constraints such as \maxorders, \maxload, and the delivery guarantee of $45$ minutes. Due to the large reduction in serviceable orders, the anomalous behavior is observed in the range $[20\%,40\%)$. We also point out that City B has the best O/Km due to the order-to-vehicle ratio being the highest in this city (Fig.~\ref{fig:dist}). On the other hand, City C has the largest WT, since the order-to-vehicle ratio is comparatively lower despite it having the largest number of restaurants.

\vspace{-0.05in}
\subsection{Impact of Parameters}
\label{sec:parameters}

\textbf{Impact of $\eta$:}
Higher $\eta$ increases the likelihood of orders being batched. When an adequate number of delivery vehicles are available, with more batching, we expect the XDT to increase (Thm.~\ref{thm:monotonicity}). On the other hand, higher batching leads to better O/Km and reduced WT.
 This behavior is visible in Figs.~\ref{fig:etavsedt}-\ref{fig:etavswt}. We note that in both O/Km and WT, the gradient of increase/decrease slows beyond $\eta=60$ seconds, which indicates that setting $\eta=60$ seconds obtains a good balance between customer satisfaction and operational efficiency.

\textbf{Impact of $\Delta$: } As discussed in Sec.~\ref{sec:agreedy}, $\Delta$ has both positive and negative impacts on quality as well as efficiency. Fig.~\ref{fig:delta} studies its impact empirically. We observe that as $\Delta$ increases, the XDT increases as well (Fig.~\ref{fig:deltavsedt}). This behavior can be explained from the fact that with larger $\Delta$, an order may need to wait longer till assignment starts. With increase in $\Delta$, the assignment time goes up and hence, the WT decreases  (Fig.~\ref{fig:deltavswt}). O/Km improves with $\Delta$ since more orders are aggregated within the accumulation window, which in turn allows clustering to batch more orders together. In terms of computational efficiency, larger $\Delta$ increases the cost of matching and batching. However, it reduces the number of windows on which assignment is performed. As visible in Figs.~\ref{fig:deltavstime}, the sweet spot is achieved at $\Delta=3$ minutes for City C and City B and $1$ minute in City A.

\textbf{Impact of $k$: } $k$ dictates the maximum degree of each vehicle in  $\fg$. All three qualitative metrics display minimal improvement with increase in $k$ (Figs.~\ref{fig:kvsedt}-\ref{fig:kvswt}). However, the increase in running time is significant in City B and City C. This indicates that limiting $k\in[100,200)$ provides a good balance between efficacy and efficiency.

\begin{figure}[t!]
\vspace{-0.20in}
  \centering
\subfigure[Extra Delivery Time]{
  \label{fig:gammavsedt}
	\includegraphics[width=1.30in]{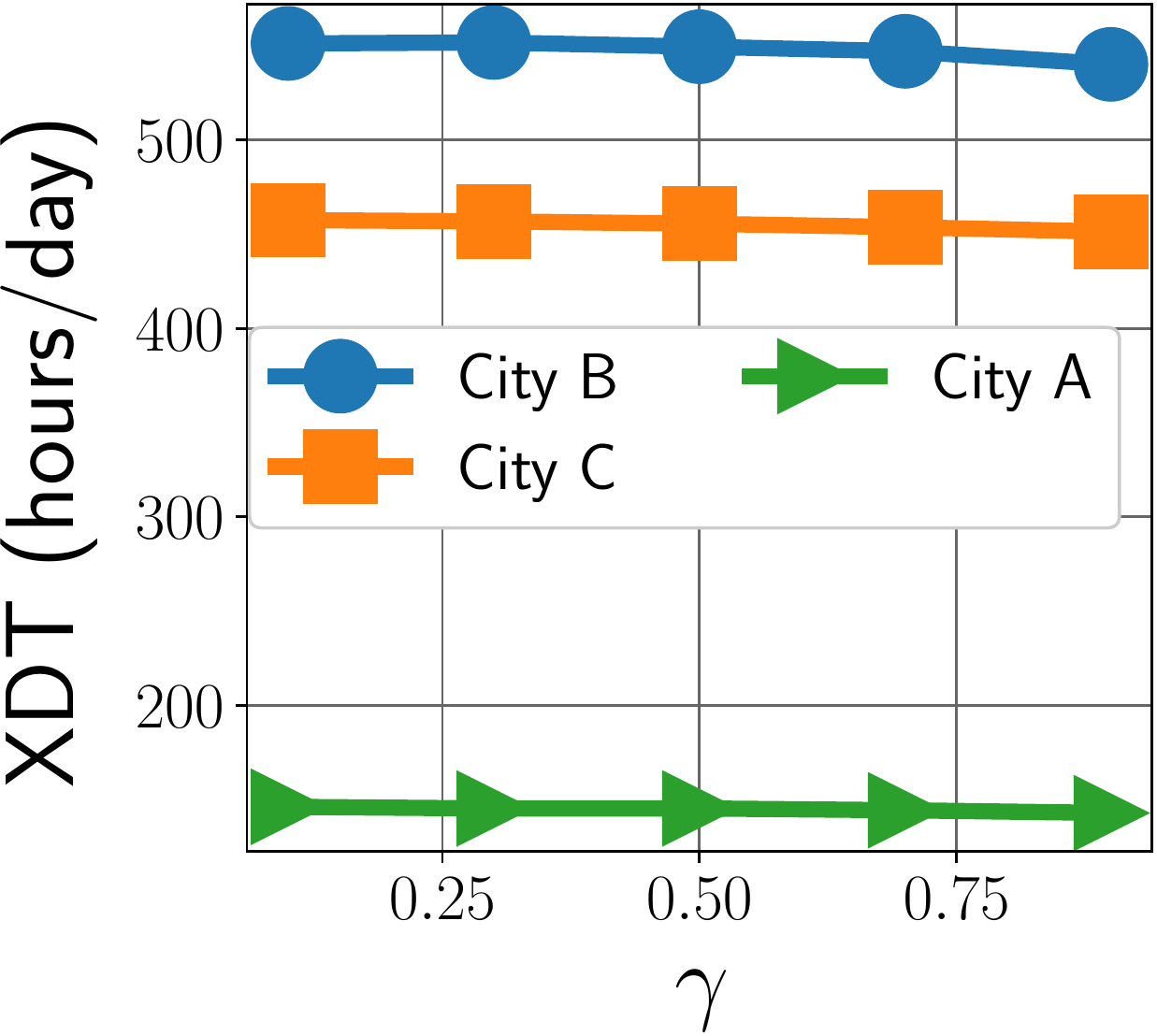}}
\subfigure[Orders/Km]{
  \label{fig:gammavsokm}
	\includegraphics[width=1.25in]{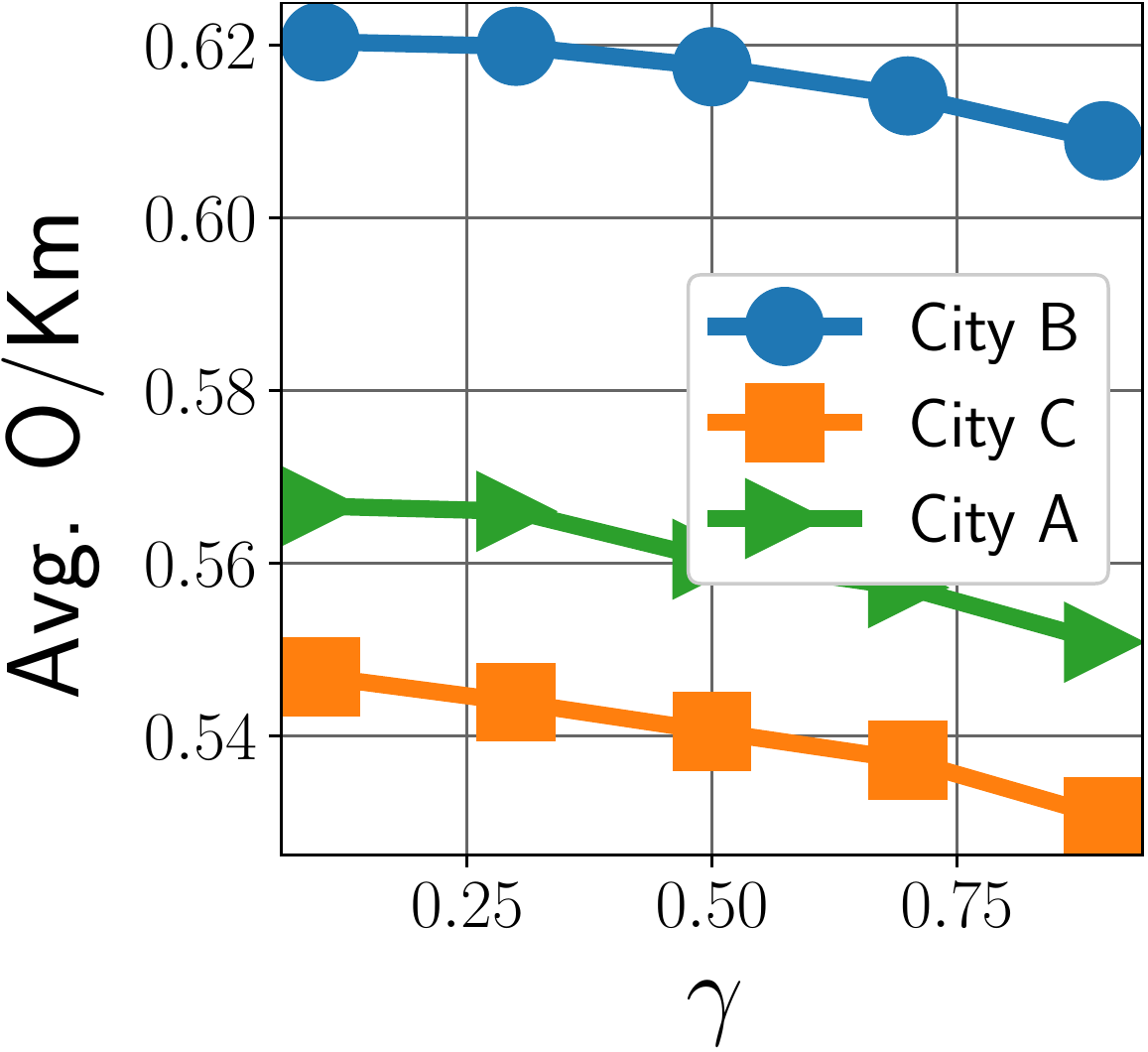}}\\
\vspace{-0.10in}
\subfigure[Waiting Time]{
  \label{fig:gammavswt}
	\includegraphics[width=1.29in]{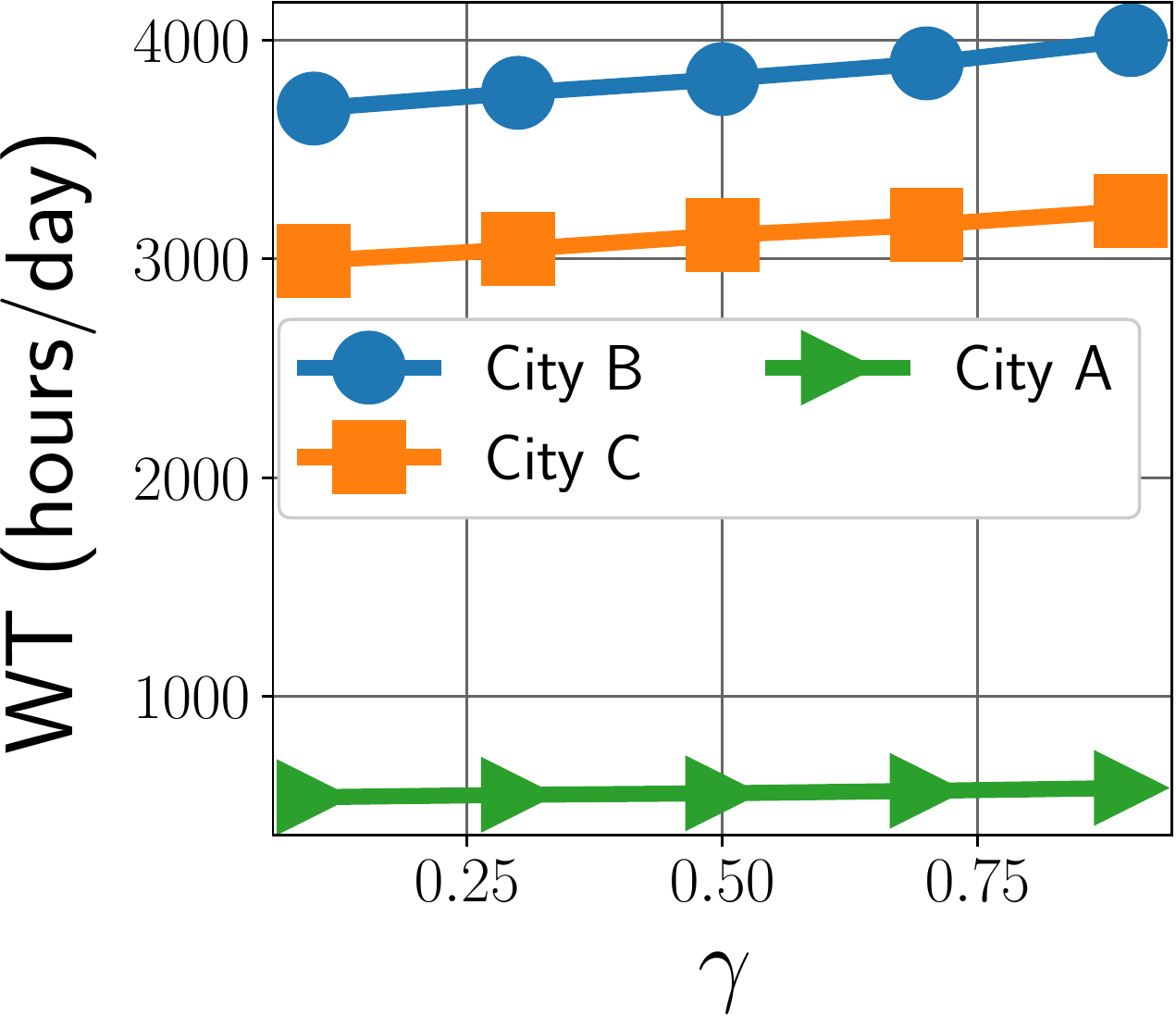}}
\subfigure[Rejection Rate in City B]{
  \label{fig:gammavsrejections}
	\includegraphics[width=1.30in]{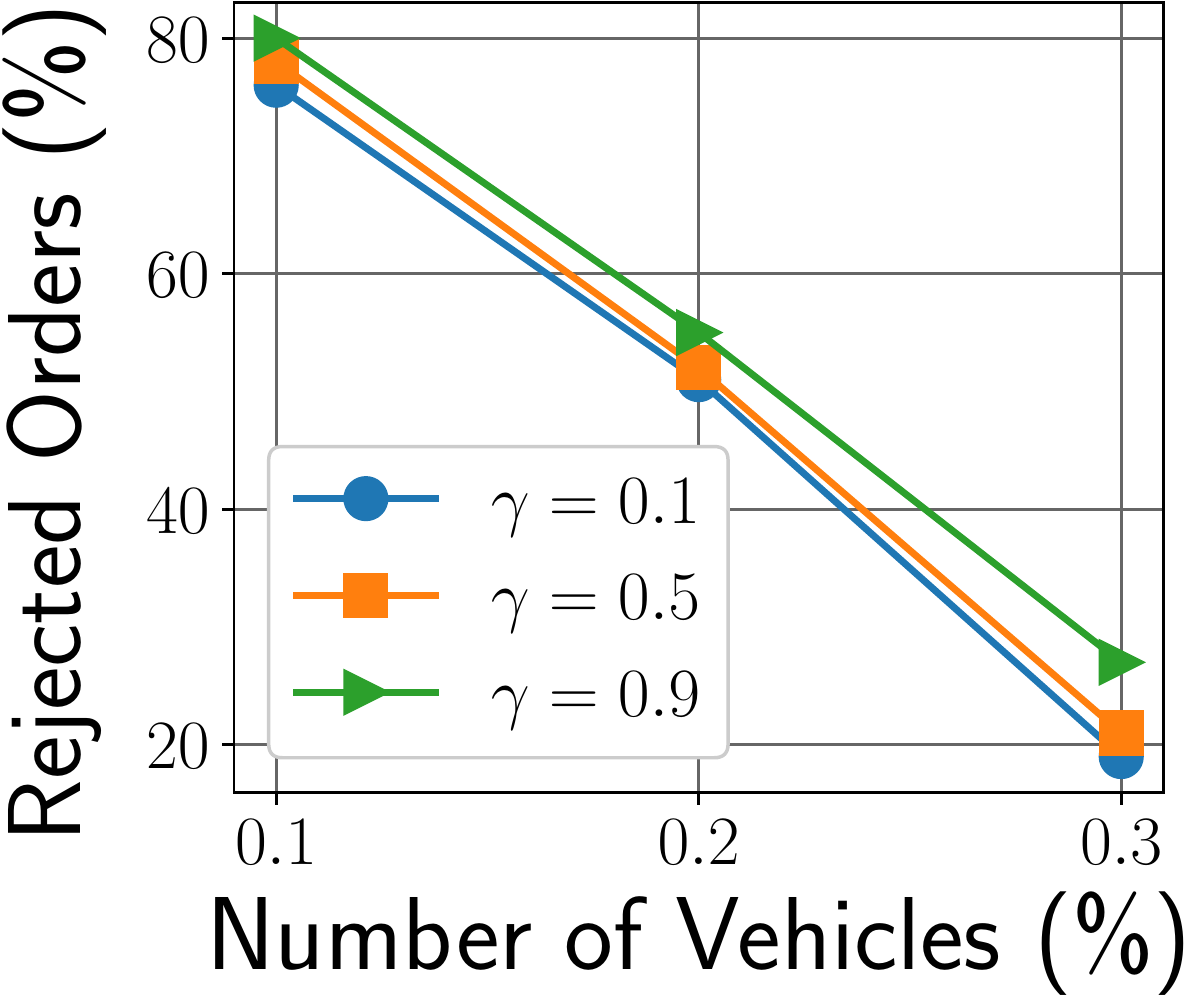}}
\vspace{-0.10in}
\caption{Impact of $\gamma$ on (a) XDT, (b) O/Km, (c) WT, and (d)  rejection rate.}
\vspace{-0.20in}
\end{figure}

\textbf{Impact of $\gamma$: }$\gamma$ balances the importance of angular distance and travel time in defining edge weights of the road network.
 Figs.\ref{fig:gammavsedt}-\ref{fig:gammavswt} analyze the impact of $\gamma$ on the performance. While the XDT remains almost unaffected with minimal decrease, both O/Km and WT deteriorates sharply. As $\gamma$ increases, a vehicle would have edges to \emph{only} those orders that originate from a node in the same direction as the vehicle's destination. This leads to reduced chances of batching, and hence O/Km decreases. Without batching, the vehicle often arrives in the restaurant early and increases its waiting time. Furthermore, when the number of vehicles is less, reduced batching leads to higher number of order rejections (Fig.~\ref{fig:gammavsrejections}). Hence, $\gamma=0.5$ is our recommendation.
\vspace{-0.05in}
\section{Conclusion}
\label{sec:conclusion}
Delivery time is a key factor in both improving customer experience as well as increasing business productivity in the food delivery business.
Despite food-delivery being a multi-billion dollar industry, there has been little research on strategies aimed at minimizing this key metric. Our contribution, \fmplus, addresses this need. Our empirical evaluations establish that {\bf (1)}~\fmplus
has an order of magnitude smaller extra delivery time compared to Reyes\cite{mdrp}, and $30\%$ less than Greedy. {\bf (2)} \fmplus leads to higher number of orders delivered per km, while reducing waiting time at restaurants for delivery personnel by $\approx 40\%$. {\bf (3)} Besides, \fmplus scales to real-world workloads on large cities. At a broader level, the proposed work empowers food-delivery companies with an in-depth understanding of the various sub-problems such as \emph{batching} of orders into groups, coping with \emph{dynamic availability and positions} of delivery vehicles and the associated scalability challenges. We hope \fmplus will provide delivery services with a platform that may be experimented upon with live data and further improved with insights obtained from its performance in the real world.
\vspace{-0.05in}
\balance

\end{document}